# Blockchain and Beyond: Understanding Blockchains through Prototypes and Public Engagement

DAVE MURRAY-RUST, Human-Centered Design, TU Delft, Netherlands and Institute for Design Informatics, University of Edinburgh, Scotland
CHRIS ELSDEN, BETTINA NISSEN, ELLA TALLYN, LARISSA PSCHETZ, and CHRIS SPEED, Institute for Design Informatics, University of Edinburgh, Scotland

This paper presents an annotated portfolio of projects that seek to understand and communicate the social and societal implications of blockchains, DLTs and smart contracts. These complex technologies rely on human and technical factors to deliver cryptocurrencies, shared computation and trustless protocols but have a secondary benefit in providing a moment to re-think many aspects of society, and imagine alternative possibilities. The projects use design and HCI methods to relate blockchains to a range of topics, including global supply chains, delivery infrastructure, smart grids, volunteering and charitable giving, through engaging publics, exploring ideas and speculating on possible futures. Based on an extensive annotated portfolio we draw out learning for the design of blockchain systems, broadening participation and surfacing questions around imaginaries, social implications and engagement with new technology. This paints a comprehensive picture of how HCI and design can shape understandings of the future of complex technologies.

CCS Concepts: • **Human-centered computing → Contextual design**; *Empirical studies in HCI*; • **Security and privacy** → *Human and societal aspects of security and privacy*; • **Applied computing** → *Digital cash.*

Additional Key Words and Phrases: blockchains, public engagement, research through design, HCI, smart contracts, distributed ledger technology, conditional giving, distributed energy

## 1 INTRODUCTION

Blockchain systems are complex. As an example, the Ethereum network [151] includes a digital ledger of transactions and other features that enable a cryptocurrency, a Turing complete programming language that runs on top of this, with its own currency, all supported by a system of miners who turn computational power into cryptocurrency, and a secondary network of organisations that help move money into and out of the network. This is a mixture of human and technological actors, with multiple, conflicting goals and reasons for participating. To make sense of the whole system requires a combination of specialised cryptographic reasoning, distributed systems thinking and

Authors' addresses: Dave Murray-Rust, d.s.murray-rust@tudelft.nl, Human-Centered Design, TU Delft, Delft, Netherlands, Institute for Design Informatics, University of Edinburgh, 47 Potterrow, Edinburgh, Scotland, EH8 9BT; Chris Elsden; Bettina Nissen; Ella Tallyn; Larissa Pschetz; Chris Speed, c.speed@ed.ac.uk, Institute for Design Informatics, University of Edinburgh, Design Informatics, 47 Potterrow, Edinburgh, Scotland, EH8 9BT.

game theory, along with a socioeconomic treatment of why and how people participate in the system, and what the potential real-world applications are. This combined viewpoint is not accessible to many people, hence multiple different understandings of the blockchain proliferate, covering different aspects. However, these systems are the subject of great hopes for radical transformation. The UK Government's Walport Report states:

> "In distributed ledger technology, we may be witnessing one of those potential explosions of creative potential that catalyse exceptional levels of innovation. The technology could prove to have the capacity to deliver a new kind of trust to a wide range of services. As we have seen open data revolutionise the citizen's relationship with the state, so may the visibility in these technologies reform our financial markets, supply chains, consumer and business-to-business services, and publicly-held registers."[147]

The current focus of much research steers towards scientific or technical solutions and systems, and the algorithmic or game theoretic functioning of consensus algorithms. Alongside this, in order to develop a clear picture of how, when and why to employ this technology, it is necessary to look at the broader context in which blockchains may be used and understood. In particular, we are interested in how we can develop public understanding of the technologies, partly in their functions, but also in their implications - what are the possibilities for a world in which these decentralised, 'trustless' systems are common? This fits with the UK government's Holmes report:

> "[Distributed Ledger Technology] presents us with an opportunity not just to consider how we might make what government currently does better, but to rethink what government can and should be doing to promote democratic engagement and the welfare of UK citizens and to stimulate and strengthen the UK economy" [68]

This sense of possibility is driven by the way that the technical potentials of blockchain systems are understood: as ways to decentralise systems, empowering people and enhancing their agency; to democratise transactions through disintermediation; to support artists by tracking sales and re-use; to develop increasingly autonomous objects that can make their own transactions; and ultimately to completely rewrite legal and social structures as formal code.

The set of speculations is broad, and many of the promises made are impossible, undesirable or both. There is a need to engage publics with these propositions, to understand and develop the new economic imaginaries [130] and socio-technical possibilities currently being thought through:

> "As with most new technologies, the full extent of future uses and abuses is only visible dimly. And in the case of every new technology the question is not whether the technology is 'in and of itself' a good thing or a bad thing. The questions are: what application of the technology? For what purpose? And applied in what way and with what safeguards?" [Walport, 147]

In this paper, we are concerned with how we can use design, with its associated toolkits of research and public engagement methods as a way to both articulate to the public some of the meanings and possibilities of this new technology, and to co-create innovation and understanding with the practitioners. We believe the interdisciplinary nature of design and HCI research offers a lens to hold new technologies to account, engage participants, and expand imagination [40]. Just as blockchain is allowing the rethinking of existing practices, we are interested in how design helps researchers and publics alike to rethink their conceptualisations (c.f. Barry et. al's logic of ontology [4]) and create space for deeper ideation and engagement. To develop this, we engage in experiential work, that brings people into the system in order to engage with knowledge. As Barad puts it: *"We do not obtain knowledge by standing outside the world; we know because "we" are of the world. We are part of the world in its differential becoming."* [3]. Through the creation of new experiences through which stakeholders can meaningfully engage, design gives a tool to navigate the space between technological hype and froth on one hand, and dystopian fears on the other.

The key contribution of this paper is bringing together and analysing a collection of work from the Centre for Design Informatics in Edinburgh University, covering projects from 2014 to the current day. This work is brought together as an *annotated portfolio* [13, 58], from which a range of themes are drawn out and analysed. This approach is warranted here due to the complexity of blockchain systems – carrying out design work in the space involves dealing with value exchange, cryptographic security, decentralised and distributed systems, network infrastructures and more. This portfolio includes work based on a range of different design methods from probes to design fictions and group workshops, with different manifestations such as artefacts, prototypes and games, spanning a range of levels of technical fidelity. Some are small prototypes, that pick up certain aspects of the technology, and articulate them through provocative objects. Some are larger studies about the relations between people and blockchains that are manifested as interactive experiences and physical things, while others are public engagement workshops, that try to translate the complex infrastructures into forms that encourage public participation.

Combining multiple projects allows for an analysis of the factors that make experiences work, including i) traditional designerly concerns such as tangibility or transparency, the use of role-play and seamfullnes; ii) specific concerns for working with blockchains around fidelity, abstraction and contextualisation; and iii) broader concerns around value exchange and re-imagining society. This allows us to paint a broad picture of the interconnected issues and concerns that arise around complex socio-technical infrastructures in a way that is not possible with individual pieces of work. We demonstrate the richness, complexity and diversity of knowledge that can be gained by applying a range of approaches to a single technological locus.

In particular, this allows us to engage with four main questions that run through the work:

- How to go about the design of blockchain systems, in particular, connecting them to real-world contexts of use and interactions with humans?
- How to involve a wider audience in meaningful critique, ideation and design of blockchain technologies?
- What aspects of blockchain technology can be generalised to support the design of other network infrastructures?
- How can design work around blockchains extend our envisioning of technological and social futures?

The paper is structured as follows: First, we introduce a short survey of relevant work exploring implications of blockchains through social studies and creative practice (Section 2). Next, we introduce the setup of our annotated portfolio and the key design methods used across the projects (Section 3). We then present the portfolio as four groups of related projects, covering i) understanding blockchains (Section 4); ii) public engagement around blockchains (Section 5); iii) co-creation of system rules (Section 6); and iv) the development of autonomous objects (Section 7). For each group of projects, we detail their context and the questions they set out to explore, and then draw out themes and understandings gained through the projects. This is then brought together into a discussion (Section 8) along the four questions given above, looking towards rethinking value and values, developing autonomous systems, understanding the imaginaries surrounding blockchain and the move towards automated societies.

Due to the length and format of the paper, we suggest some reading journeys for exploring different aspects of the work, as it is not necessary to read it as a complete text. We would suggest:

- For those new to Blockchain and HCI, the background and related work (Section 2) provides an overview to the core technologies and methods, which should enable an exploration of individual projects of interest.
- For HCI researchers looking to engage with blockchains and DLTs, the text that accompanies each block of the portfolio (e.g. Sections 4.1-4.7) surfaces key HCI issues alongside the projects where they emerged.
- For engineers and creators of blockchain systems, the discussion (Section 8) and in particular the sections on learning about design around blockchain systems (Sections 8.1 and 8.2) give an overview of key strategies and point back to the projects of interest.

## 2 BACKGROUND AND RELATED WORK

To set the scene for our annotated portfolio, we start by introducing some key terms, features and debates regarding blockchain technologies and decentralisation which have set the scene for our work since 2014. We then give a brief survey of the wide range of artistic and design-led practice which have expanded the understanding and imaginaries of these technologies and infrastructures. Finally, and crucially, we identify specific areas of prior work in HCI and design that this portfolio builds upon and extends.

### 2.1 Introducing Blockchains and DLT

As a class of technologies, blockchains and DLTs have been extensively analysed and summarised in a range of disciplines, both technically, and as ‘disruptive’ ‘general purpose’ technologies which are envisaged to transform whole industries [29, 80]. As a very brief sample, Swan [133] provides one of the earliest overviews of how the technology that emerged as a cryptocurrency in 2009 through the Bitcoin Blockchain [107], could become the basis for all manner of contracts, and ultimately a source of distributed governance. Swanson [134] details key distinctions between permissioned and permissionless ledgers. Tschorsch and Scheuermannn have conducted a technical survey of decentralized digital currencies [142] while Garay et al. [57] provide a canonical analysis of the Bitcoin protocol. As a less technical, but more comprehensive general overview, Rauchs et al. [120] offer a conceptual framework of Distributed Ledger Technologies in an effort to refine terminology and bring order to a vastly expanding field of applications with quite varying qualities. Finally, there are also a plethora of domain-specific summaries, such as Dunphy et al.’s overview of blockchain applications for identity management [35].

It is a challenge to give a concise and complete account of a blockchain system, and the gap between ‘a blockchain is just a database that can only be appended to’ and a full technical knowledge of the details is large. The Ethereum network [151] consists of: i) a distributed ledger that records transactions immutably; ii) a consensus algorithm for determining which transactions should be honoured; iii) a Turing complete programming language for creating programs to be executed in a distributed manner; iv) a cryptocurrency that makes use of the distributed ledger, with a secondary currency to manage execution of distributed programs; v) a collection of miners, who enable the network by contributing processing power; and more. In order to orient the reader to some of the key features of blockchain technologies, we provide an overview diagram and glossary of key terms in Appendix 10, but the takeaway here is that the system is composed of many different actors and technologies, operating at different levels with different logics.

Beyond these more technical definitions, we wish to briefly address some of the significant debates and imaginaries that have fuelled research, hype and discussion of blockchain technologies. It should also be noted that this work took place before the explosion and crash in the price of Bitcoin in 2020/2021, the growing awareness of the ecological impact of proof-of-work mining, or rise of 'NFTs' with Beeple selling an artwork for $69M and Tim Berners-Lee selling an imaging of the world wide web source code for $5.4M[82, 144].

Most evidently, the emergence of Bitcoin in the wake of the 2008 financial crisis has inspired waves of envisioning alternative digital currencies and the future of the global financial system. Kow and Lustig [88] recount competing visions within the Bitcoin community about how to ensure the sustainability, prevalence and accessibility of the network. Through eight 'MoneyLab' conferences, the Institute of Network Cultures has produced two compelling 'readers' [63, 93] charting the vast possibilities of cryptocurrencies, from Bitcoin Maximalism (where Bitcoin ultimately replaces fiat currency), to 'common-coins' for communities of shared values such as FairCoin [1], and more corporate visions such as Facebook's 'Libra' network [2].

Besides currency applications specifically, DLTs have spurred the conception of various 'token economies', where actions and behaviours on a platform are regulated and incentivised by a token or coin, with various embedded rules and possible values. In the first instance, token economies offered a vehicle for initial investment in blockchain-based start-ups through controversial 'Initial Coin Offerings' (ICOs). However, platforms such as Brave [3] demonstrate how token economies might be configured to revolutionise internet publishing. Such visions are compelling, but also speak to interactions that are deeply 'financialised' [143] and premised on individuals acting in rational and economic self-interest.

A striking and confounding feature of blockchain technologies is how they simultaneously appeal to deeply libertarian ideals [81] of independence from state governance, as well as more socialist envisioning of equitably shared commons [25, 100, 113]. Similarly, there are deep ideological tensions between wholly independent and 'permissionless' networks developed outside of institutions and state regulation (e.g. Bitcoin) and corporate-led initiatives where the technology is used to agree or enforce industry standards (e.g. IBM Hyperledger, Consensys).

Distributed Autonomous Organsiations (DAOs) are an interesting example of this: autonomous code that can carry out financial trades or other actions without the need for or possibility of human intervention. One of the highest profile of these, simply named "The DAO" gathered up $250M of Ethereum tokens, 14% of the entire supply, but was quickly exploited, and the money stolen. This resulted in a 'hard fork', where a consensus emerged and participants in the Ethereum network collectively decided to 'roll back' all transactions and erase this bit of history from the supposedly immutable ledger [37].

The potential use of blockchain technologies to manage personal identity data [34, 35] demonstrates some of these tensions. In opposition to nation states, or global technology companies, some look to the distributed consensus mechanisms of blockchain technologies as a platform for 'self-sovereign identity' – where one's identity is issued and controlled only by the individual themselves, without relying on an external authority to prove this identity. These visions dovetail with aspirations for a more open, peer-to-peer and distributed web [9, 125, 140]. Ideologically, one's identity can therefore never be taken away or discredited by an authority. However, other decentralised identity schemes take a different path; they seek to make use of distributed ledgers as a 'tamper-resistant' records, through which trusted authorities can share their affirmation or 'attestation' of different identity attributes. While both systems afford the individual a more portable record of their identity, this second system is heavily predicated on the sharing and reinforcement of specific identity standards within existing regulatory frameworks. A particularly striking example includes the World Food Program's 'Building Blocks' program which encodes the identity of refugees in camps through a retina scan in order to facilitate cash assistance [22].

[1] https://fair-coin.org/
[2] https://libra.org/en-US/white-paper/
[3] https://brave.com/

Similar aspirations for 'proof-as-a-service' and the trusted exchange of credentials extend to registries of goods and global supply chains. It is envisaged that through chains of tamper-resistant attestation a provenance of goods could be achieved and bring greater transparency. However, while blockchain protocols may ensure that data shared 'on-chain' is transparent and perpetual, these systems ultimately depend on faithful representation of real-world 'off-chain' assets. Somewhat paradoxically for a supposedly 'trustless' technology, a deep trust is required across institutions to make such infrastructure work.

These are only brief examples of complex debates, however they serve here to illuminate paradoxical and contested character of blockchain technologies and their applications. This ambiguity is only heightened by the abundance of visionary 'White Papers' in contrast to the sparsity of widely used applications. More than 10 years on from the emergence of Bitcoin, and despite vast investments, the industry arguably remains embryonic. And while a great deal of critical work in STS, political economy and infrastructure studies have analysed the prospects of these technologies [30, 70, 76, 88, 90, 100, 113, 152], there is still much to understand about how they might be designed and experienced by end-users and the implications of their widespread adoption in everyday life.

### 2.2 Expanding Public Imaginaries of Blockchains and DLT

Taken together – much of this prior work reflects the very broad and systemic thinking that underpins the emergence of this technology and the grapples in various domains to contemplate the applications of such a disruptive technology. However, where there is far less prior work is in the ways people will actually interact with and experience these technologies. The sector is replete with visions, white papers and illustrative product videos; working examples and meaningful real-world interactions (especially beyond cryptocurrency trading) are considerably more sparse.

Artists have often gone further in this regard. In 'Artists Re:Thinking the Blockchain', the artists collective Furtherfield write:

> "There is a curious equivalence between art's speculative abilities, to play with fact, fiction, and abstraction, and the blockchain's own chimeric character. Both art and the blockchain grapple with the instability of authorship and authenticity." [18]

Indeed, it is striking how much more 'real' and tangible some artistic projects have been, than many aspiring blockchain start-ups. Furtherfield's edited collection [18] demonstrates artists natural desire to make and craft in order to understand new technologies. In doing so, many projects invite attention to specific features of the technology, and offer entry points for a general public to grasp their significance.

Several projects focus on the nature of 'mining', and specifically the extreme energy costs involved in doing so: Lindley's 'CryptoHeater' [92] attached a mining rig to a radiator to generate heat; Oliver's 'Harvest' [4] brought together wind turbines and miners to create Zcash from atmospheric movement; and Bittercoin uses a calculator in order to be the worlds worst miner [5].

Projects have also drawn attention to the human roles in what are often perceived as purely technical infrastructures: Dovey's 'Respiratory Mining' [6] uses crypto-currencies to investigate the role of the body in emerging financial systems and how the body can perform computational processes by using human respiration to mine crypto-currencies; Smith's BlockMirror [7] combines the viewer's reflection with the act of mining currency to illustrate our value potential within this system. It hints at a possible future where all aspects of our lives and our attention are commodified and mined for currency.

[4]https://julianoliver.com/output/harvest
[5]https://escuderoandaluz.com/2016/03/03/bittercoin/
[6]https://maxdovey.hashbase.io/Respiratory_Mining/

Moving away from explicit critique of the technology itself, some of the most provocative artist work in this space has explored possible economic reconfigurations. Furtherfield's 'Artists Re:thinking the Blockchain' collects multiple approaches including art and speculative design [18], in particular the Plantoid - a blockchain based lifeform [50] and Terra0, a forest that is attempting to buy and own itself [127]. This sense of autonomy and ownership is also seen in more plausible systems, such as the Fairbike hire bikes that own themselves and commission more as needs arise [97]. The DAOWO workshops have provided a specific venue for exploring the ways that blockchains and critical artistic practice reflect and shape each other[8].

While many of these projects are of artistic or speculative nature, remaining abstract, provocative concepts, they all powerfully reflect and demonstrate the rich imaginaries that circulate through blockchain technologies. These particular artistic instantiations, cut through purely theoretical or technical imaginaries, and present compelling propositions. However, these provocative projects only focus to a limited degree on engaging publics and exploring their actual interactions and experience with blockchain technologies.

The work we present in this annotated portfolio is deeply inspired (and has sometimes been undertaken in collaboration) with many of these artists. However, in our work we have prioritised the practices of design and HCI research alongside public engagement. In this sense, our research draws richly from artists' approaches, but crucially seeks to investigate and test out many of the underpinning theories and technologies in these works.

### 2.3 HCI and Designing Interactions with Blockchain Technologies

The HCI community has approached studies of blockchain technologies in a number of ways. One strand of work on Bitcoin and alternative currencies taps into a rich history of studying money, finance and peer-to-peer exchange [7, 15, 16, 19, 48, 49, 86, 87, 89, 114, e.g.].

Other work has looked more specifically at the Bitcoin community itself, with particular interests in infrastructuring [88]. In particular, this work demonstrates the material and social factors shaping the development, use and implications of what is often prefigured as a purely technical intervention [76, 81, 86, 95]. In this respect, the HCI community, as ever, plays a role in articulating the importance of the human in the loop.

HCI and particularly design-led researchers have also sought to understand the opportunities and implications of blockchain technologies in specific domains, such as: education [123], charity and philanthropy [44, 45, 102], identity management [23, 34, 35, 154], supply chains and transport [51, 77, 78, 115, 136], shared commons and civic participation [25, 39, 96].

With a focus on particular applications and communities of use, this work has also tended to adopt more participatory, creative and bottom-up approaches that endeavour to involve end-users in understanding and informing the design of this new class of technologies, such as performative works and design fictions [e.g. 83]. Such approaches are a stark counterpoint to the often far-removed and top-down envisioning prevalent in most blockchain start-ups.

Through a survey and typology of blockchain application areas for HCI researchers, Elsden et al. [40] also propose the roles that HCI researchers may be well placed to play in advancing the field. They propose the HCI community should hold these technologies to account, engage participants around the technology, bring critical design knowledge and practice to bear on blockchain systems, and expand the imaginaries around blockchains.

[7]http://www.dominicsmith.info/the-block-mirror/
[8]http://www.daowo.org/

This is a wide brief - from technical understanding, to design and theoretical imaginaries, to engaging with participants and end users on the ground. The diversity of such approaches was embodied in a CHI 2018 workshop, on 'HCI for Blockchains', captured in four manifestos [42].

In presenting this annotated portfolio, we endeavour to demonstrate a range of projects that do one or all of these things. The projects we discuss provide exemplars for the HCI community, and by looking across a number of projects, we show several different approaches, as well as accumulated learning that can guide HCI and design researchers in navigating research projects on these complex technologies.. Our work also cuts across a number of the domains above - and we see that although specific domains differ, similar questions and challenges emerge. By looking at all of the work as a whole, we can articulate the design strategies that work well for engaging participants, the particularities of designing around blockchains, and synthesise a picture of the social impacts of distributed ledger technologies.

## 3 AN ANNOTATED PORTFOLIO OF BLOCKCHAIN IMAGINARIES

The projects that we discuss here have emerged from a series of collaborations with partners and organisations from both industry and academia including experts in computer science and cryptography as well as fields of HCI and business and organisational studies. The central strand is understanding the relations between publics and blockchain technologies - starting from a core understanding of how blockchain technologies work, but working towards a concern for social worlds, how these technologies will manifest and be experienced in everyday life, engaging with "the potential for surprise, imagination, and creativity, which is immanent in the openness of each moment of experience" [153, p.184]. As such, a primary goal of our work is to engage various publics and offer non-experts the opportunity to experience, ideate and think creatively in relation to blockchain technologies.

The way of working here is broadly within the domain of*research through design* (RtD) a design-led approach, which generates knowledge by a design-led approach, where knowledge is generated through all stages of a design process, from initial ideation, through to the deployment of functioning prototypes, probes and research products [55, 60]. Our participatory approach to RtD particularly emphasises engaging with a wide variety of stakeholders and publics, alongside the more traditional RtD focus on materials and objects.

This approach is used within studies of human computer interaction (HCI) [155], especially when looking at work 'in the wild' – aiming to study technology use in real-world contexts with likely end-users [8, 122]. The type of research is often iterative in nature; active, participatory, playful and performative [60] and its main aim is to allow publics to experience potential novel technologies and their implications through experiential and embodied rather than purely academic or technical ways, as Frayling put it: "How can I tell what I think till I see what I make and do?"[55]. As researchers and practitioners who are focusing on engaging multiple audiences with concepts of blockchain technology, design and creativity play an important role in creating engaging experiences for our participants. And we as designers and HCI researchers agree that we should "take pride in [design's] aptitude for exploring and speculating, particularizing and diversifying, and – especially – its ability to manifest the results in the form of new, conceptually rich artefacts" [60].

There is a key challenge in presenting this kind of work as much of the value is embodied in specific objects and situations. Many design researchers and academics have grappled with the disconnect of experiential depth and its translation into written 'knowledge'[60, 131], e.g. exploring other forms of media [69, 94, 146]. One approach emerging from these debates, the *annotated portfolios* introduced by [13] and [58] aims to present practice as a multi-layered, cumulative collection of research artefacts to draw out shared concepts in a "descriptive, yet generative and inspirational fashion" [13] that go beyond individual project findings. This allows a combination of empirical studies, observational approaches and critical design work to be brought together – a necessary practice to cover the combination of formal concepts, behaviour understandings and speculation necessary to engage with emerging infrastructures such as blockchains:

> "If a single design occupies a point in design space, a collection of designs by the same or associated designers – a portfolio – establishes an area in that space. Comparing different individual items can make clear a domain of design, its relevant dimensions, and the designer's opinion about the relevant places and configurations to adopt on those dimensions." [58]

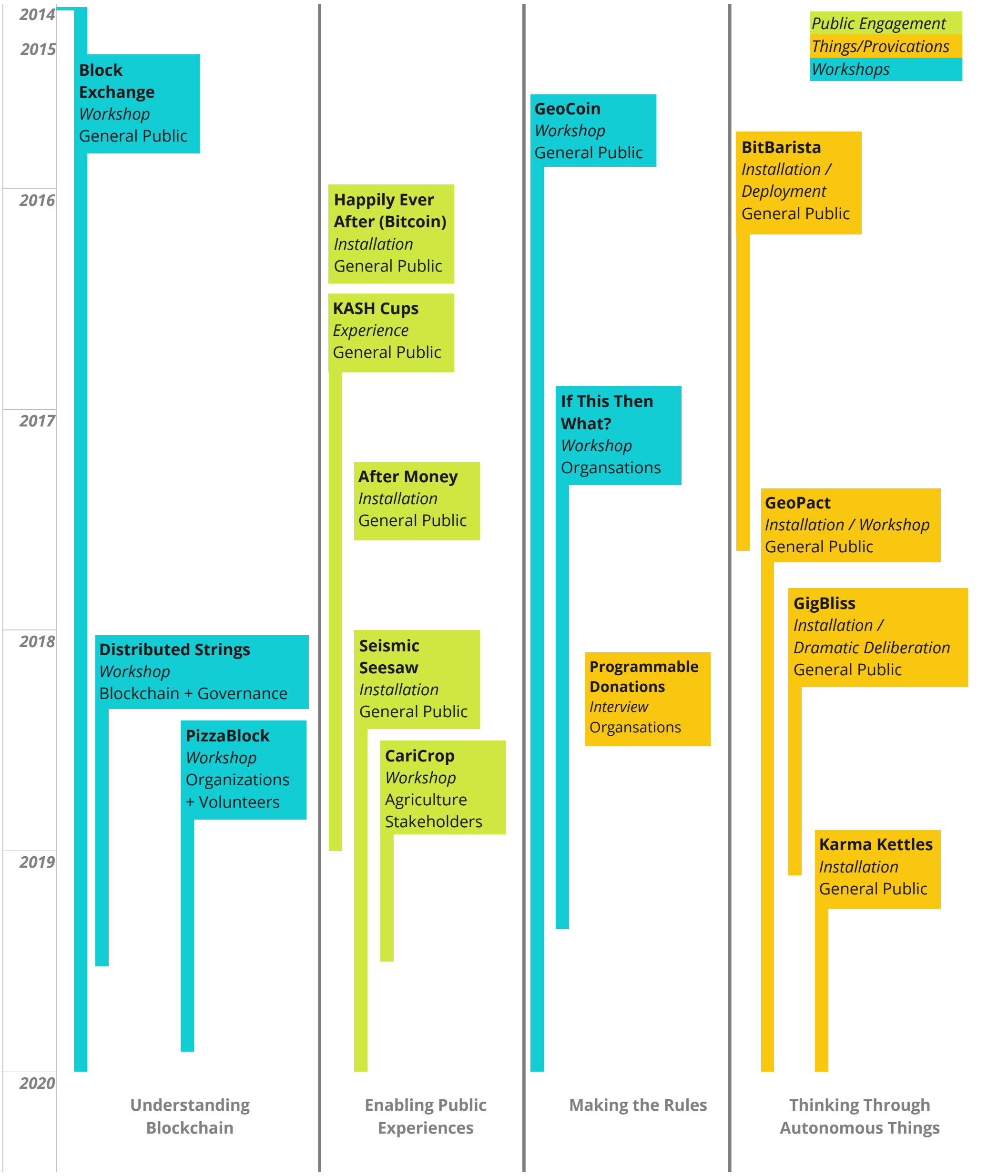

Fig. 1. Timeline of projects in this portfolio, categorised by the type of activity - public engagement, provocations and workshops.

In order to work with different communities, situations and research directions, a wide variety of design-led approaches are deployed. The common thread is engaging people with these new and complex technologies in accessible and experiential ways, looking beyond current technical capabilities or issues (e.g. scaling, mining costs etc.) and critically reflecting on the broader socio-material practices, societal impact and meaningful future use of blockchain technologies. As such, we have been inspired by a range of methods, while at the same time responding to contexts, partners, aims and audiences of each project. Our key methods can be roughly summarised as follows:

- Ideation and creative thinking, where we stimulate divergent thinking in new areas by using ideation cards to suggest possibilities [64], bodystorming [111] for situated ideation, and unfinished software [108] to support participants developing their own ideas about novel infrastructure.
- Material interventions that use a variety of objects and tasks in order to understand participants sociologies, including technology probes [72], design probes [145] and cultural probes [59].
- Provocations that help to generate new ideas by finding ways to articulate new concepts or defamiliarise existing ones, such as physicalising data [79] or other concepts, or using provotypes – provocative prototypes – to elicit responses [11].
- Speculative works, whether design fictions or speculative designs, that paint pictures of possible worlds in order to support new ways of thinking [2, 10, 33, 38].

To help make sense of the multiple facets of this work, we have structured the portfolio in four sections, which draw out different concerns, with a narrative thread through them. The work roughly follows a trajectory from simple through to complex, and from didactic to open-ended. We have purposefully not followed a chronological order here (although this can be seen in Figure 1) to draw out shared commonalities and valuable narratives for HCI.

The projects (Figure 1) have been thematically grouped into four areas, each of which builds on the previous ones:

- Section 4 looks at workshops that develop public understanding of concepts behind blockchain technology, from basic concepts (BlockExchange) through abstract notions (Attaching Strings) to specific use cases (PizzaBlock and Caricrop).
- Section 5 looks at creating accessible public experiences that illustrate implications of blockchain developments, through rethinking marriages (Happily Ever After (Bitcoin)), work (After Money), donations (Seismic Seesaw) and socialising (KASH Cups).
- Section 6 explores how to engage people in creating the rules that systems work by, from abstract concepts (IFTTW) through how charities work (Programmable Donations) and creating geolocated currencies (GeoCoin)
- Section 7 develops the possibilities of autonomous objects, for location aware objects (GeoPact), understanding supply chains (BitBarista) and negotiating with energy grids (GigBliss, Karma Kettles).

To present the annotated portfolio, each project is described through a double page spread combining text and images – these have been made visually distinct from the body text for clarity. The analytic body text for each section sets up the context and intentions surrounding each group of projects, presents shared analysis over the projects, and motivates the next section. The first stage of analysis was carried out on individual projects, as those involved drew out key learnings and outcomes for that particular piece of work. These were then taken into discussion among the authors of this paper, to work towards a coherent set of themes for each group of projects. Based on these, as well as a holistic overview of the whole portfolio, the authors developed the final discussion and key points. This means that the paper can be easily skimmed, by flicking through the pictorial pages, or a particular group of projects can be explored in more depth.

## 4 UNDERSTANDING BLOCKCHAINS

Blockchains rely on complex cryptographic protocols, but also require systems-level thinking across economics, mathematics and distributed technologies. To explore the full potential of these technologies requires new ways of thinking, for example, considering radical new forms of governance and organisation, set in opposition to traditional infrastructures and institutions (e.g. the banking sector). This can lead to unearthing and reconsidering long held and forgotten assumptions about how whole sectors of societies and economies function, such as unravelling the tacit understanding of money and currency, revealing it as an abstract structure imposed on our innate understanding of value and the need to exchange things.

However, working through ideas such as Merkle trees and hash functions is challenging to many audiences, and is a barrier that limits who might be able to engage with the possibilities of these technologies. We require ways to explain core blockchain concepts in a way that supports non-experts in imagining future possibilities and implication without having to understand all of the implementation details. Design methods can help both with ways to develop the technical understanding and with the defamiliarisation [5] of existing practices, giving space to rethink them with new metaphors.

In this section, we look at projects that abstract and distill key aspects or qualities of distributed ledgers, whether simplified metaphors for how blockchains are built or materialising concepts such as self-sovereign identity.

Several of these were driven by working with non-academic partners in a wide range of domains, who were both excited, and sceptical of the hype surrounding blockchain technologies, and often began from a very limited conception of its potential scope and applications. The four workshops tackle different aspects of blockchain technologies in order to have relevance in different domains with different stakeholders. They are structured to allow a detailed exploration of the questions at hand, typically lasting 2-4 hours with a group of participants. As well as helping develop understanding, these workshops are designed to open up discussion around alternatives to a centralised status quo, supporting a grounded imagining of future possibilities.

**BlockExchange** (p. 13,14) starts with the fundamentals of making peer-to-peer transactions on a public ledger [101]. It is intended to provoke new thinking of value exchange and lead participants to consider what might happen if money is no longer the mainstay of value exchange. The workshop has now been run many times for a broad range of audiences and has been adapted to a variety of different contexts, including an online pack of workshop materials and instructions, that can be picked up and used by others.

**Attaching Strings to Distributed Systems** (p. 15) invites the exploration of distributed networks through material entanglements. Decentralised Autonomous Organisations (DAOs) are financial organisations composed entirely through code - typically smart contracts running on a blockchain. As such, they offer interesting possibilities around transparency and fairness, while raising challenges around relations, control and decision making [109, 110].

**PizzaBlock** (p. 17) explores how a distributed ledger, maintained by a decentralized network following a specific protocol, can support self-sovereign identities [119]. In most cases when we make claims about our identity (e.g. our age, our address, a qualification) we rely upon records generated and maintained by a third party, for example a government, bank or technology company. Self sovereign identity systems aim to carry out the same job without the need for a central authority. By using blockchain technology, this data can be made more tamper-proof, and potentially allow the individual greater control and portability of their data between multiple parties.

**CariCrop** (p. 19) was designed to critically understand how Distributed Ledger Technologies (DLTs) could be designed to support socio-economic development in rural communities particularly focusing on developing countries and the Caribbean region [115]. CariCrop envisioned the use of Blockchain to ensure that transactions would continue to take place within a trusted infrastructure while farmers waited for de facto payments to take place.

These four projects and workshop methods each tackle different aspects of blockchain technology and aim to make such complex and technical features more accessible and understandable to a range of stakeholders. The key aspects we have addressed here are concepts of peer to peer value exchange, publicly accountable ledgers, distributed network systems and self-sovereign identity management.

# BLOCK EXCHANGE

Using Lego and math problems to collaboratively simulate a blockchain and explore the possibilities of decentralised trading

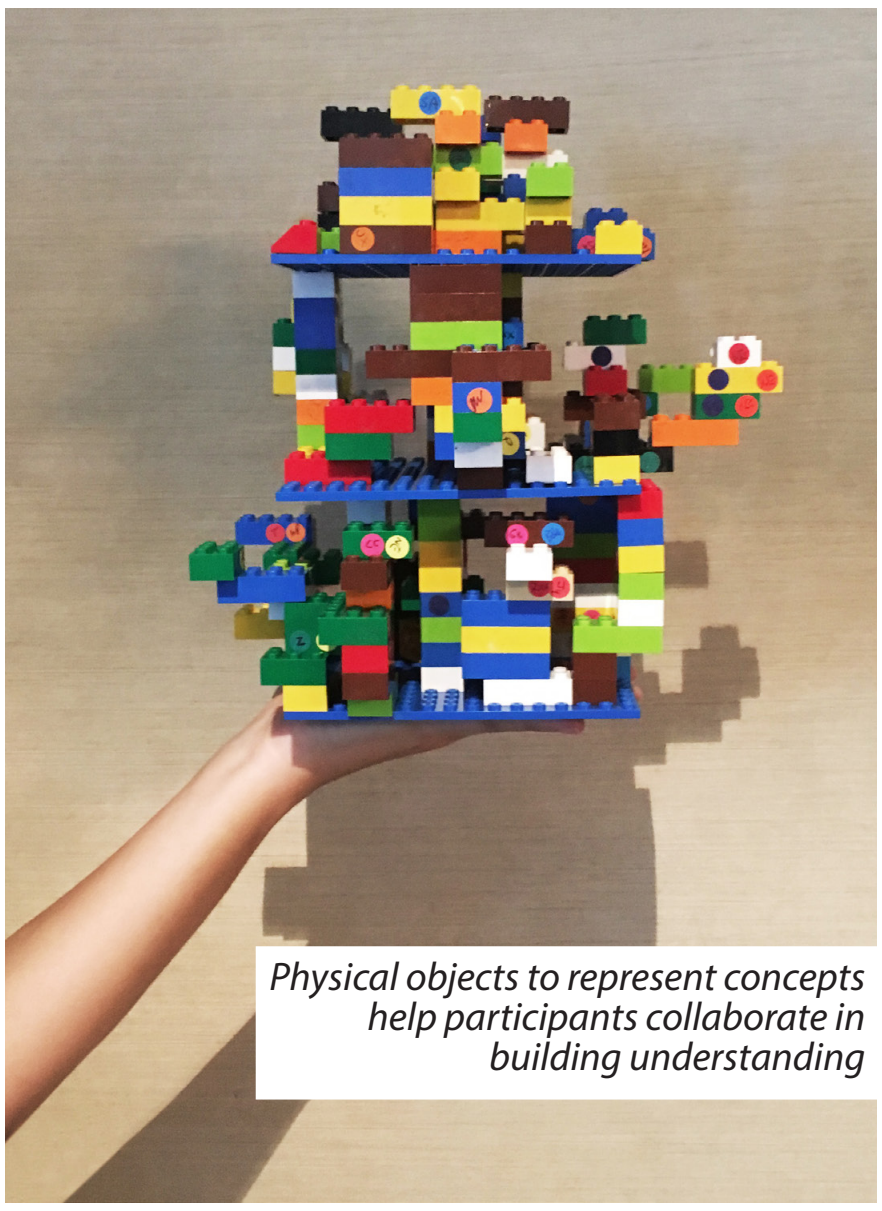

*Physical objects to represent concepts help participants collaborate in building understanding*

Lego stack representing the record of transactions on the Blockchain
Image: Bettina Nissen

**Tangibility**

**Abstraction & Focus**

**Value Exchange**

*Defamiliarising the concept of money...*

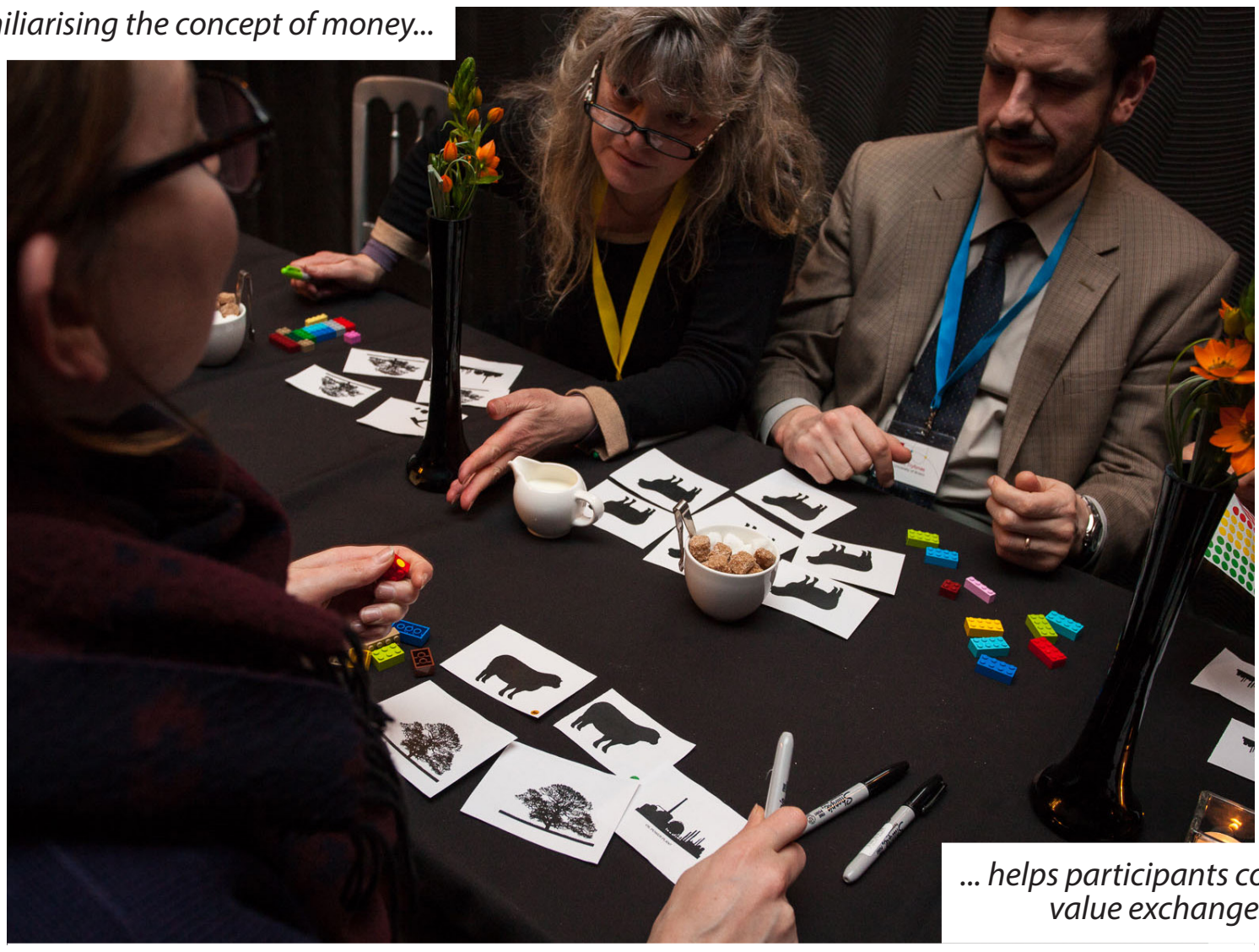

*... helps participants consider value exchange afresh*

"Block Exchange workshop in 2015 Collider. Image credit: Lindsay Perth'

# *Block Exchange - Details*

## Aims and Context

In the Block Exchange workshops, participants experiment with peer-to-peer trading using a Lego blockchain to record their trades in a game-like activity.

This allows them to explore some of the basic principles of how Blockchain works, but more importantly to anticipate the social, technical and economic opportunities that Distributed Ledger Technologies, such as Blockchain, may offer. It is intended to provoke new thinking of value exchange and lead participants to consider what might happen if money is no longer the mainstay of value exchange.

The workshop provides a light overview of the concepts as introduction to the technology and uses Lego bricks to represent intangible aspects of Blockchain mechanisms.

## Experience

Block Exchange is played over 3 rounds. Participants start by trading 'resource cards' such as water, wheat and oil, recording their transactions by labelleing Lego blocks with initialled stickers and fixing building them into a Blockchain. Alongside the trading, a group of participants act as miners who compete to solve a mathematical puzzle - once this is solved, the round of trading ends. In the second round, a scenario highlighting scarcity is introduced, leading to a change in the value of one or two of the resources. This introduces the concept of the market and how changing values affect desirability, moving away from the intrinsic value of material goods and gaining value becomes an end in itself. The third and final round has varied with different iterations of Block Exchange, but often ends with the resource cards being taken away and participants asked to trade anything they can think of, imaginary or real. This is intended to undermine the concept of the market that manifests in the second round and stimulate thinking on the nature of value in itself, leading participants to question what they really want and how they might find it through new models of peer-to-peer trade. Following this central activity with the Lego blocks, participants are asked to develop and produce novel products or services that use blockchain in new forms of value exchange. It culminates with a session on innovating and presenting new ideas and discussing potential opportunities.

# Attaching Strings to Distributed Systems

Using tangible materials to explore Decentralised Autonomous Organisations - physical networks to unpick complex abstract structures

**Tangibility**

**Visibility & Transparency**

**People and Machine**

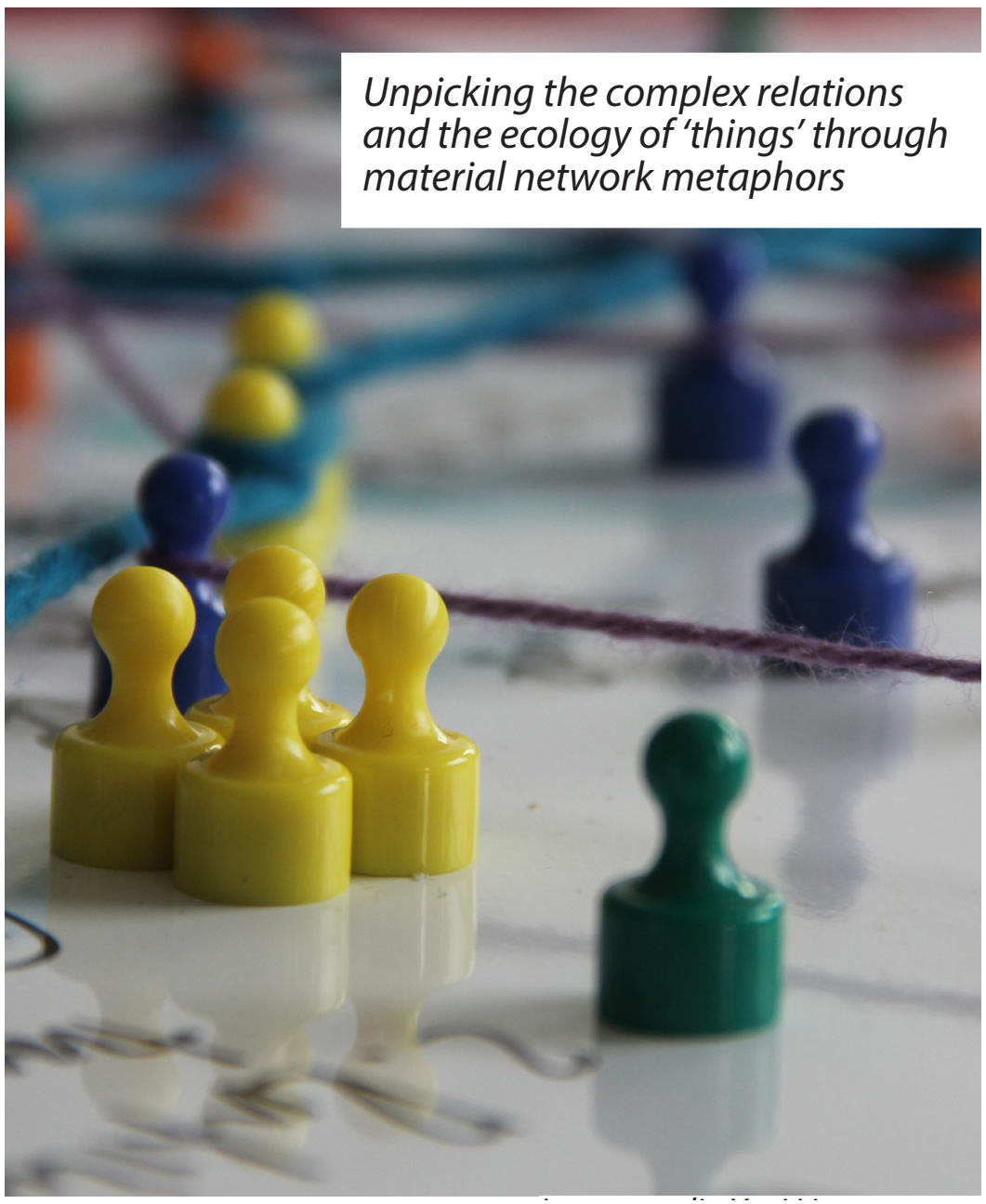

*Unpicking the complex relations and the ecology of 'things' through material network metaphors*

Image: Yuxi Liu

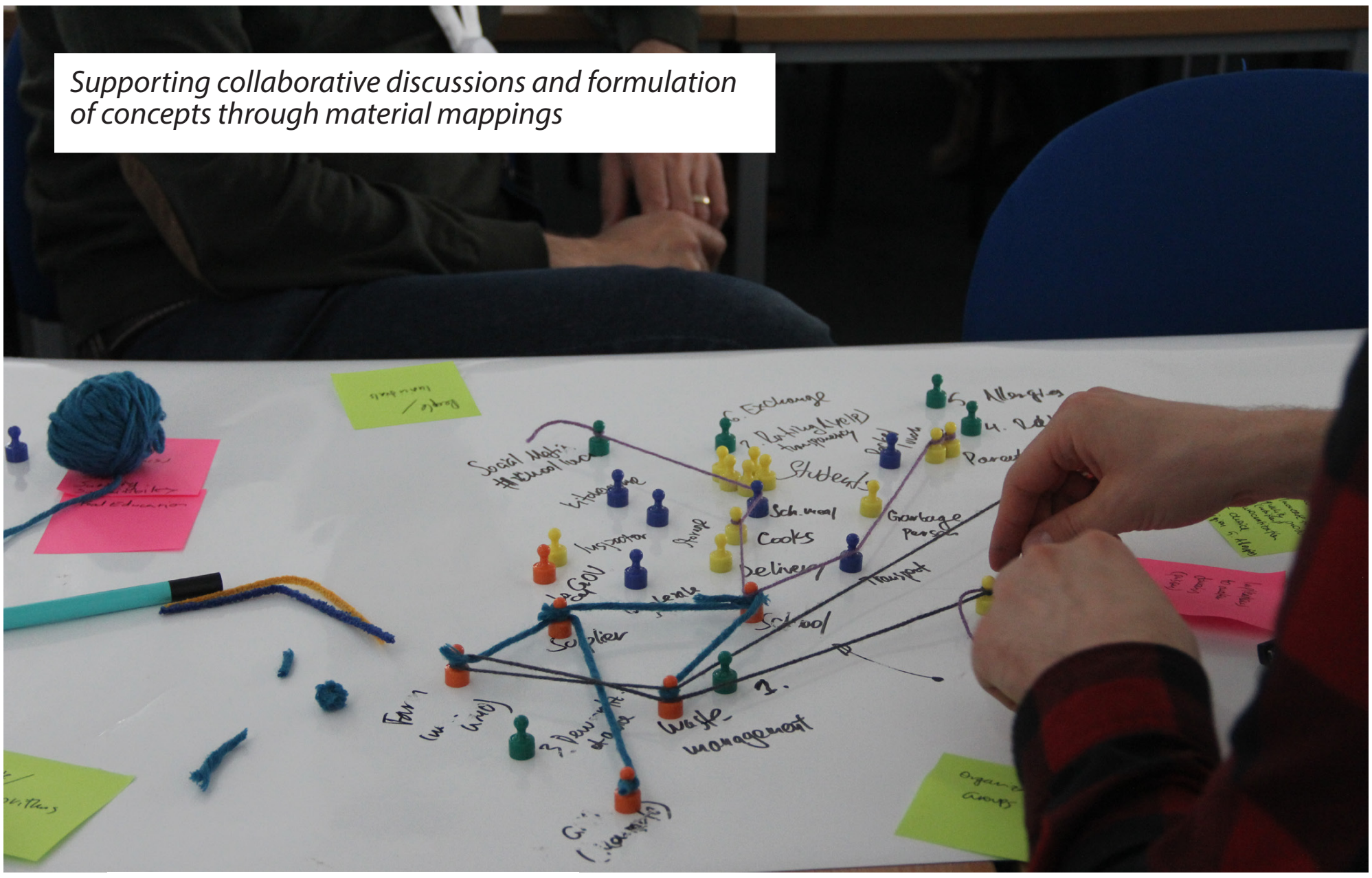

*Supporting collaborative discussions and formulation of concepts through material mappings*

Image: Yuxi Liu

# *Attaching Strings to Distributed Systems - Details*

## Aims and Context

Decentralised Autonomous Organisations (DAOs) are financial organisations composed entirely through code - typically smart contracts running on a blockchain. As such, they offer interesting possibilities around transparency and fairness, while raising challenges around control and decision making. This workshop activity explored the opportunities and challenges of distributed autonomous organisations (DAOs) through tangible materials. Participants collaboratively designed new DAOs by identifying key actors and their relationships. They used a range of materials to co-construct a tangible map of their new Distributed System, with strings and threads to represent the different connections. The central purpose of this process is not to create realistic maps of potential DAOs, but to structure and support critical debate within the group on issues of ownership, power, governance, agency, materiality and politics.

## Experience

The workshop begins with an initial discussion to understand current work relating to DAOs, in which potential application areas for DAOs are identified. Participants then form groups, select an application area, and work together to collaborative design a new DAO within that area. Participants are asked first to identify key actors in their DAO and choose materials and objects to represent them, and then to discuss the relationships between them and choose further materials to represent these connections, finally using these materials to construct a tangible map of the DAO. Throughout, the workshop focuses on three key topics to guide discussions:

- Developing notions of what a DAOs is, through thinking about how to design them, in order to designing and defining DAOs;
- Ownership, power and governance, in particular how existing structures may be challenged or reproduced;
- Agency, materiality and politics - how are the different components of a DAO connected, and what power do they have?

At the end of the design process, each group presents their concepts to the workshop as a whole, using their maps, and the workshop concludes with a discussion to draw together new thinking around implications and considerations for the design of DAOs.

# PIZZABLOCK

Creating great pizza as a way to understand self-soverign identity management

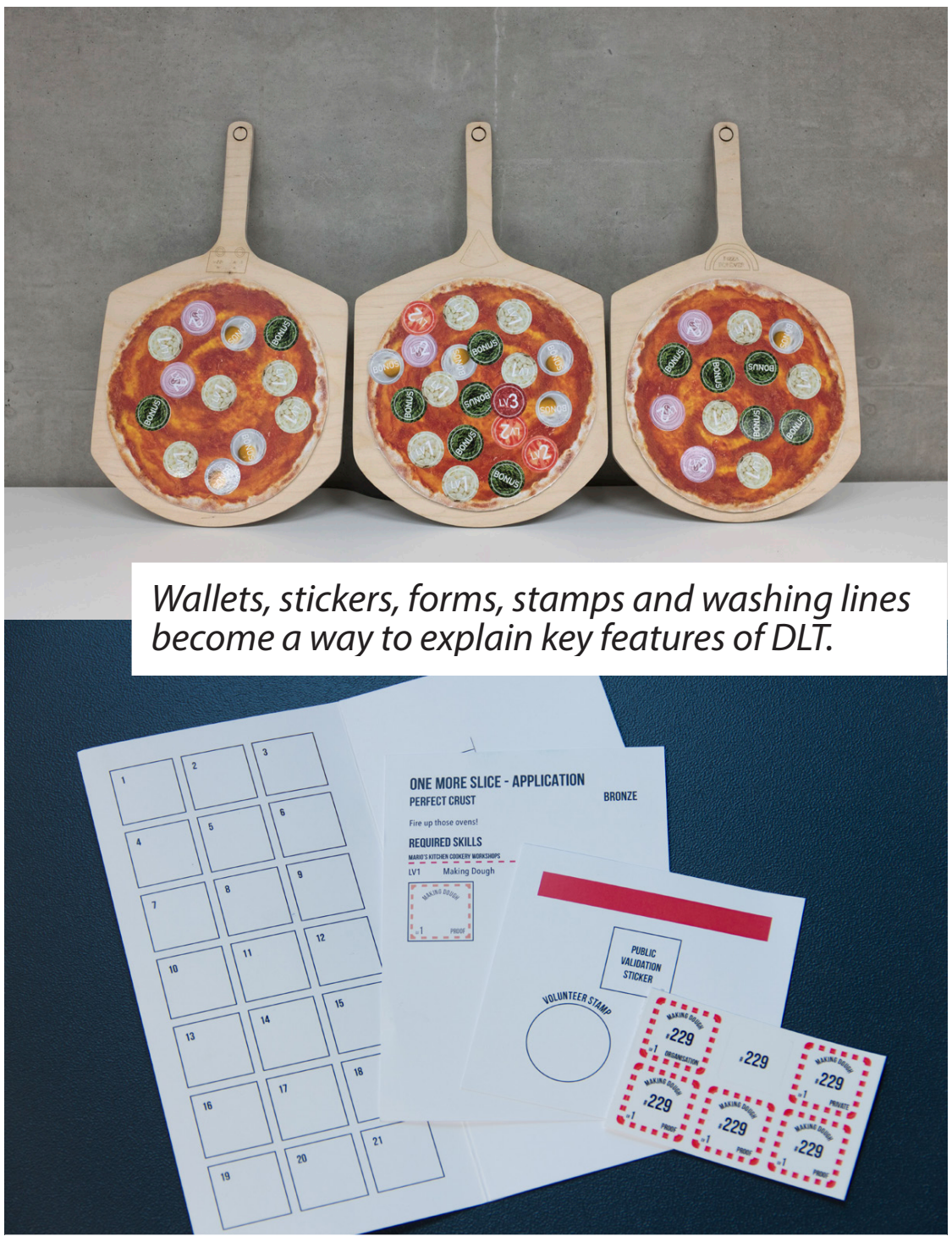

*Wallets, stickers, forms, stamps and washing lines become a way to explain key features of DLT.*

Top: PizzaBlock paddles used to keep score during the game. Bottom: Wallets, forms, receipts and stickker sets from PizzaBlock . Image credit: Jonathan Rankin.

**Roleplay & Collaboration**

**Fidelity of Technology**

**Rethinking Society**

*Roleplay demonstrates the distributed labour required to run a DLT*

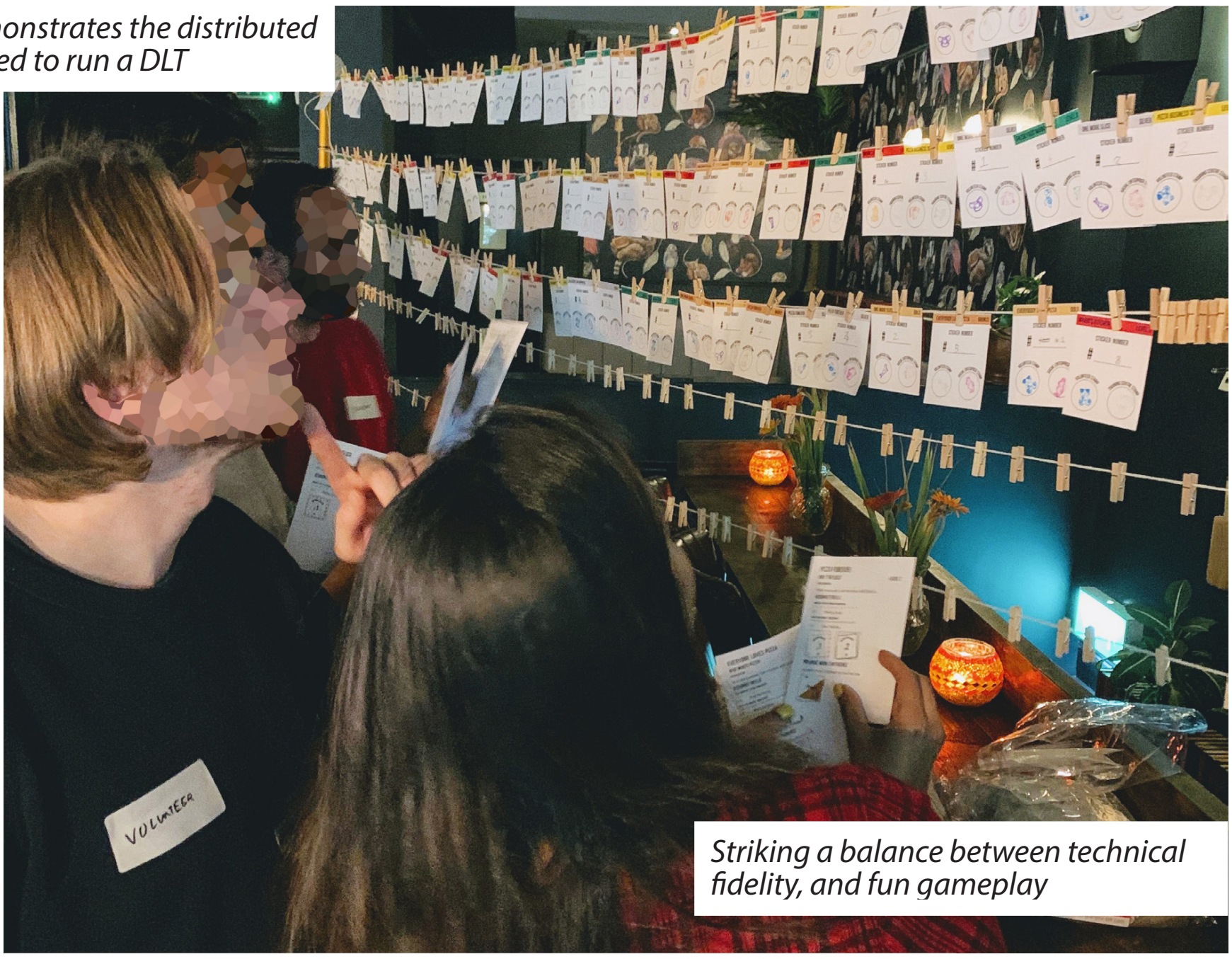

*Striking a balance between technical fidelity, and fun gameplay*

‘Participants using the ‘public ledger’ in a Pizzablock game in a London bar. Image Credit: Andy

# *PizzaBlock - Details*

## Aims and Context

PizzaBlock is a workshop and collaborative game to explore a specific set of blockchain applications related to decentralized and 'self-sovereign' identity management. In most cases when we make claims about our identity (e.g. our age, our address, a qualification) we rely upon records generated and maintained by a third party, for example a government, bank or technology company. Self sovereign identity systems aim to carry out the same job without the need for a central authority. Participants are given a social mission to 'improve the local pizza scene by working as volunteers, training centres and social enterprises. Using a range of bespoke artefacts, they physically enact all of the steps required to record and verify important transactions with each other through a distributed ledger. By playing PizzaBlock we introduce between 10-20 participants to decentralised identity management systems and produce a number of physical artefacts which relate back to the core features of DLTs for identity management.

## Experience

Here, Social Enterprises have a mission to improve Edinburgh's lack of good pizza, but they must find Volunteers, who have the right skills for each task. Volunteers earn these skills from Training Centres. By playing PizzaBlock we introduce between 10-20 participants to decentralised identity management systems and produce a number of physical artefacts which relate back to the core features of applying DLTs to identity management. Examples of key features and physical artefacts include:

1. Transactions between volunteers and organisations are recorded on a tamper-proof public ledger, represented by a washing line, so anyone can verify claims.
2. When Volunteers learn a skill, or complete a task, they are awarded a set of uniquely numbered stickers, one which is published to the public ledger (washing line), to support claims they make about their identity.
3. Volunteers maintain an individual ledger or 'wallet', which gives a CV that can be selectively shared showing skills learnt and tasks completed.
4. Social Enterprises have a record of volunteer's completed jobs and the skills each volunteer required for a specific task rather than a whole database. This minimises data retained about volunteers and promotes sharing.
5. The game is ultimately governed by the players themselves as they take on more roles and are then able to 'attest', sign and prove the skills they have learned, and tasks that have been completed.

# CARICROP

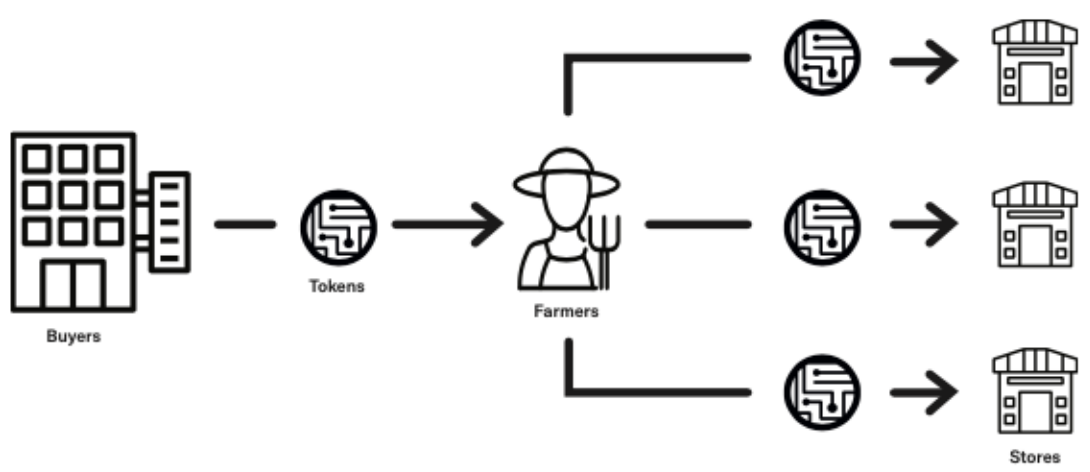

**Visibility**

**Value Exchange**

**Rethinking Society**

Speculative prototyping of decentralised currencies to engage stakeholders around values and equality

*Using prototypes to discuss issues of fairness and equality with stakeholders with competing values and attitudes?*

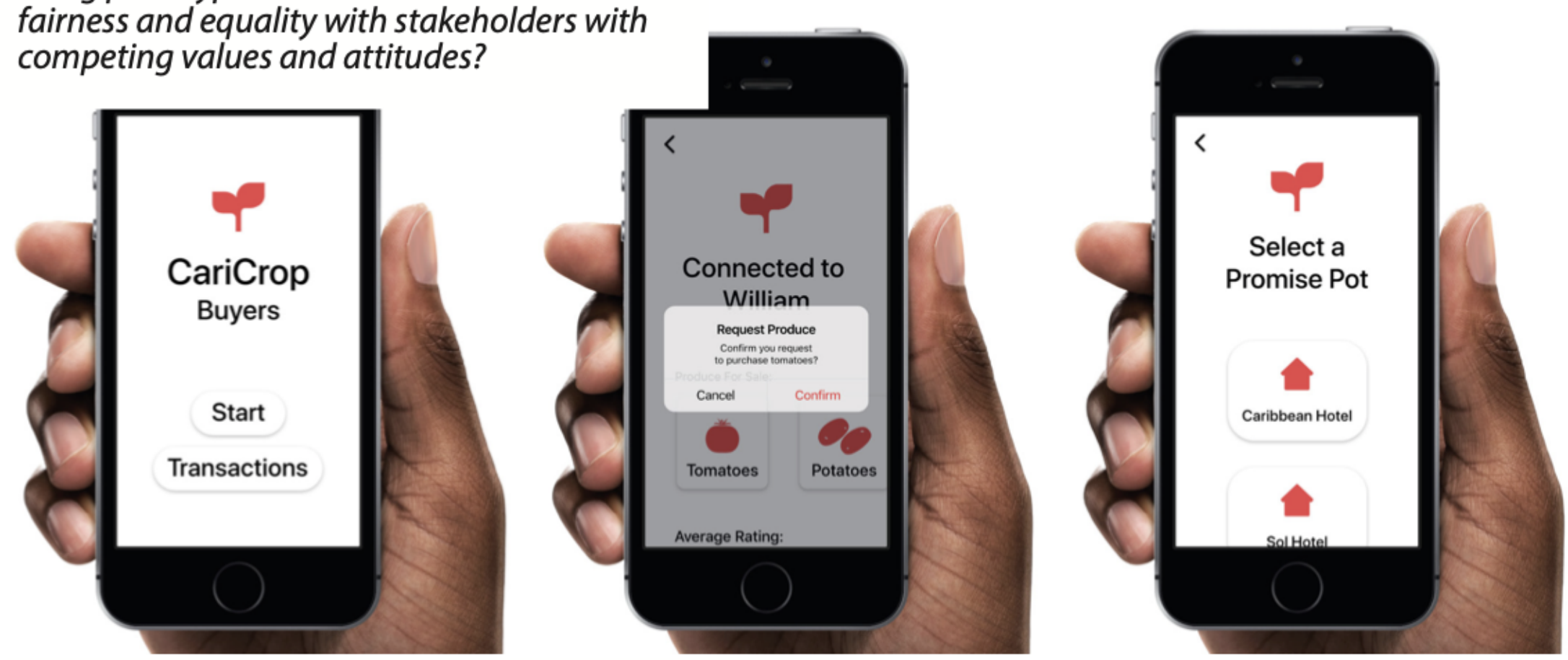

CariCrop interface

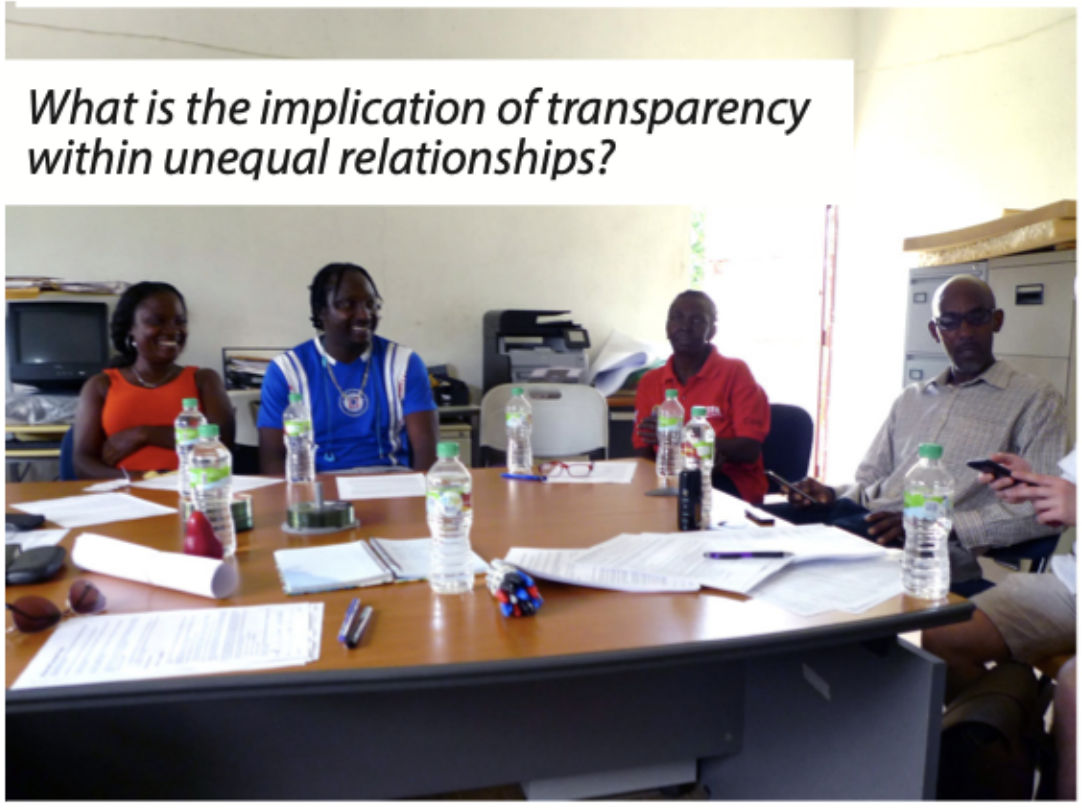

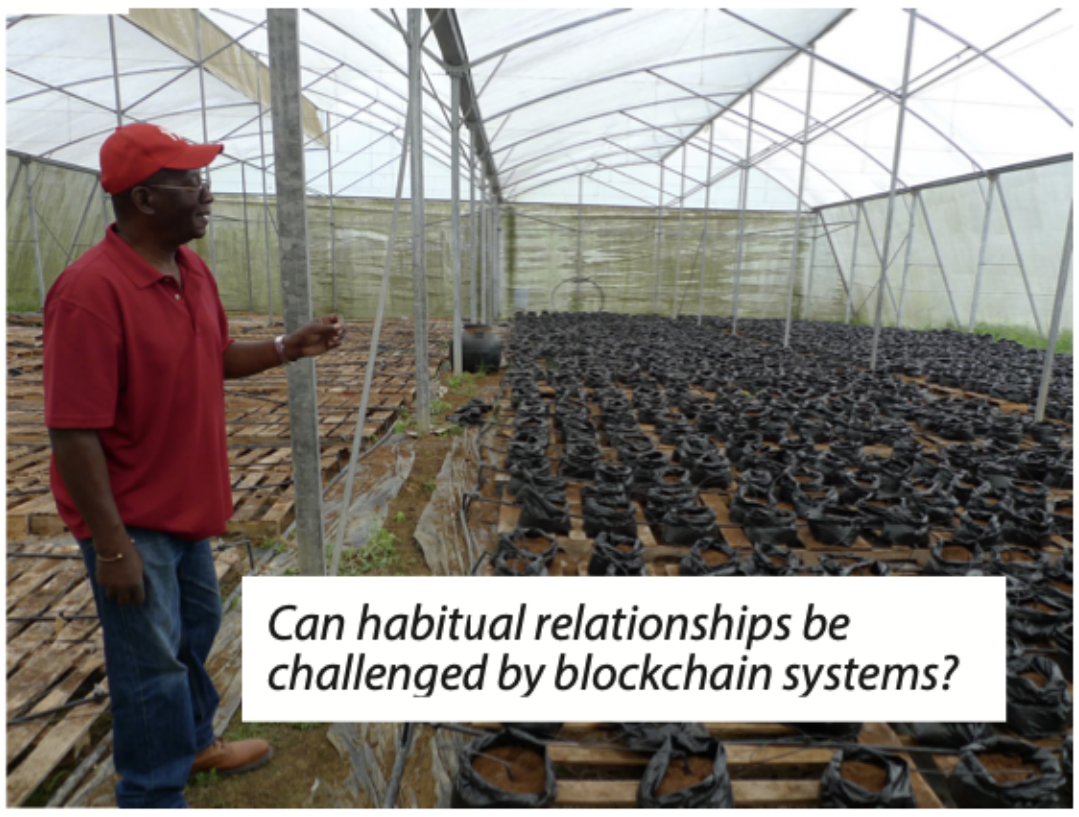

Workshop with farmers

# *CariCrop - Details*

**Aims and Context**

CariCrop is a platform that explores the possibility of using blockchain to mediate payments to farmers in the Caribbean. Delay of payments makes farmers highly vulnerable and unable to plan for future harvests. Among rural communities in the Caribbean and elsewhere, the trade of perishable goods can be surrounded by insecurity: the produce needs to reach the final consumer before deteriorating, while payments are sometimes delayed until businesses have made some profit, which poses challenges not only to farmers but also to businesses that trade with farmers, who start to operate through informal agreements of trust. CariCrop provides a bridge-currency that operates based on the agreement that a buyer will transfer money to farmers once they have made enough profit to cover for the original purchase. While waiting for payment, farmers can use this currency at input and general stores, and these transactions are securely tracked on a Blockchain.

**Experience**

*CariCrop envisioned the use of Blockchain to ensure that transactions would continue to take place within a trusted infrastructure while farmers waited for de facto payments to take place. The concept proposes a bridge-currency that operates based on agreements that a certain monetary transaction will occur in the near future. When a transaction is made:*

1. *Payment is agreed by both parties, and the amount is released immediately in the bridge-currency.*
2. *The farmer can use this currency at input shops and general stores while payments are securely tracked on the Blockchain.*
3. *When the buyer guarantees enough funds and is able to pay for the purchased produce, the de facto payment is released.*

*A smart contract follows the digital chain of transactions and distributes the money to all individuals and establishments on record, guaranteeing the delivery of money to all new owners. Hence, the burden of holding the loan is distributed across many stakeholders, to whom the original buyer now owes different amounts of money.*

*CariCrop was developed into an app that was employed in several workshops with farmers and other stakeholders to discuss the impact of using such a technology.*

## 4.1 Common Themes

These workshops were conducted with a range of stakeholders, most often with individual organisations, who were keen to develop an understanding of blockchain technologies and their potential within their organisation. The primary aims were to develop understanding and stimulate and support ideation, to open out creative thinking with blockchain technology and its possible applications for non-specialists. This often led to considering the deeper concepts that the emergence of blockchain technology has provoked: around systems of governance and value exchange that challenge the status quo; uprooting existing bastions of power and trust; and creating new power dynamics through decentralising and democratising services currently monopolised by small numbers of individual organisations.

In order to engage people with this complex mix of socio-technological concepts these workshops have employed a number of strategies. Across the four workshops, we draw out some of the key strategies, techniques and concerns here.

## 4.2 Roleplay and physical enactment

Role play is a key device to help participants understand systems by pretending to take part in them. Simple interactions such as exchanging a resource card for Lego bricks, or recording a new pizza making skill with a unique set of stickers and stamps provides a gateway to considering the possible dynamics and outcomes of emerging practices around Blockchain systems. With PizzaBlock, a starting point in a structured activity around making pizza helped participants understand how trust can be derived through creating and maintaining a public ledger of transactions in the game.

The materials and the mapping process in Attaching Strings supported collaborative discussions around distributed concepts as participants were able to externalise their ideas and share them with other group members as they progressed through the mapping process. Making ideas physical, even with simple materials, provided discussion points, and helped participants to formulate complex concepts. With CariCrop, roleplaying around the app highlighted the habitual relationship among stakeholders that could compromise the transformative power of the distributed ledger technologies. Enacting the system helped not only to connect with the future possibilities, but also highlight the frictions in getting there.

## 4.3 Tangibility

Tangibility and material properties are important to creating compelling experiences. Attaching Strings focuses on relations between entities, and uses materials to represent different qualities of relationship; Block Exchange uses the possibilities of familiar blocks to represent technical operations; and Pizzablock uses collectible stamps and stickers to create a high fidelity replication of a distributed ledger.

The use of simple physical metaphors was crucial – Lego could represent many of the operations that make the blockchain work (mining, blocks, atomic transactions) while staying simple and familiar. It was important to make sure that we had a small, coherent set of concepts to represent, to avoid baffling participants, while also having enough richness to support further thinking. Working from the physical objects, and starting with highly structured activities gave a space for participants to engage: following the instructions, placing the blocks together, and they could then relate this to terms they had previously encountered.

Some material qualities were important: with Attaching Strings, repositionable markers (e.g. magnetic pins and post-its) allowed participants to reconfigure them as the map developed, while yarn, edible shoelaces and wire mesh were useful to indicate different types of relationship such as rigid, temporary, natural, transparent or grid-like; in Pizzablock, ink stamps represented immutable signatures, with information hung on a washing line to make it visible to all players at once. This is in contrast to the intensely fungible nature of Lego blocks, relating the material properties to the aspects of the systems being explored.

### 4.4 Focus and Abstraction

Focusing the workshop on a single aspect of the technology in each case gave a solid grounding for discussion, but still allowed wide ranging discussion of concepts and implications. BlockExchange focused very specifically on the public and immutable nature of peer to peer transactions while consciously abstracting other aspects of the technology. This allowed participant experiences and discussions that are target on specific applications or implications of the technology rather than attempting to explain the entirety of such complex systems. In Attaching Strings, the abstractions provided in the workshops allowed participants – experts as well as novice – to ask many what-if questions, and collaboratively unpack the concepts and meaning of DAOs. While CariCrop was potentially less abstract than Attaching Strings, the focus on farmers trading information for record-keeping and establishing trust helped participants to develop the implications that access to this data must be carefully considered. In many of the works, metaphors played an important role. In Attaching Strings, physicalising the idea of networks helped to unpick the complex relations and the ecology of 'things' which DAOs might encompass. Through this process participants develop a deeper understanding of DAO's, their structures, uses, implications, and the flow of data and value through these systems.

### 4.5 Technical Fidelity

A key axis through the workshops was the level of technical fidelity, balancing the complexity of concepts with highlighting the key aspects of systems.

With BlockExchange, we set out to defamiliarise currency, revealing it as a socially-constructed abstract system imposed on our innate understanding of value and the need to exchange things. Inevitably, this involves a simplification of Blockchain technology, in favour of its operation as a disintermediator. The workshop is not designed as a comprehensive or wholly accurate explanation, but by encouraging free thinking and discussion, participants could speculate about what might happen if we move beyond traditional currencies. However, more technically savvy participants pointed out some of the shortfalls due to levels of abstraction. Pizzablock, in contrast, aimed to more closely reflect details of how distributed ledgers worked, pushing at the level of fidelity possible while keeping the game enjoyable to play. By doing this, we could then use each of the artefacts in the game, (personal wallets, public ledgers, stickers) to explain and reflect upon specific aspects of blockchain-based identity management. The intended audience and aim of the research or activity are essential here in determining which approach to take and when to commit to high fidelity translations of the technology or when to focus on simplicity of interaction.

### 4.6 Building Reflection

One of the motivations behind the workshops is to see how far participants can develop their reflections on the technologies and implications, through their participation and discussion. With PizzaBlock, for example, it became apparent to participants that new kinds of labour that would be involved in decentralisation, as the whole network needs to work together to produce and maintain a faithful record of events. Similarly, while fairness is accepted as the aspiration for the design of some systems such as DAOs, it became clear to participants that fairness is contextual, and cannot be integrated into systems without the active involvement of those affected. In general, making the systems tangible and working through the activities helped participants frame crucial questions such as: how do such immutable systems support human adaptation and creativity? Who has the power to resolve unforseen issues in the code? What rules need to be programmed and who gets to make the decisions? And in a distributed system, where is legal responsibility held?

### 4.7 Summary and Directions

Whilst all four workshops focus on the production of new ideas resulting from both the challenge to current thinking and possibilities presented by role-playing through new technological possibilities, ultimately these workshops are more about opening up conversations in this new area, and can be seen as a first step in considering concepts such as currency, value and governance afresh.

Overall, the workshops help clarify – but also move beyond – the challenge of defining things in a fast moving area. This meant focusing not just on Blockchain, but considering the nature of *value* and the mechanisms of its exchange, moving towards exploring the power relationships within socio-technical systems and the governance and control of those systems. The workshops create space to think through things in a highly collaborative, social situation, and this is often as important as developing a technical understanding. Much of this work used features of blockchains as a springboard to discuss wider concepts – a recurring theme throughout the portfolio.

Workshops provide spaces for detailed collaborative exploration of ideas. However, accessibility is limited, they are time consuming to run, so there is a limit to how many people can be engaged. In the next section, we look at moving from the workshop setting into more public spaces, with projects based around provocative user experiences that can take place in shorter periods of time. This allows publics to rapidly engage with the broader implications and potential effects of blockchain technologies for their own and others' lives.

## 5 ENABLING PUBLIC EXPERIENCES

There is a clear need to raise awareness and support grounded debate about the possibilities of blockchains. A wide range of organisations, governments and companies are pursuing strategies that include blockchains and decentralisation as part of their makeup [e.g. 65, 66, 103]. Methods that can quickly reach a large number of people and raise an awareness about how such technologies may impact society help to create a public that can engage more fully in decisions about their futures. In particular, experiential methods have a long history of helping publics to engage with on multiple levels of depth and complexity [85].

The projects in this section aim to provoke responses, encourage debate and challenge perceptions through real-world, relatable experiences. In contrast to the more educational approaches above (Section 4) that convey conceptual knowledge about new technologies, in this section we move these interactive installations out of the academic and industrial context to engage new, non-specialist public audiences from all walks of life, backgrounds and age groups. Public spaces and events, such as festivals, exhibitions and conferences, offer a powerful platform to reach diverse audiences. Complex technical systems such as blockchains are difficult to explain through language alone. Hence, we designed playful and accessible experiences for several contexts, offering an interesting and engaging entry point into the multi-layered technologies.

The pieces all start with relatable, even mundane, everyday life activities, such as getting married or drinking coffee. These trigger interactions and incentives to provide a gateway for people to imagine a potential future scenario or narrative in which blockchains could change current exchange systems and societal norms. By crafting participatory experiences that introduce new ideas in familiar contexts, publics were guided from an accessible performative interaction towards a provocative or challenging narrative exploring the implications or applications of this technology.

We discuss four interactive installations, each exploring an area where blockchain technologies may affect our future society, engaging non-expert public audiences with aspects of blockchain technologies in playful, provocative and experiential ways.

- **Happily Ever After (Bitcoin)** critiques traditional social norms around marriage, contrasting it with the possibilities of dynamic smart contracts. It was developed from an outcome of the GeoCoin workshop (Page 42) called *Handfastr*[9], that re-imagined marriage as a time-limited agreement. It was based on a discussion around how smart contracts are different from traditional legal contracts, in particular rapid deployments and programmatic conditions.
- **After Money** reveals current and potential future practices and perceptions of value exchange. Premiered at the Edinburgh Festival Fringe 2017, it is presented as an interactive installation and associated mobile app that asks participants to trade various things for sweeties – work, personal data or a variety of more or less fictional currencies.
- **Seismic SeeSaw** is an interactive poster that embodies escrow transactions, exploring the possibility of 'conditional charitable donations'[10]. The Seismic Seesaw was built as part of the OxChain project [44], which sought to explore the use of blockchain technologies in the charitable sector [28] and the possibilities of programmable money [45].
- **KASH Cups**[11] work as an exposition of how design and technology can reconfigure the representation and flow of value, investigating ideas of social currency and algorithmic rules. The collection of NFC enabled coffee cups can be used in conferences or other settings to investigate value exchanges around sociality, networking and coffee.

Although these projects seem to differ in their manifestations, from simple coffee cups to interactive marriage installations, each experience aims to challenge a broader non-specialist public audience's perception and understanding of current value systems to playfully engage them in critically thinking through the influences such new technologies could have on their own future lives.

[9]https://aftermoney.design/handfastr/
[10]https://www.youtube.com/watch?v=WiDY0Uea0Qw
[11]https://www.designinformatics.org/research_output/kash-cups/

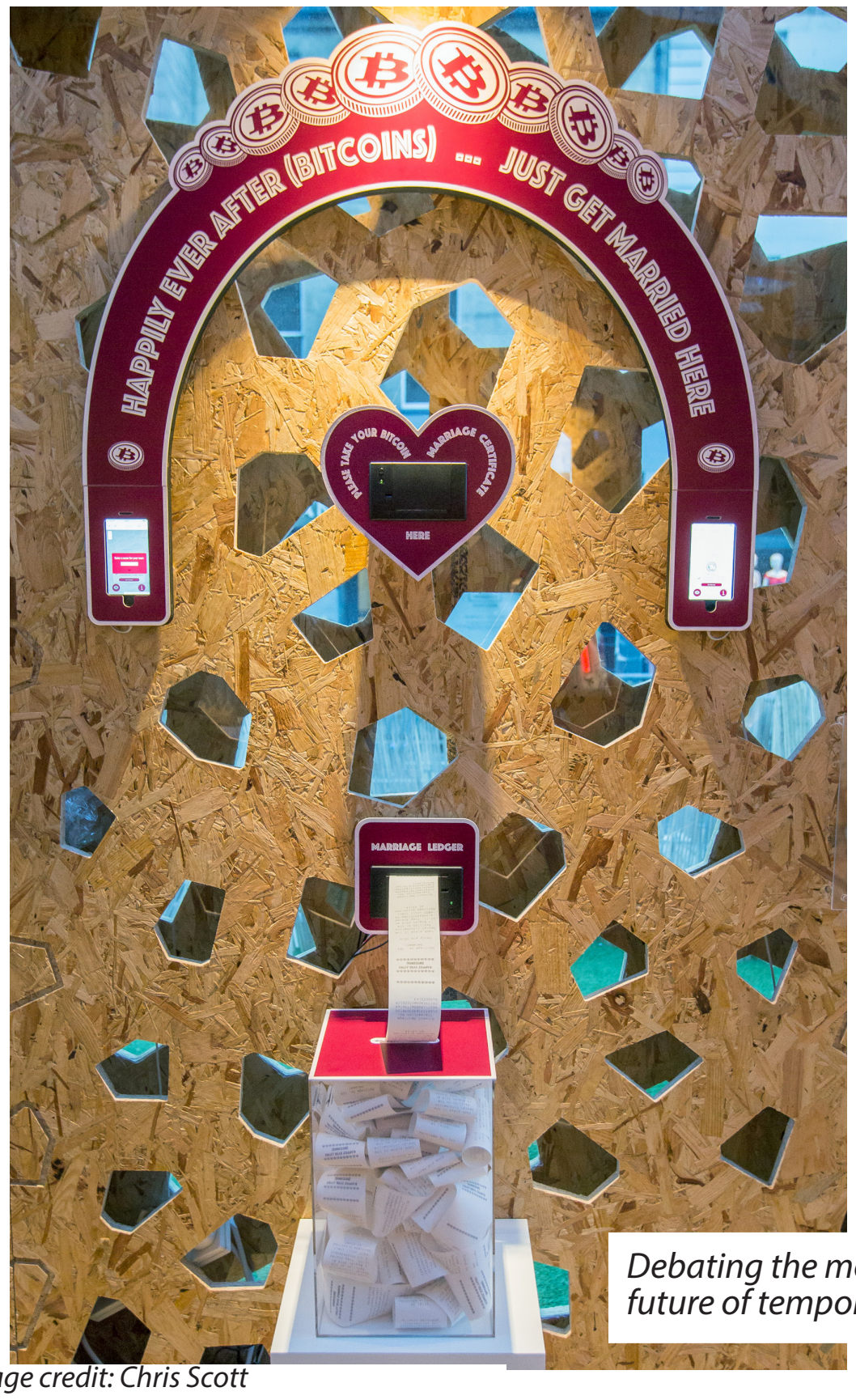

# HAPPILY EVER AFTER (BITCOIN)

Have you ever wanted to join your partner in holy bitcoin matrimony? Or wanted to get married just for a vacation? This project uses blockchains to think about social structures

**Visibility & Transparency**

**Abstraction and Focus**

**Rethinking Society**

*Debating the meaning of a blockchain union and the future of temporary contracts with permanent records*

*Image credit: Chris Scott*

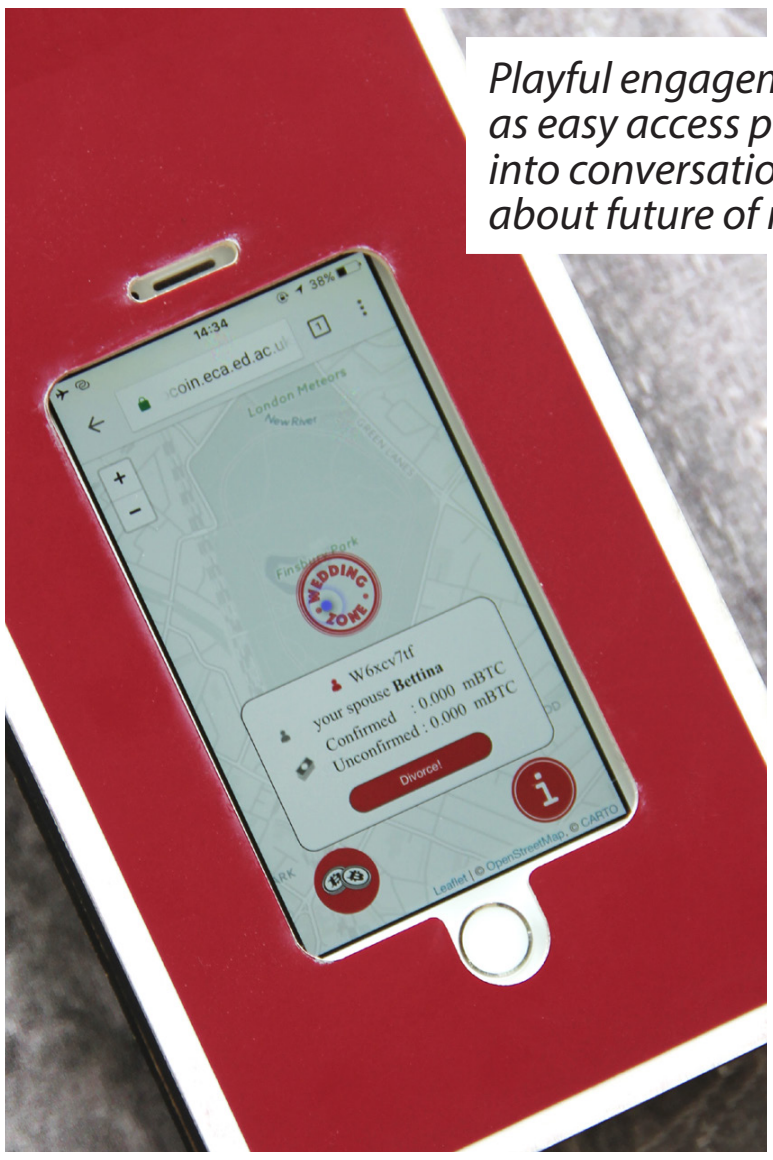

*Playful engagement as easy access point into conversations about future of money*

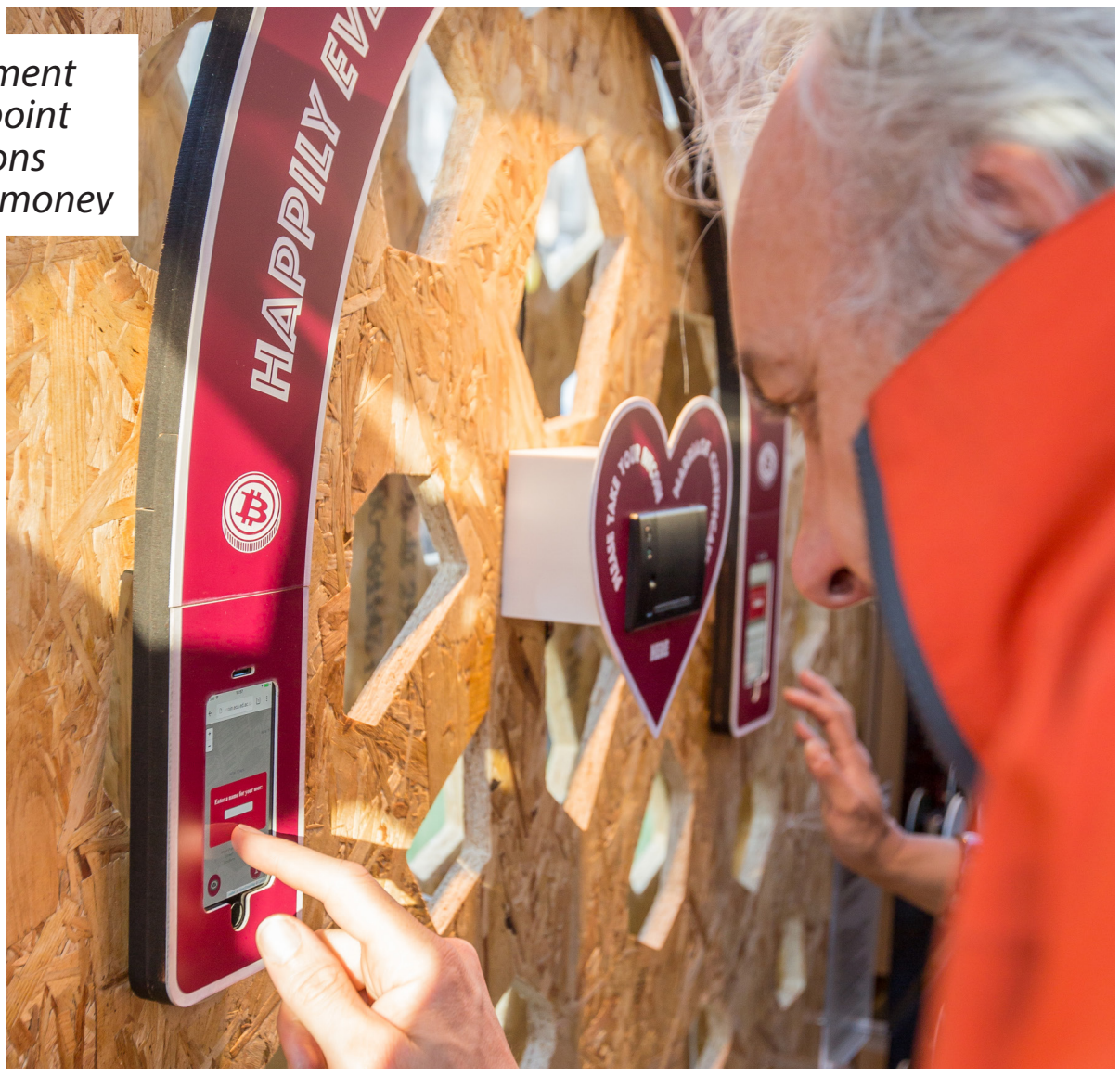

Marriage wallet interface

Image: Chris Scott

# *Happily Ever After (Bitcoin) - Details*

**Aims and Context**

The Happily Ever After (Bitcoin) project create a new kind of blockchain marriage. This interactive installation allowed public visitors, blessed with a small amount of bitcoin, to merge their digital wallet with co-visitors or strangers for the short duration of their nuptials. It was based on a discussion around how smart contracts are different from traditional legal contracts, in particular rapid deployments and programmatic conditions. We aimed to highlight how we negotiate roles and values through looking at new forms of short-term partnerships and how we could update traditionally fixed legal contracts into mobile, contemporary, digital agreements for a range of applications and communities.

It engages a very general public around questions such as: how can we negotiate new forms of contracts and transactions? What new forms of contract will we be able to design in the future? Who will benefit from such contracts and how will these affect the relationships and partnerships we build?

**Experience**

The installation was based around an app, and the Ethereum blockchain. Any two people could come to the stand, and be quickly guided through making an accepting a marriage proposal. When a proposal was accepted, the romantic agreement was sealed in the blockchain, and the newly-weds gained access to a shared bitcoin wallet that could only be used for a limited time, and only when they were in the same location. Two printers in the centre of the installation would print a receipt or marriage certificate, one copy the newly-weds to take home, and another as continuously printed ledger of all marriages, displayed in a clear container for all to see.

This demonstrated how we could update traditional fixed legal contracts into mobile, digital arrangements.

Happily Ever After (Bitcoin) was installed at the Furtherfield Gallery, London, and at a pavilion in the Edinburgh Arts Festival.

# After Money

Would you swap your friends for a sweetie? AfterMoney lets you make trades - swapping sweeping for cryptocurrency, or personal data for imaginary tokens - to explore the future of transactions

**Visibility & Transparency**

**Abstraction and Focus**

**Rethinking Society**

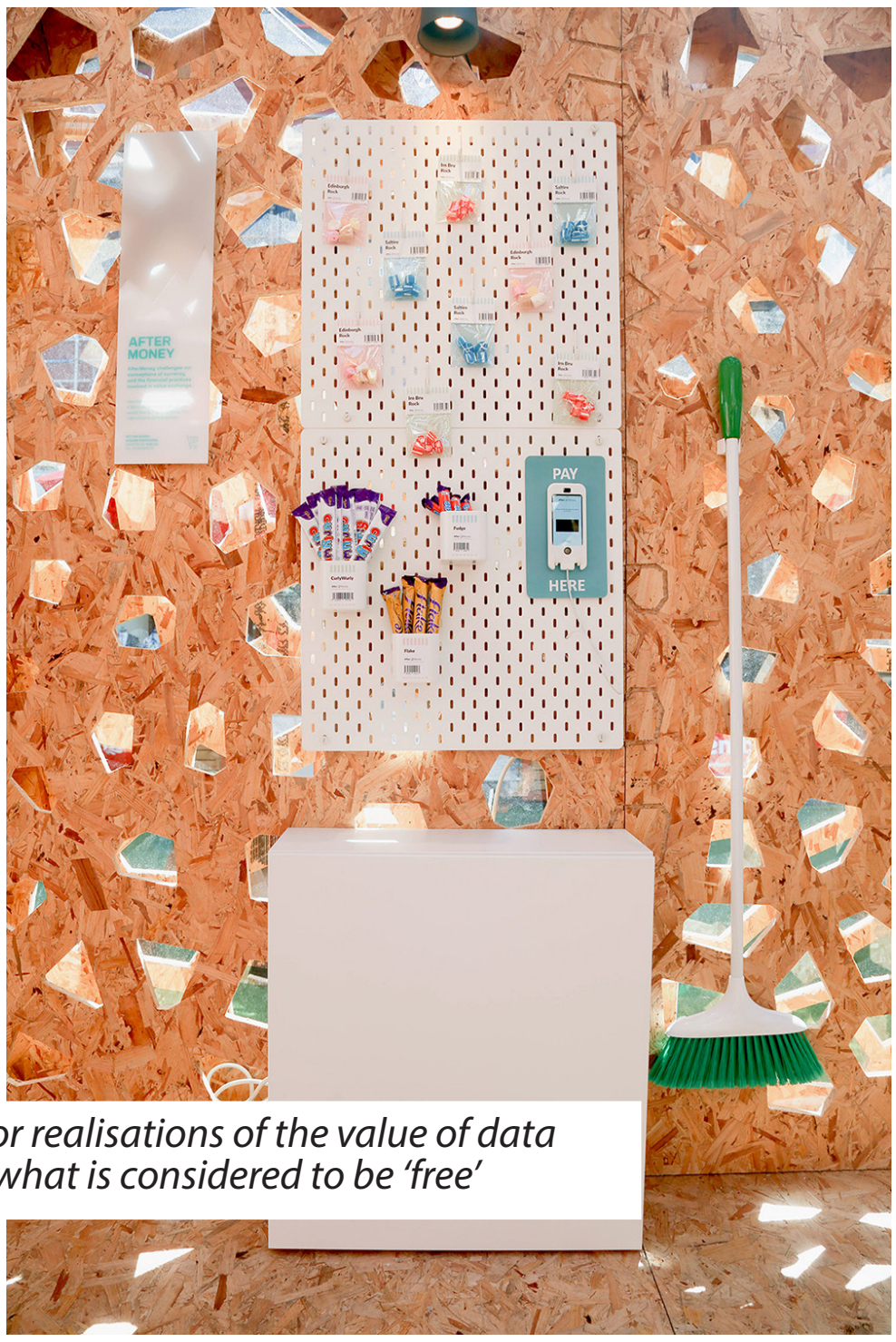

*Visitor realisations of the value of data and what is considered to be 'free'*

*Image credit: Yuxi Liu*

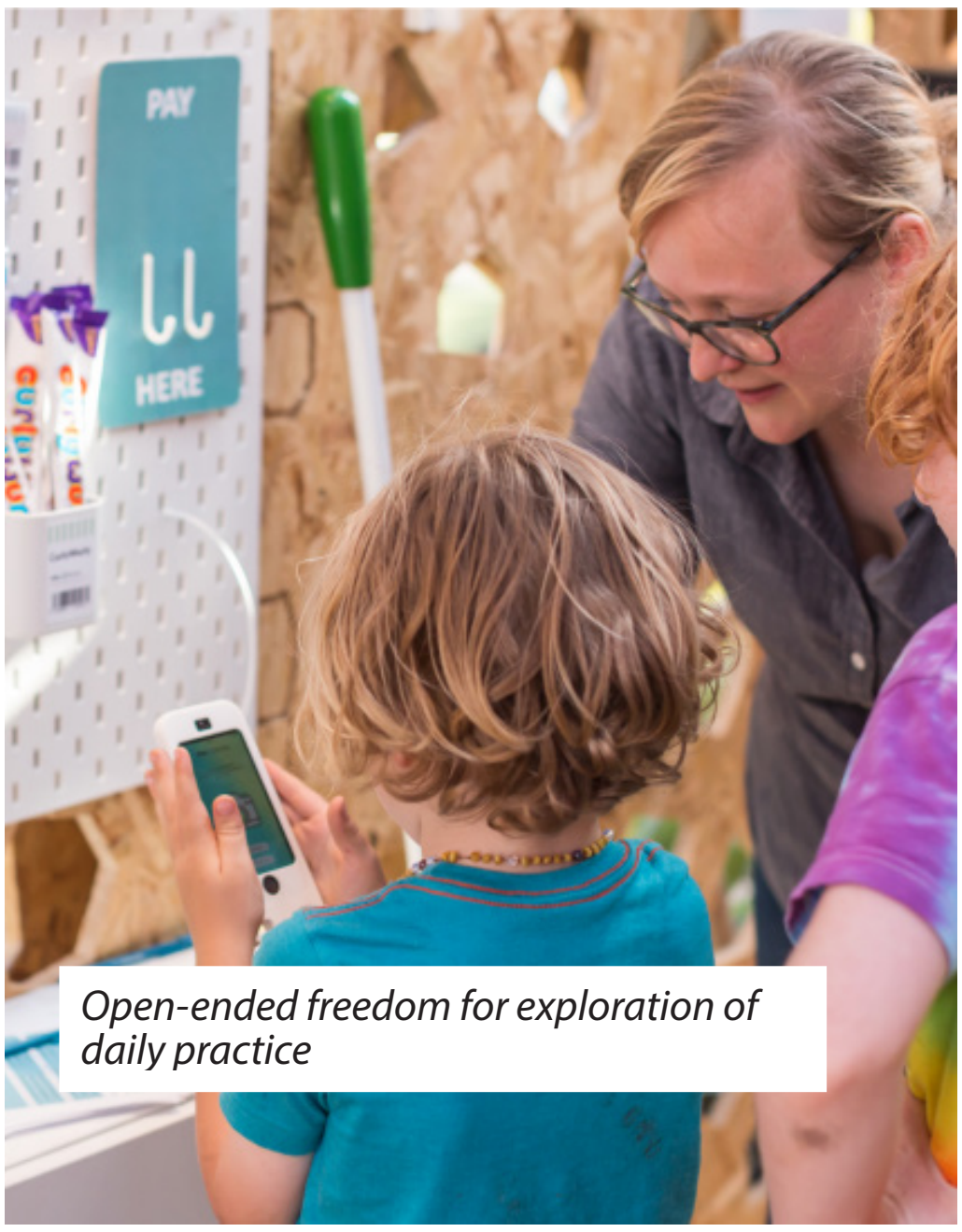

*Open-ended freedom for exploration of daily practice*

Young members of the public exploring alternate currencies.
Image: Chris Scott

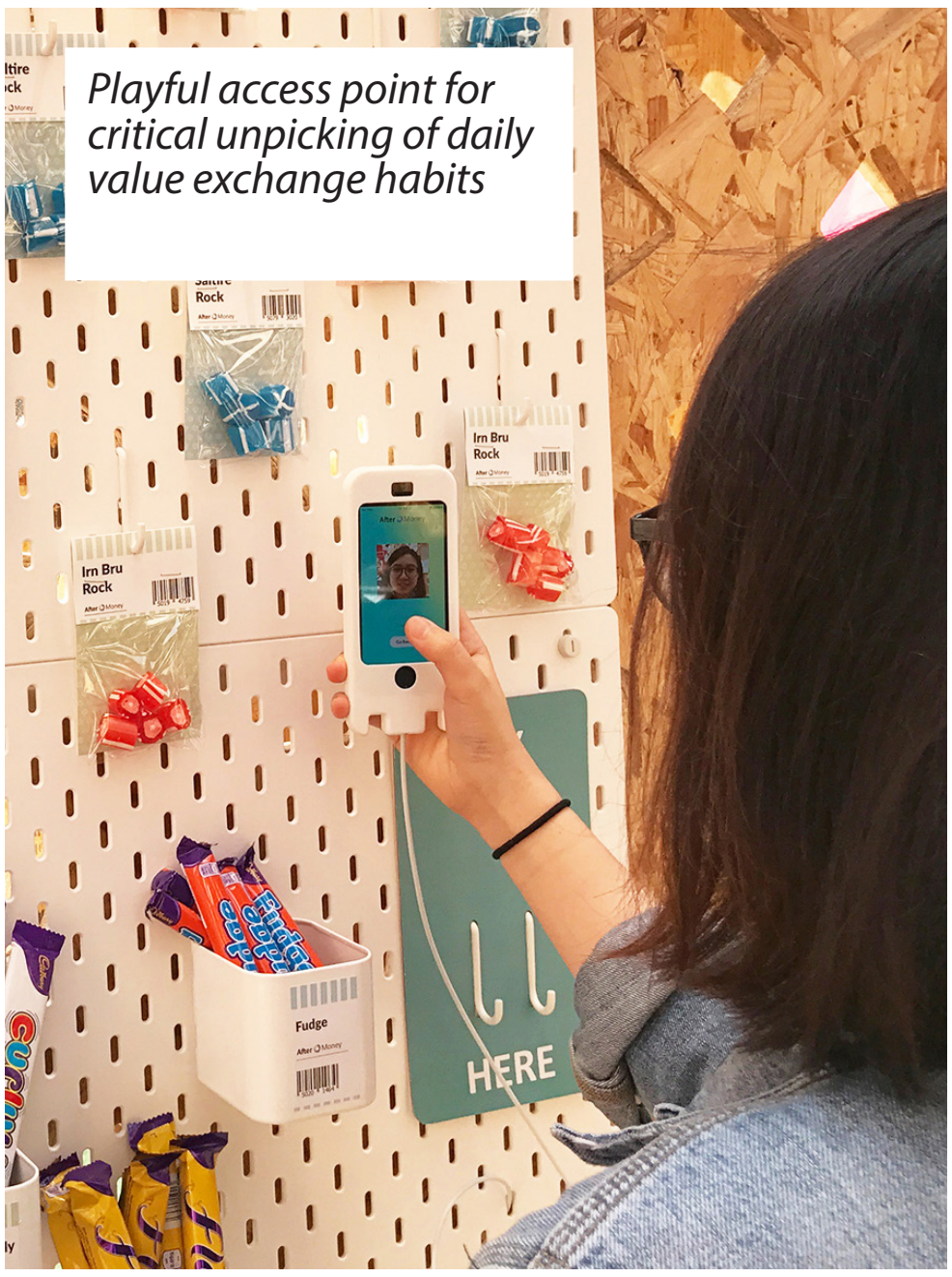

*Playful access point for critical unpicking of daily value exchange habits*

Swapping selfies for sweeties
Image: Yuxi Liu

# *After Money - Details*

## Aims and Context

The AfterMoney installation asked participants to choose what to trade for sweeties - they could use money, whether traditional, cryptocurrency or fictitious; they could sweep the floor for a set amount of time, or dos some exercise; alternatively, they could share personal data such as selfies, Facebook friends, their fingerprints or GPS data. AfterMoney highlights the tacit practices of how we trade, barter and share value across multiple forms of currency. The installation and mobile app reveals and crystallises some of the increasingly common non-monetary exchanges of value in a growing digital economy. This allows the public to question what constitutes currency today and in the future, and how this will be affected by increasingly precarious employment situations due to automation and a growing gig-economy. This app and installation aimed to challenge the perception that we currently only pay with one type of currency and bring awareness to the fact that we are constantly exchanging different types of value amongst ourselves and with digital businesses and services.

## Experience

The AfterMoney app appeared like a regular barcode scanning app to check and show the price of a given commercial product -- sweets, in this case -- which visitors could then purchase. The user was asked to select a payment method: money, time or data. When using money, choices included current fiat currencies, cryptocurrencies or imaginary 'in-game' currency similar to those in online games. When paying with time, the user was asked to sweep the floor for a specific amount of time using a sensor-enabled broom, but other options also included helping a neighbour, giving someone a lift or exercising for 15 minutes. Data payments included taking a selfie -- without any indication of how it would be used -- or sharing daily step data, a facebook friend, their phone's GPS data, their fingerprint or a tweet. Although none of the pictures were saved for ethical reasons, the app still gave the impression of storing real user data.

# SEISMIC SEESAW

Give where it's most needed - Seismic Seesaw allows "programmable donations" that are only activated when earthquakes happen

**Tangibility**

**Visibility**

**Moments & Seams**

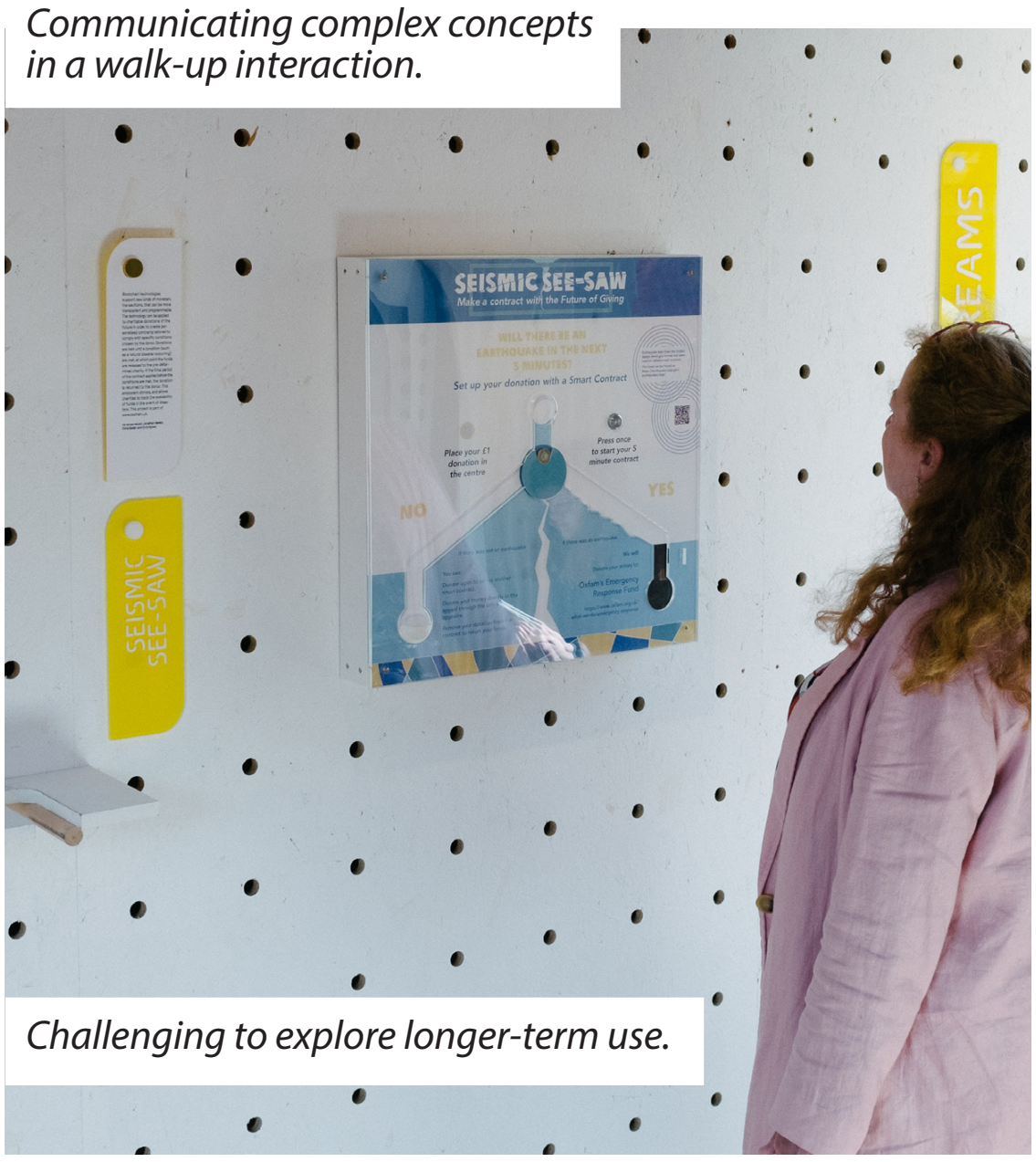

A member of the public waiting for the Seismic Seesaw in a street exhibition during the Edinburgh Fringe Festival . Image Credit: Jonathan Rankin.

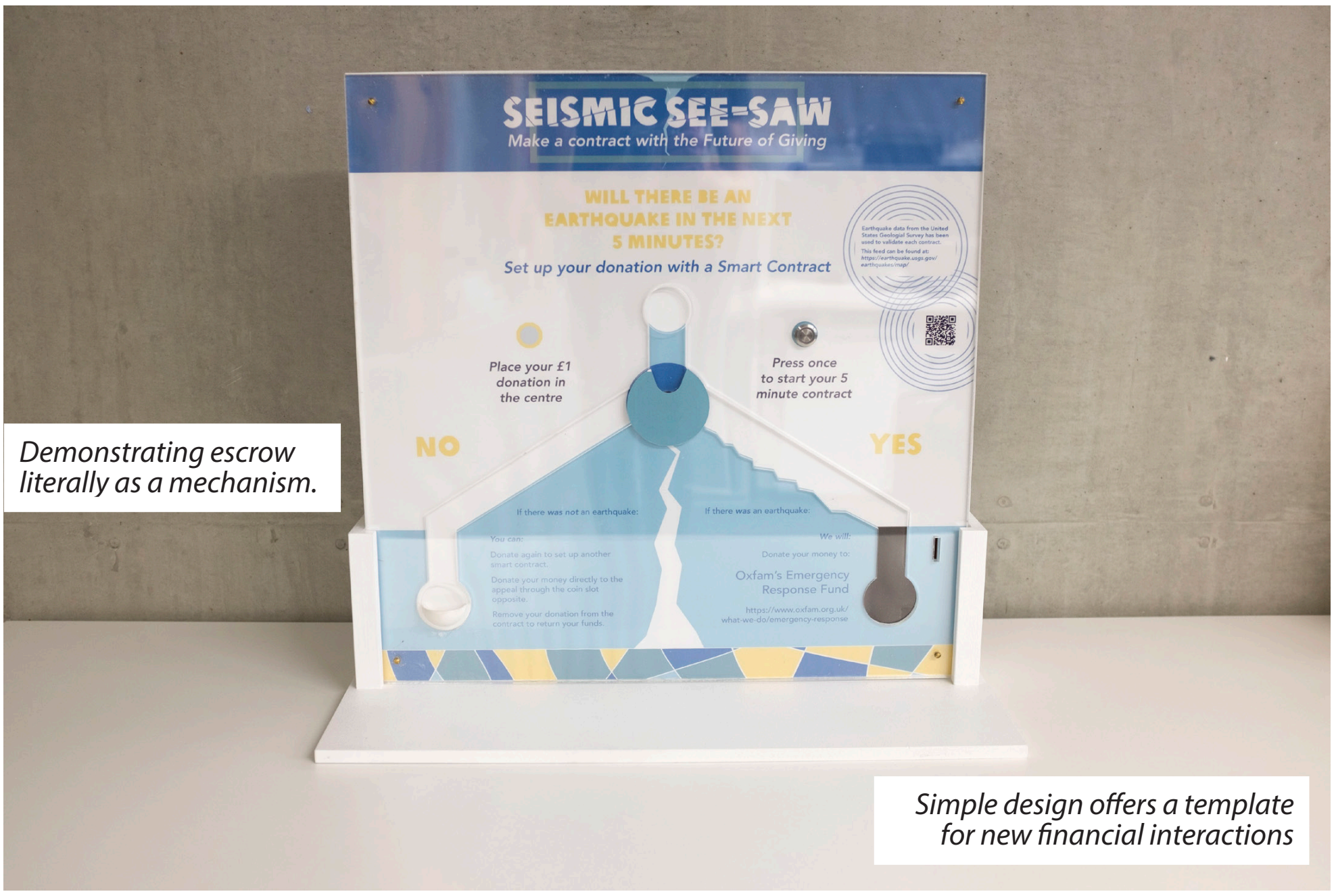

The Seismic Seesaw: an interactive mechanical poster behind perspex, . Image Credit: Jonathan Rankin.

# *Seismic Seesaw - Details*

## Aims and Context

Seismic Seesaw explores the possibility of making donations that are activated under certain conditions. In this case, giving to an Emergency Response Fund, with the donation only being taken if there is an earthquake in the given time. If there is no earthquake, the money is returned to the donor. Escrows, also known as trust funds, are a form of intermediary, where assets are held upon some conditions, before being transferred to another party.Through working with Oxfam, our project partners, we saw blockchain based escrows as a powerful tool to track the status of various donations, while also raising the prospect of attaching conditions to when and how a donation is given.

This artefact provides a tangible, transparent embodiment of financial escrow. While escrows are traditionally managed through legal contracts and financial institutions, smart contracts offer the means to rapidly create transparent and automatic forms of digital escrows, radically remediating relationships between donors, charities and beneficiaries.

## Experience

The Seismic Seesaw is a simple interactive mechanical poster that invites users to conditionally donate £1 to an Emergency Response Fund. Once donated, the donation is visibly fixed in place; neither the donor nor the charity can reach it. If there is an earthquake of any magnitude in the next 5 minutes anywhere in the world, according to live US Geological survey data, the money is given to the Response fund. Otherwise, the money is returned to the donor who can take it back or use it to make a traditional donation

# KASH Cups

Pay for your coffee by meeting new people! Exploring transactions with NFC coffee cups, seeing how design and technology reconfigure value exchanges

**People & machines**

**Contextualisation**

**Value Exchange**

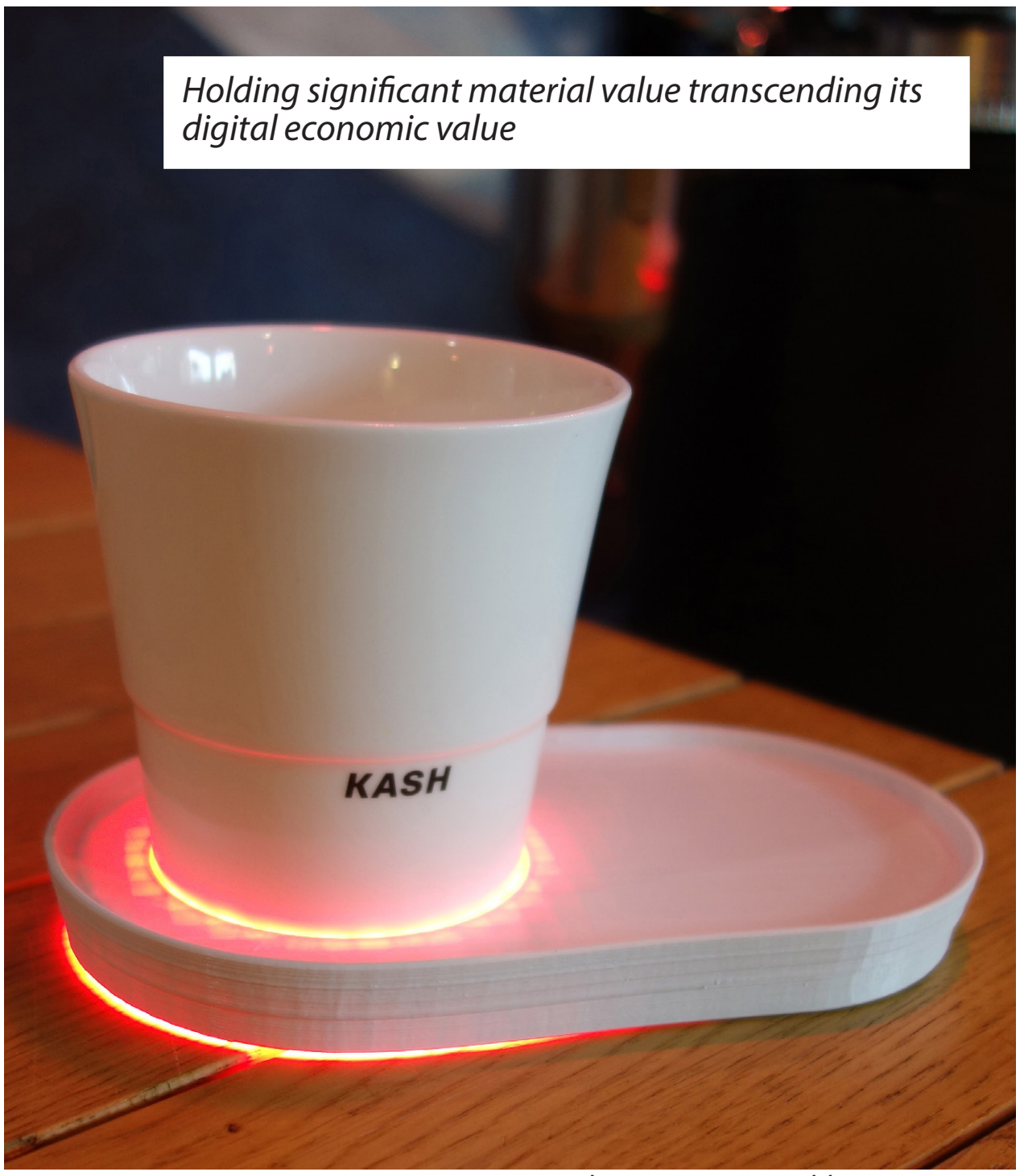

KASH Cup charging user credit. Image: Jane MacDonald

Participants selecting KASH Cups at a workshop. Image: Jane MacDonald

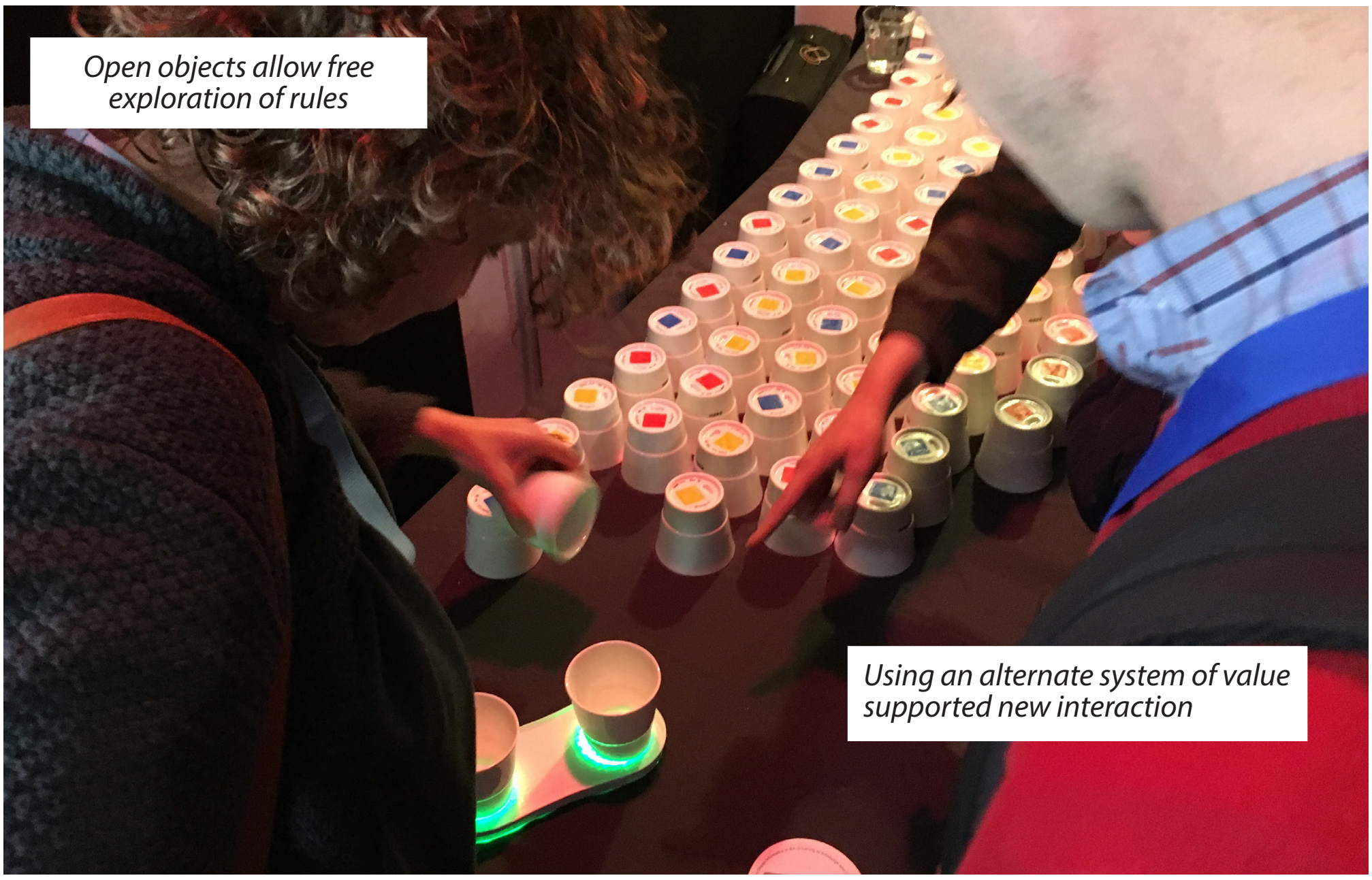

# *KASH Cups - Details*

## Aims and Context

KASH Cups asks what happens when objects have their own digital identity, and can participate in value exchanges. Each cup has a digital wallet which is credited whenever a participant socialises with a new person, demonstrated by putting their cups next to each other on a smart saucer to earn credits for coffee. This intervention into the social conventions of a conference complicates the value constellation surrounding free coffee, by adding more values to the traditional service - because the software requires two cups to be placed on the saucer before adding credit, participants are forced to collaborate with another delegate in order to receive coffee. The system is is build on simple technology - the saucers simply report presence and identity, and display colours to indicate outcomes. This means that it can can be programmed in various ways to open up the space for developing rules around exchange: for example, being able to steal credits when talking to people, or sharing personal information.

As a whole, the project looks at how interaction design can engender different value interactions by intervening into familiar social practices.

## Experience

KASH Cups have embedded NFC chips that connect to with 'smart saucers' that can sense the presence and identity of one or two cups. Connected to its owner, each cup has a digital wallet that is credited whenever a smart saucer detects its owner socialising with a new person. This credit can then be used by placing the cup on the barista's smart saucer, which checks the cup's balance and debits the cup in return for coffee.

Originally deployed during Dutch Design Week 2016, the cups have been used to explore different kinds of value exchange at international conferences and events.

### 5.1 Common Themes

The previous section focused on workshop environments where people dedicate time to understanding, learning and ideating with blockchain and distributed ledger technologies. In contrast, the projects presented in this section are focused on rapid engagement with a general public through interactions and provocations. There is move towards crystallising interactions around a single key concept that can be communicated extremely directoy, but which is nonetheless a seed for more wide ranging discussion and reflection.

As such, the projects had varying degrees of fidelity to actual blockchain technology, which is a key axis for designing engaging activities (Section 4.5). In several cases, the presence of a blockchain was not necessary for the interaction e.g. Happily Ever After (Bitcoin) used Ethereum wallets until currency fluctuations made this untenable. However it was the experience itself that was crucial, distilling complex concepts into rapid interactions with a simple pivotal transaction that moves from current practices to novel forms of currency and contracts.

### 5.2 Reaching wider audiences

All of these projects seek to engage the public on varying levels, using design strategies to bring them into technological discussions. The KASH Cup system establishes a material, encrypted, algorithmic exchange and social interactions, through which participants find value manifest in many forms. Familiar with the incentive of getting good coffee, and content with talking to people, the system was both highly engaging and simple in its interaction. The Seismic See-Saw communicates the complexities of programmable donations in a single-walk up interaction. This presented temporal challenges – we required a condition to be set for the donation which was reasonably frequent and predictable, and could occur within a few minutes. After Money showed that playful interaction offered visitors the opportunity to understand convenience, privacy and current data sharing behaviour in the context of economic practices. With Happily Ever After (Bitcoin), as well as groups of friends and couples, a father joined his children aged 3 to 12 in Bitcoin matrimony, while explaining the technological developments that will directly impact their future interactions with money. In all cases, the use of objects that embody concepts helped a wide range of people engage with complex technical issues.

### 5.3 Strategies for engaging publics around blockchain

Across this set of projects, several strategies proved powerful for public engagement and offer designers tools to connect people with complex concepts:

- **Playfulness and defamiliariastion** allowed an easy access point into conversations for non-tech audiences to discuss the future of money, banking, currency. In Happily Ever After (Bitcoin) participants explored the conceptual space of marriage, rethinking traditional institutions, and marrying the Means of Production or a Table being wed to a Chair. Other participants married their friends or family members while debating the meaning of a blockchain union and the future of banking and currency. The Seismic Seesaw took the familiar experience of putting a coin into the charity box, and then defamiliarised this with an interaction that makes that donation conditional on world events.
- Whether donating to charity or drinking coffee, **seamfulness** [21] created moments when someone has to decide what currency to pay with, or learn a particular ruleset, exposing the hidden architectures. It is this moment that creates space for thought and allows a move from everyday practices and their structures into speculative realms.
- Controlling **visibility and transparency** foregrounded crucial parts of the interactions. The value of the Seismic Seesaw is to be able to see the moving parts, and understand, experience and trust the transaction taking place, even if the background mechanisms are elided for the sake of simplicity. In KASH Cups, users could only see their credit when they checked their status, which sets up different interactions compared to systems where value is individually or publicly visible. Happily Ever After (Bitcoin) used a transparent box to collect the marriage receipts, making it highly visible that the contract was being recorded, and it could be seen but could not be tampered with.

- The previous section surfaced the theme of **tangibility** (4.3) as a way to engage with abstract ideas. This re-occurs here, for instance Happily Ever After (Bitcoin) used tangible artefacts to highlight the idea of temporary contracts that had a permanent record - a key principle of smart contract systems. Beyond this, it became clear that the **materiality** of the components affected the experience, and the physical embodiment of digital objects has an effect on the way that they are used. As weighty, uniquely designed artefacts, the KASH Cups held significant material value for some people, who took them away to use after the event despite the fact they wouldn't work as digital objects outside the interaction. The receipt-like nature of the printouts from Happily Ever After (Bitcoin) connect to transaction records, while the playful embodiment of ideas in Happily Ever After supported an exploration of themes that were otherwise difficult to engage with.

### 5.4 Quantifying behaviours

In some of these installations, aspects of the interactions were recorded for both practical and conceptual reasons which allowed another level of analysis and exploration of the publics' behaviours. The main aim of Happily Ever After (Bitcoin) was to record participants short-term contracts which were printed and publicly shown as 'proof of the marriage'. These recordings of names in each union also offered additional insights into playful behaviour of participants not just joining other humans in a union, but exploring marriages to animals or things. While playful, some more conceptual marriages to things or abstract concepts such as 'The Means of Production' or 'Time' offered us an oversight of how the general public engaged with concepts of marriage and union when it was removed from the traditional models. Similarly, for the KASH cups interactions to function the backend database was collecting the numbers of cups which were used together. In the conference setting where each conference attendee had their specific numbered cup, patterns emerged such as people who were very busy meeting other conference attendees, or other who were just using their cups as icebreakers. While this was in no real terms an analysis of people's intentions or actions, the quantitative nature of an individual cup's 'transactions' could however be seen as indicative measure of networking behaviour – or caffeine addiction.

Due to its nature, AfterMoney supported a more detailed quantitative analysis than other projects as its specific intention was to record people's transactions and their attitudes towards data, time and money as different forms of currency. In this case the transactional data allowed us to analyse and indicate that an equal split of participants preferred sweeping for their sweets or taking selfies. Participants were less likely to pay with a fingerprint and least likely to give up a facebook friend or their phone's browser history. A questionnaire included in the experience highlighted that 81% of participants considered the sweet to be either free (38%) or cheaper (43%) than in regular monetary exchanges. While treating data collected from public engagement activities such as these with care or additional robust analysis processes, the potential for not only engaging the general public in socio-economic thinking but gaining more detailed quantitative insights into their practices and behaviours is a powerful one. Most of the projects here were not intentionally aimed at quantifying behaviour or even collecting particular data. However it is notable that by creating these public experiences it allowed us to get some initial insights into how such installations could potentially offer additional quantitative basis for analysis of behaviour, use or perceptions in general or specific future socio-economic contexts.

### 5.5 New forms of value

The development of blockchains centred around exchanging currency – purely financial value transactions, despite their encoding in a system that brought together a collection of value*s* around autonomy, decentralisation and secrecy. However, the projects here used the exchange of financial value as a gateway to asking what other kinds of value could be traded. AfterMoney and Kash cups looked at shifting value practices and explored the kinds of value that could be exchanged, wether data, labour or sociality. Seismic SeeSaw looked at alternative structures for monetary exchange, giving rise to a new sense of what the value proposition around donating could be. In this case, looking at the values that were exchanged and constructed rather than the technical operations of the blockchains underlying the exchanges provided a clear entry point for publics.

Using alternative systems of value, supported new interactions. By nudging people to interact with a stranger at the conference to gain credit, delegates enjoyed the opportunity that the cups gave them to break the ice, and had a starting point for the conversation. Whilst the perceived 'labour' of the participants to earn credit for the coffee was the act of talking to a stranger at the conference, in most cases this 'friction' had a significant social value in itself - that of meeting new people. The playful uncovering of current non-monetary value exchanges (such as sharing data) in an experiential manner, offered audiences a playful way to understanding non-traditional forms of value. It helped non-specialist publics to act as engaged economic agents - making choices about what values to exchange and how to balance them. Overall, by holding up a collection of different value systems, people were able to make their own decisions and start to navigate what would otherwise be an abstract space by making situated value judgements.

New forms of value gave rise to new forms of contracts, both social and legal: Happily Ever After (Bitcoin) was a way to re-examine traditional marriage contracts as dynamic and precisely specified; Seismic SeeSaw offered new contracts for giving that connected donations to evidence of real world events – namely an earthquake. This possibility of using new technology as a platform to rethink society is extremely powerful - it engages critically with the promises of the technology, subverting the hype in service of creative thinking.

### 5.6 Summary and Directions

These projects used a range of design strategies, in particular seamfulness, playfulness, defamiliariastion and control of the visibility of information to create public engagement around blockchains. This leads to a focus on what it is that blockchains enable, rather than how they do it.

This also illuminates some of the difficulties of working with blockchains, in particular questions of inflexibility, time and how to mediate between the blockchain and the physical world. The escrows in Seismic Seesaw are valuable as a transparent enforcement mechanism, but the 'smart' contract remains entirely dependent upon the actors and data to which it is connected, which cannot be simply programmed out. While some envision vast DAOs, governed by a web of smart contracts, in practice their pre-determined and inflexible nature requires contracts to be simple and predictable. These projects raise questions about how to see smart contracts as part of a much larger socio-technical ecosystem, from the source of the data used to validate a contract, to the wider infrastructure that acts on behalf of the contract.

As well as exploring possibilities and limitations of blockchains, these projects opened up research directions that give rise to other projects in this portfolio. The openness of KASH Cups allowed an experimentation with alternative rulesets around the same physical objects. Similarly, Seismic Seesaw raised the question of how the concept of 'escrow', once internalised, could be applied to other situations. Together, as part of engaging with possible futures, this leads to the question of who gets to write the rules which underpins the projects in Section 6. There was also an engagement with the idea of autonomous objects, with marriages being formed with a family's dog or tree *w6xcv7tf* of the autonomous Terra0 forest[127]. This leads to an exploration of the idea that non-human actors can own wallets and start joining (socio)economic unions for mutual benefits, laying the groundwork for Section 7 to interrogate relations between humans and more or less autonomous objects.

## 6 MAKING THE RULES

The previous section on creating experiences described what happens when people are confronted by rules made by others. However, distributed ledgers are often touted as being able to open up the process of making rules, so communities can decide how they want to operate, leading to questions of what happens when people start to make the rules themselves. Where Bitcoin formalises a system of IOUs, other systems such as Lazooz [12] use a tokenized ledger to support social ridesharing, Mycelia [13] uses blockchains to manage music rights and so on. As with other aspects of distributed cryptographic systems, this builds on a rich history of work around how to structure, support and constrain interaction. Here, we start to see institutions emerge - collections of protocols that shape interactions by specifying what actions can be taken and what the implications are [106, 126].

A common thread through this work is the idea that the rules are decentralised. We are interested in approaches that investigate what happens when participants can create their own protocols and contracts. Writing any kind of contract or interaction protocol is a challenge, from conceptualising an idea of what should happen, through encoding this formally and then creating the community or social structures that mean people actually engage. If 'anyone' can write smart contracts, how can we ensure firstly that an inclusive group of people can participate in the writing, and secondly that an even wider group can understand what a contract means, have a sense of the consequences, and be able to make informed decisions about their participation.

This ties in with our attempts to foster a nuanced engagement with this technology, in a way that is connected and integrated with actual life. Participation and creativity are powerful drivers for deep engagement: when people start trying to design systems, it is a chance to think through in detail how the networks would operate. In particular, prototyping a system raises questions about how that system would interact with the rest of one's life, so frictions and problems become more apparent.

This section looks at three projects that take a participatory approach to the design of smart contracts, guiding participants to conceptualise and prototype ideas with varying degrees of fidelity:

- **The If This Then What?** (IFTTW) cards are based on a common design methods of ideation cards [64, 74] and are used to imagine and explore the possibilities of smart contracts. Inspired by graphical programming systems such as If This Then That [14] and Scratch [99], they express a simplified logic for smart contracts, using modular cards to build If...Then... or When... Then... statements in the context of collaborative ideation workshops.
- Building on the concept of programmable and conditional donations that emerged during the OxChain project[15], **Programmable Donations** used engagements with a partner NGO [44] and emerging blockchain applications in the wider humanitarian sector [28] to identify and explore particular possibilities for programmable donations through interviews, workshops and enactments. Outcomes from this led to a real-world trial with Oxfam Australia of a 'Smart Donations' Ethereum App [141].

[12]http://lazooz.com decentralised transport platform, now defunct.
[13]http://myceliaformusic.org/
[14]https://ifttt.com/
[15]https://oxchain.uk/

- **GeoCoin** [108] sets up an open bodystorming experience that leads into the ideation and prototyping of geolocated currencies. Participants use a smartphone app to explore smart contract architectures connected to their own city surroundings, and design their own new forms of value exchanges. GeoCoin was used in a series of workshops [24, 108] leading to a various concept designs, including Happily Ever After (Bitcoin) (Page 25).

All of these projects grapple with ways that the complex and often bewildering potentialities of new technology and abstract ideas can be used as a public playground, allowing those with no experience to rapidly understand the system and imagine new rules through participation.

# IF THIS THEN WHAT?

Design cards for smart contracts - developing ways for non-programmers to describe the logic of distributed systems

*Creates a basis for discussing logic structures for interactions*

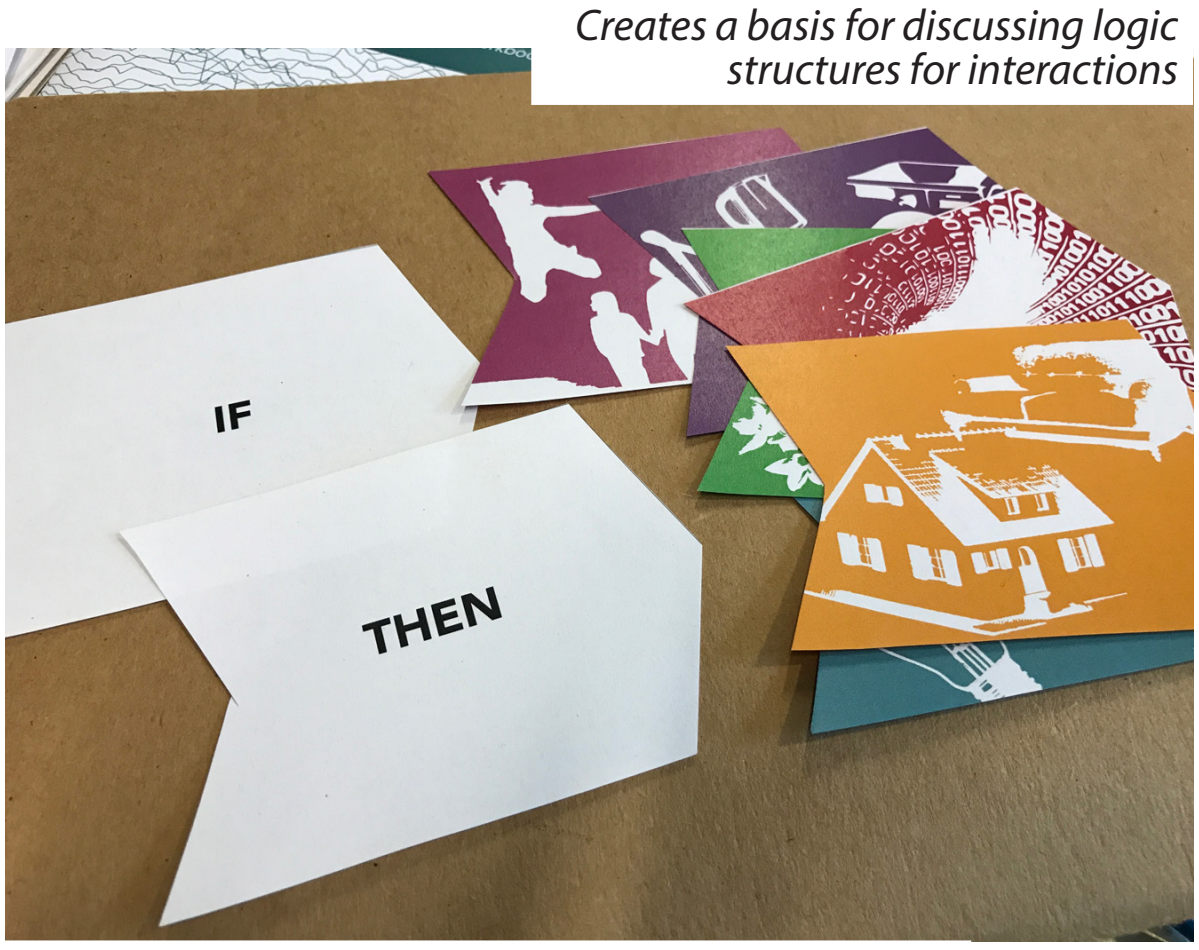

First iteration of IFTTW cards.

**Moments & Seams**

**Abstraction & Focus**

**People & Machines**

*Simple representations of logic are open to different levels of abstraction*

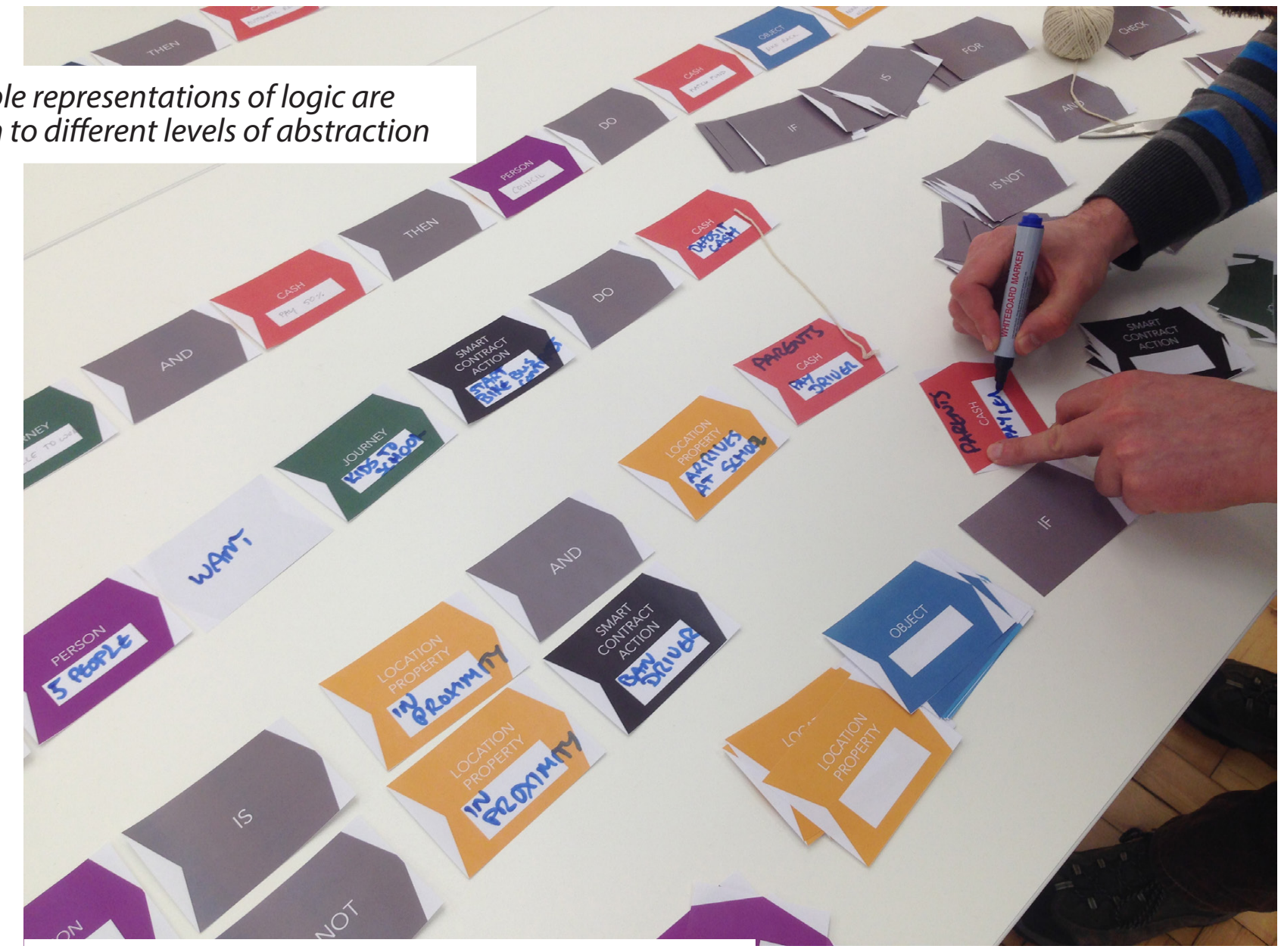

Second iteration of IFTTW cards.

# *If This Then What? - Details*

## Aims and Context

The If This Then What? cards scaffold development of smart contracts - distributed programs that run on blockchains. They wrap up key concepts around how to structure conditions and effects in a set of physical cards and a workshop structure. Participants with little or no background in programming or blockchains can ideate possible smart contracts and design blockchain applications and create visual prototypes of their ideas.

Smart contracts embody a lot of technical complexity - distributed, trustable computing with message passing, cryptographic functions and models of execution cost. However, many potential applications can be broken down into connections between certain conditions that may be met, and the actions that result. By creating a simplified, tangible version containing these conditions, actions and connectives, the IFTTW cards provide a framework for thinking and collaborating. They provided a structure for contract design, raising the level of detail enough that participants have a closer relationship to what would be needed in a real application. This meant that participants got a high level introduction to programming concepts, saw the challenges of grounding formal systems in the real world, and worked with the idea of 'self executing' code, that can run independently of the originator.

## Experience

The IFTTW ideation cards are used within collaborative workshops to imagine and explore the possibilities of smart contracts. They provide a framework for thinking both by providing a structure for contract design and by raising the level of detail enough that designers have a closer relationship to what would be needed in a real application. Participants were asked to speculate on potential smart contracts, and then try to create their structures using the cards

The cards went through a series of iterations from abstract pictorial forms to more specific descriptive cards, in response to what happened in the workshops. In addition to logical connectives, the first iteration of the cards included pictorial concepts grouped into high level categories: People, Things, Places, Nature, Resources and Currency, selected to cover key design concerns. Later iterations included more directed categories -- specific locations, modes of transport -- and more logical connectives such as 'is', 'is not', 'check', 'record', to enable participants to explore the potential mechanism more fully.

# PROGRAMMABLE DONATIONS

Using familiar objects to explore the possibilities of smart contracts for donating to charity in response to world events.

*Cards and leaflets support ideation...*

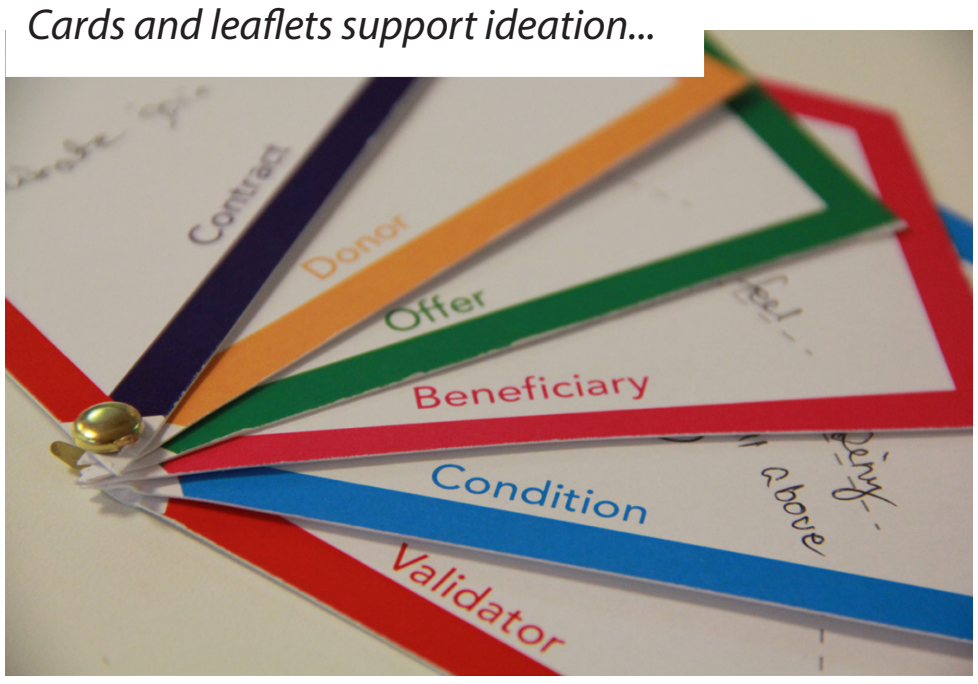

*... but conditions quickly become complicated.*

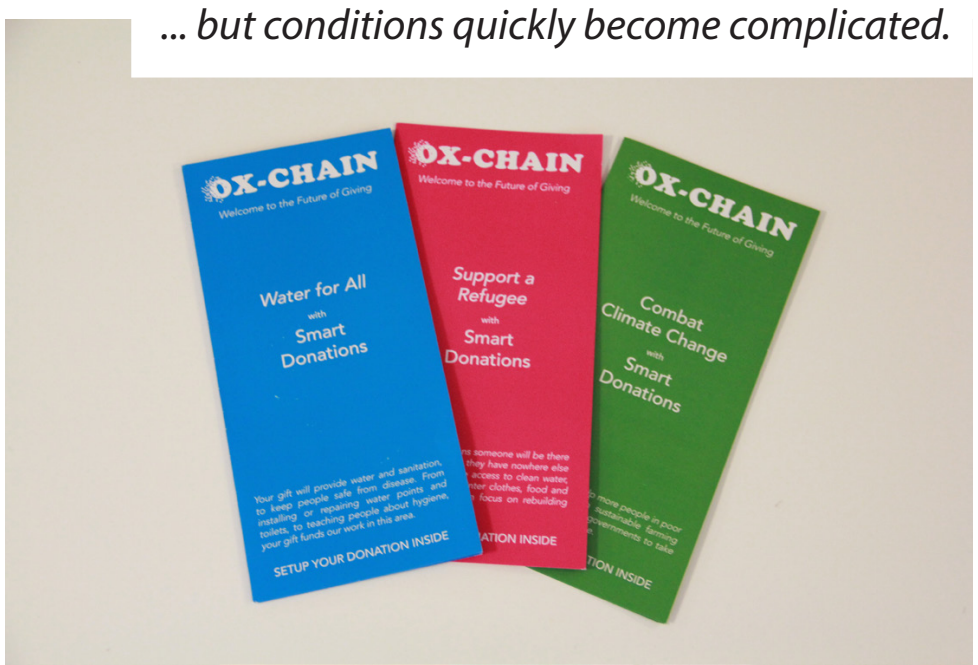

Top: Original workshop materials. Bottom: Glossy leaflets inviting speculation about Programmable Donations as a service.

**Abstraction & Focus**

**Rethinking Society**

*Imposing conditionality challenges the very nature of giving*

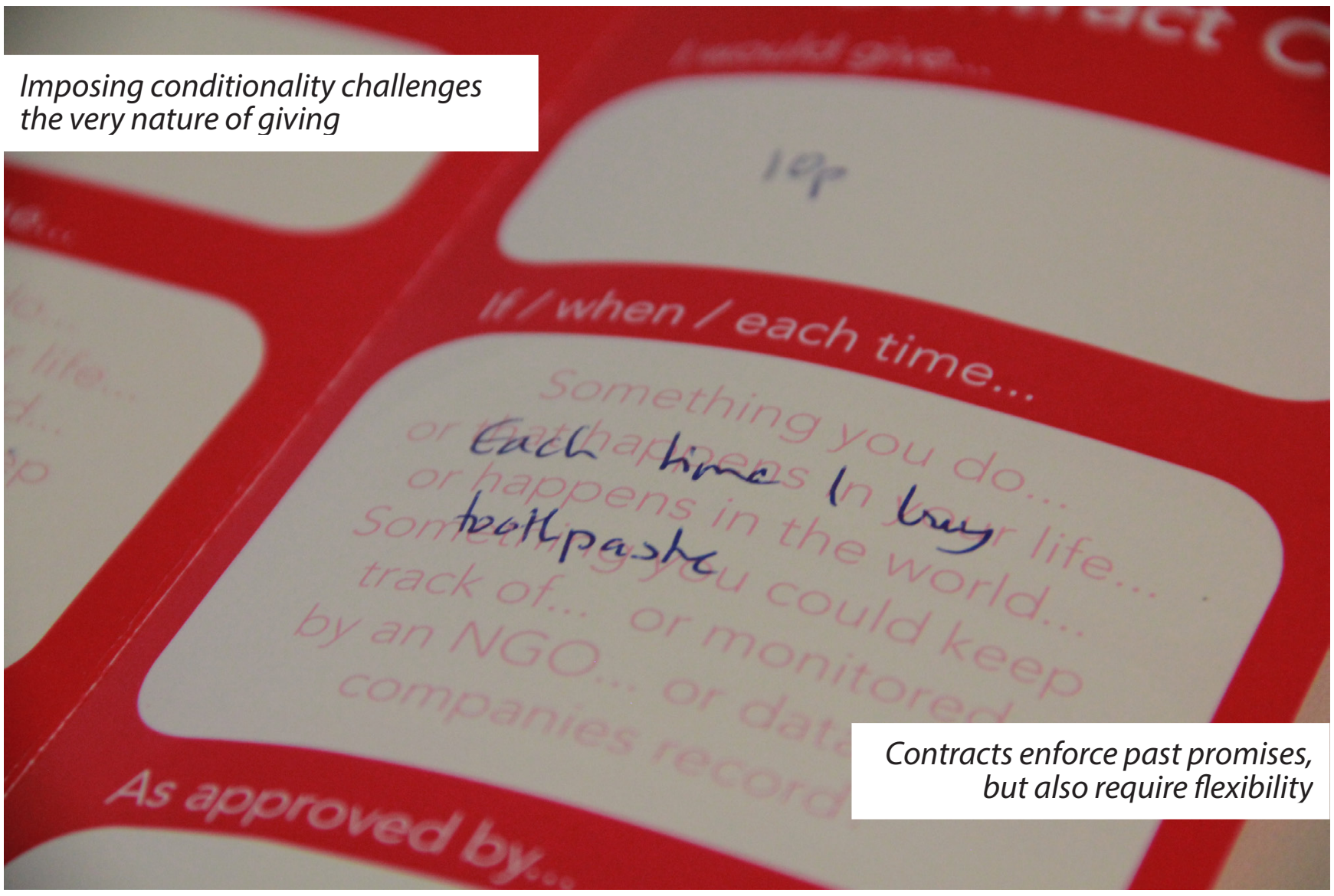

*Contracts enforce past promises, but also require flexibility*

Close up of one participants programmable donation for a refugee project.

# *Programmable Donations - Details*

## Aims and Context

Programmable Donations explores what it means to allow people to make and enforce very specific rules and conditions about how and when they give to charity. In this project, we co-speculated with a range of charitable donors the kind of rules and data they would seek to underpin 'programmable' donations - ways to give charitably that are dependent on events in the world. We set out to investigate new interactions and relationships with charitable giving that could be facilitated through blockchains or Decentralised Ledger Technologies (DLTs). The specific functioning of each donation could be governed by a smart contract, meaning its operation would be pre-determined, automated, independent, and reliant upon specific data inputs. For example, a donor could set up a contract, where they would donate £1 to the Royal National Lifeboat Institution each time a lifeboat was launched on a rescue mission from a nearby lifeboat station.

## Experience

Through a series of cards and glossy leaflets resonant of traditional charity appeals and services, we co-speculated with a range of charitable donors the kind of rules and data they would seek to underpin programmable donations. The cards and leaflets invited participants to envisage their own contracts, and articulate for various causes: an amount they would give, conditions they would set, and data they would trust to 'validate' those conditions. For example, participants imagined donating to water charities each time they went swimming, or donating to slow climate change if average temperatures in a location were rising. Through these simple speculative activities, participants could envisage wholly new relationships and values to be expressed in the way they would give to charity.

# GEOCOIN

Programmable location based currency - design your own geospatial financial interactions

**People & machines**

**Contextualisation**

**Rethinking Society**

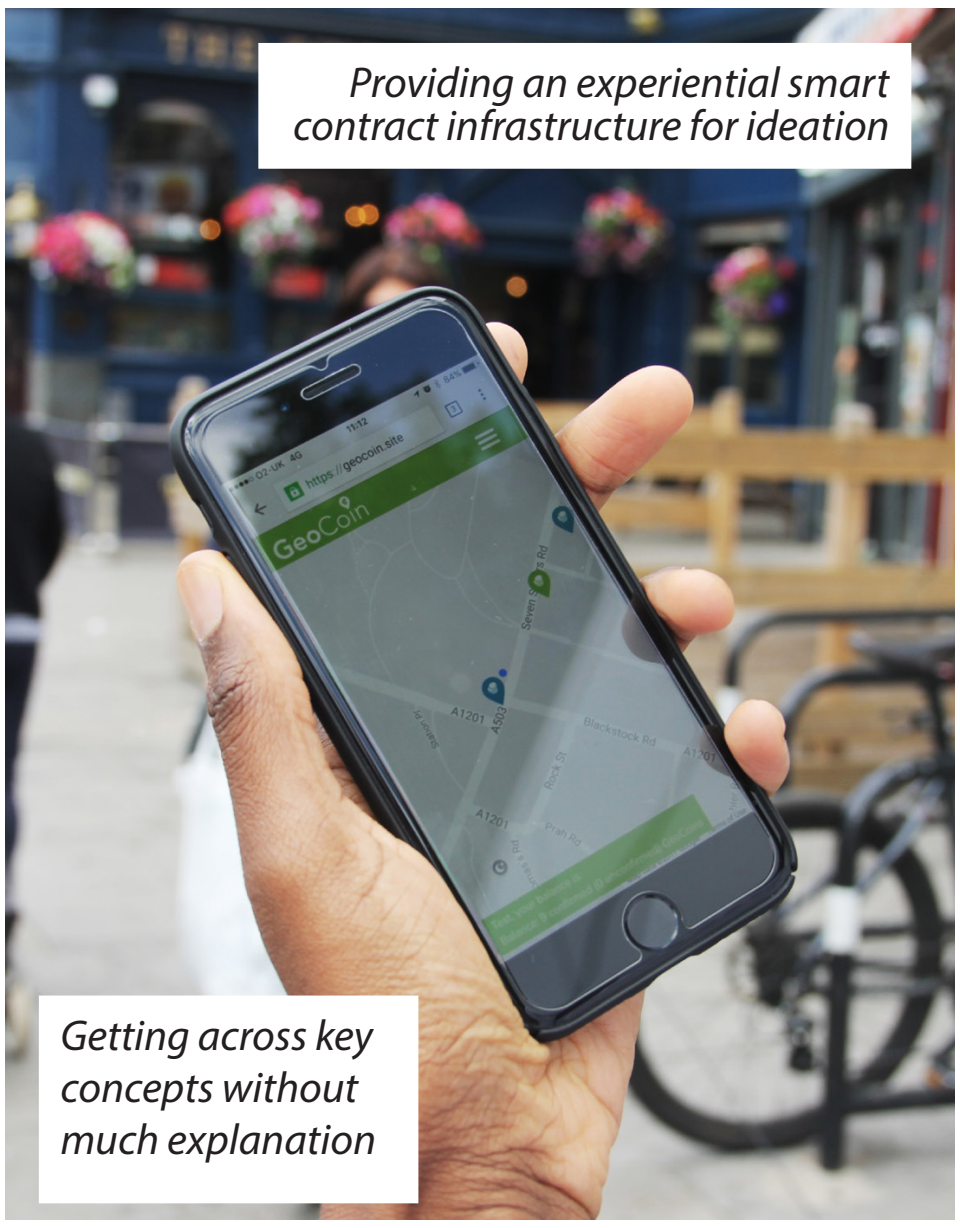

GeoCoin Interface in action

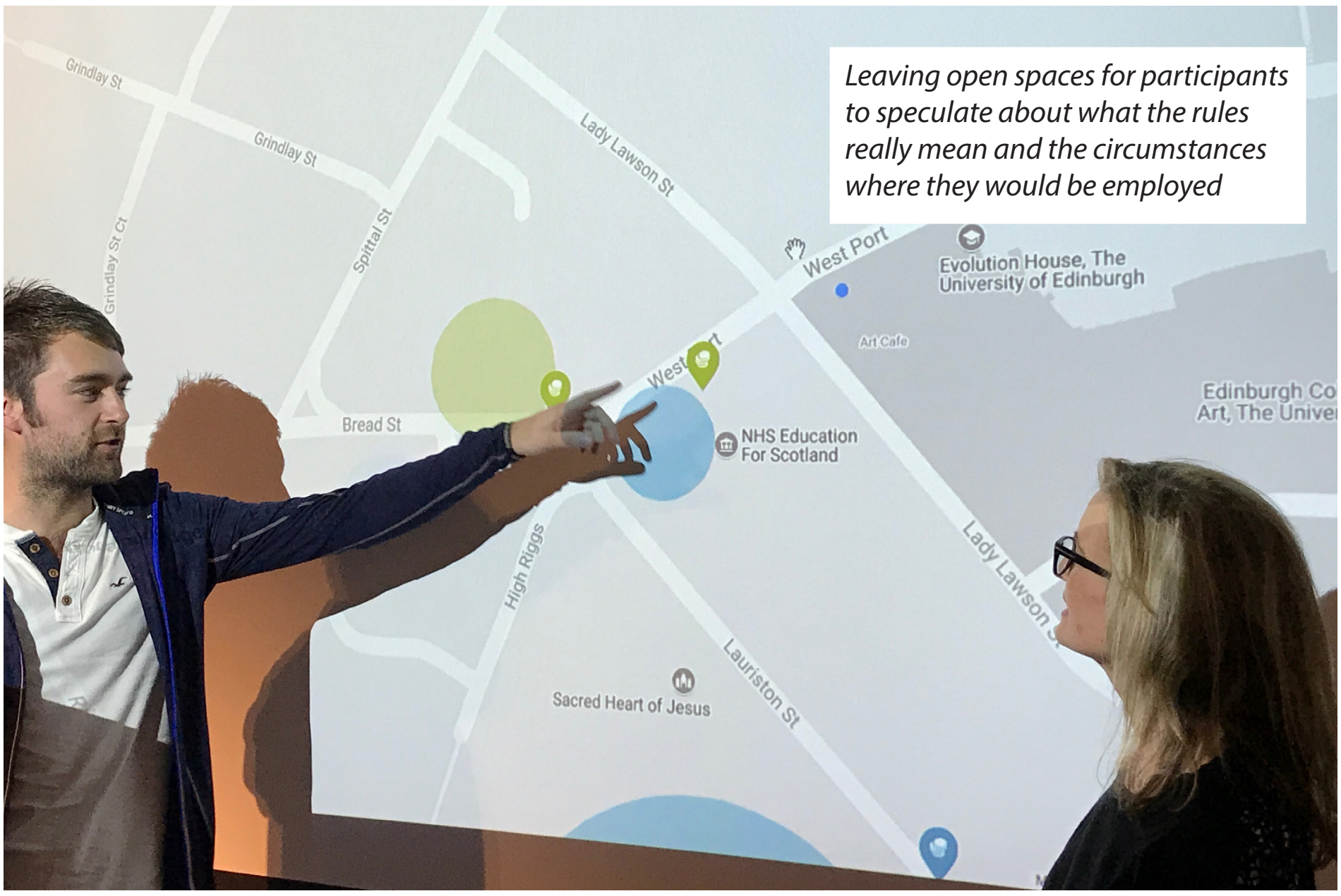

Developing possibilities around GeoCoin interface in a workshop

# *GeoCoin - Details*

## Aims and Context

GeoCoin is a location-based platform for experiencing and ideating with smart contracts. In collaborative workshops with GeoCoin, participants engaged with location-based smart contracts, using the platform to explore digital 'debit' and 'credit' zones in the city. Smart contracts are attached to GPS locations, and participants interact with them using a smartphone app that shows the contracts on a map as they move through space. The app maintains a digital currency wallet for each person, so that participants can see changes to their balance in real-time. Debit and Credit zones can be easily administered through a web interface allowing workshop organisers and designers to quickly set up new experiences alongside participants.

GeoCoin demonstrates how an experiential prototype can support understanding of the complexities behind new digital infrastructures and facilitate participant engagement in ideation and design processes. It is a location-based platform for embodied learning and speculative ideating with smart contracts in the city.

## Experience

In practice, this means that participants use an app on their phone to use a geolocated currency, that allows for structures like 'debit' zones that charge the participants while they are there, or 'credit' zones where the participant is paid for the presence. Participants were guided through:

- an initial understanding aspects of cryptocurrencies, blockchains and smart contracts using the BlockExchange workshop
- a pre-made experience using GeoCoin, where they explored a map configured with a collection of credit and debit zones, that were open to interpretation—tolls on bridges, promotions from supermarkets, privatisation of public space and so on
- ideation around potential applications for geolocated smart contracts.
- collaboration to realise prototypes of their ideas.

These collaborations created systems such as a participatory budgeting application that allowed anyone to geolocate a project that would then be funded based on how much time people spent in that space, and a short-term wedding app that formed the basis of the Happily Ever After (Bitcoin) project.

## 6.1 Common Themes

The work in this section is focusing on making new rules for decentralised systems. Beyond previous workshops and public experiences, these projects support people to think through the real implications of how lives might be governed by computational systems, and gets into questions about who can make and change the rules. This computational governance is primarily explored here through the lens of smart contracts. Like much blockchain infrastructure, they have subtle and interesting properties, for example: they are immutable, which means they can be a foundation for shared trust, but at the same time cannot adapt dynamically to changing circumstances; they are strongly internally consistent, but have trouble connecting to the outside world; they expose the difficulty of encoding even relatively simple sounding ideas, both technically and conceptually. The projects here explore the notion of technical fidelity, looking for the right levels of abstraction to explore concepts of interest.

## 6.2 Levels of Detail

In each of these projects, finding the right levels of constraints and detail was important; while abstract images helped to open spaces for thinking, in the end it was more valuable to give concrete use cases or application contexts to help ground the work in real concerns and plausible applications. The initial IFTTW cards were somewhat ambiguous, in order to give participants a range of possible interpretations and meanings to foster creativity [61]. However, this created a lack of focus, and a distance from the technological possibilities. In response, a second iteration of the cards were developed for a more specific application area – transport – with more directed categories and topics: specific locations, modes of transport, transactions, behaviours and incentives. The workshop structure was refined: rather than using the cards for conceptual ideation, they were used to refine an idea through visual prototyping.

The use of application domains flowed through the other projects, as a way to engage with abstract ideas. Programmable Donations was squarely focused on humanitarian aid. By situating GeoCoin strongly in terms of the city, audiences ranging from arts organisations, artists, designers to academics, industry and blockchain experts could collaborate, using the smart contract infrastructure as a place to connect their ideas.

Across the projects, finding a core set of simple rules that mediated between technological possibilities and participant understanding was a key enabler for creative thinking. The initial GeoCoin experience only had three types of action - zones that continually credited or debited user accounts, and 'single use' zones that would make a one-off payment to the first person to enter. While this is simplistic on the surface i) it gets across key concepts without too much explanation and ii) it leaves open the space for participants to speculate about what the rules really mean and the circumstances where they would be employed.

## 6.3 Relating imaginaries to computational thinking

Each project manifested computational logics in a manner designed to be accessible. However, there are various levels and types of understanding that participants bring to bear, resulting in different relations to the underlying logics.

Even with the refined IFTTW cards, there was a spectrum between thinking of the cards as an imperative program unfolding over time and seeing them more as a box and arrow diagram. While using the cards helped the participants to both develop computational thinking and get a sense of the complexities and difficulties involved in creating their potential systems, participants paid varying levels of attention to the order and structure of the cards, the plausibility of the connections to the real world and the operation of the system as a whole. Ideas for a conditional or data-driven donation were easy to envisage with Programmable Donations, but much harder to calibrate, as participants navigated the complexities of formalising a contract to account for the uncertainties of dat to day life. GeoCoin took a slightly different approach, in passing on the creation of an underlying logic of location-based currencies to a practiced developer. This meant that participants could interact with a realised system of abstract rules without a specific purposes. The playful investigation of these computational rules then facilitated discussion and imagination about how and when such rules could be meaningfully used – while also not hiding problematic spaces or applications.

### 6.4 Infrastructural Experiences and Social Implications

These projects work with varying levels of experience. GeoCoin in particular borrowed elements of design methods such as probes and experience prototyping, as it offers an open-ended experience for participants. While experience prototypes [14] and probes [72] are often focused on gaining understanding of participants' perceptions, GeoCoin attempts to provide an informed experience of smart contracts, inviting participants to intervene in or extend the system. The aim was to mediate learning so that participants would feel empowered to apply smart contracts creatively. Overall GeoCoin presents an experiential platform as open, 'unfinished software' [108], supporting understanding and facilitating engagement in ideation and design with smart contracts and location-based infrastructures for value exchange.

Using real infrastructure was surprisingly powerful, and the increased fidelity (Section 4.5) showed up potential issues quickly. In several workshops, the participants with newer phones were able to report their locations more rapidly and accurately, which meant they could claim locations first. This helped participants to think about what happens when systems function imperfectly and the inequalities that are likely to occur. While Programmable Donations did not build out the infrastructure, the materiality of the presentation was key to developing a sense of how the experience would play out in the world.

It is seemingly attractive to plan out how and when to give to a charity, but when can we really comfortably make such firm commitments in advance and what are the implications of doing so? Programmable Donations showed how imposing conditionality and rules challenged the very nature of giving. Once conditional, donations began to resemble forms of insurance or tax, and suggested new obligations that might be placed on a donor or charity. Sometimes these obligations were an effort to attain accountability, for one's own behaviour, or a charity's actions, but on other occasions, they occluded some of the original intentions of supporting a particular cause. Working through the IFTTW cards to set conditions and possibilities let participants ask what-if questions based on other possible logics, and develop a picture of how people might be affected by rules of a given or new the systems. During GeoCoin workshops many participants experienced loss of agency due to internet connectivity issues or other technological limitations in real time. This was a powerful experience raising discussions around the correlation between access to technology and access to resources, associated privilege and social implications of designed rules, who sets and controls them.

### 6.5 Temporality and Trust

Working through these experiences helped participants to identify key issues around system deployments. Temporality came to the fore, as participants in GeoCoin contrasted immediate experience and feedback with the delay in blocks being verified on the chain, illuminating particular technical infrastructural details. Programmable Donations quickly raised the questions of what kind of limits should be placed on donations, and how far into the future could participants attempt to set rules or conditions? Between this and developing self-executing code with IFTTW, we highlighted the tradeoff that the very power of smart contracts is their ability to enforce firm commitments over prolonged periods – a kind of 'slow technology' [67] - but as times or circumstances change, a rigid one and people may need more flexibility.

Similarly, the importance of trust emerged through the projects, with leaky infrastructures in GeoCoin asking how well we really know where people are located. The validity of Programmable Donations entirely hinges on receiving a trusted data input to notify the escrow if the conditions have been met. For many donors, it was challenging to specify who exactly should be trusted to arbitrate rules and conditions and how.

### 6.6 Summary and Directions

Through the experience, appropriation, design, modification, and testing of participants' own concepts, they were able to learn and express new understandings about their environments, social contexts, and economic and political concerns in relation to smart contracts. These exercises led to the design of diverse distributed-ledger applications, for time-limited financial unions, participatory budgeting, and humanitarian aid.

The work is not completely tied to smart contracts, but as with many other projects, provides a springboard for thinking critically about power, control and governance. There was a range of granularity in the responses of the participants, from very high level conceptual works independent of the characteristics of the contracts, through to work that meticulously built up different parts of an infrastructure with a solid grounding in technology.

There are tensions between trustworthiness and flexibility, between freedom of engagement and the constraints of shared rules. Working through potential sets of rules has proven to be a useful way to get across the difficulty of designing good systems, and the extent to which rules encode values. A key notion here is *uncomfort* - when positive pictures of systems operation meet the difficulties of translating the messiness of the real world into code, it challenges challenges assumptions and expectations, and forces participants to reconsider these technologies in practice.

In Section 7, we present several projects that look at objects embodying these rules – physicalising aspects of formal or imagined systems, and giving some power to act on their own as well as highlighting the infrastructures governing them.

## 7 THINKING THROUGH AUTONOMOUS THINGS

Many of the pieces so far have been concerned with how humans experience or understand blockchain technologies in different ways, whether their basic functions, their social possibilities or the collaborative development of rules and activities for their use. A theme that has been arising from several of these projects is the relationships between humans, the objects around them, and the wider infrastructure in which these systems are embedded. In particular, distributed ledgers, through their codified rules and machine friendly APIs offer the possibility for non-human things to have greater agency in the world, in particular by having their own wallets and being able to enact financial decisions. This means that a coffee machine can manage its own money[118, 136], or a bicycle can hire itself out with no intermediaries.

In this section, we discuss four projects, with similar underlying themes: what happens when objects can act for themselves, and how do we go about designing these systems? The development of autonomous things is not in itself a new idea. From cybernetics through to driverless cars we have seen increasing physical agency given to things, while work in autonomous agents has looked at how software can make decisions and interact with others.

The following projects explore different aspects in this space:

- **GeoPact** looks at creating location aware objects that participate in smart contracts; it occupies a different level of technological readiness from some of the other systems. The project was directly motivated by work on GeoCoin [108], and intended to address some of the questions and possibilities raised in that project, by asking what might be needed for a mature system.
- **Bitbarista** is a Bitcoin enabled coffee machine that explores cryptocurrency, autonomous objects and supply chains; it has its own Bitcoin wallet, and negotiates with its users about the supply chains of the coffee they consume. It is robust enough that it has been used in two studies [118, 136] that explored interactions with the machine and perceptions of new models of buying coffee in lab conditions and when deployed in a working office for a period of several months.
- **GigBliss** [116] is a series of speculative hairdryers designed to discuss the impact of algorithmic transactions in a context of distributed energy provision where devices can mediate new forms of value and profit. The hairdryers can trade energy with a smart grid, and use smart contracts (e.g. on an Ethereum blockchain) to define who has control over the energy supply and who will profit from its trading.
- Extending the research initiated with the GigBliss hair dryers, the **Karma Kettles** [117] mimic a scenario of energy storage, consumption and sharing in Distributed Energy systems. Through improvisation with actors and roleplay, the Kettles explore how to design complex infrastructures through deliberation and engagement.

Joined together, the design and development of these autonomous things intends to assess and discuss how decentralised systems and autonomous behaviours can affect wider infrastructures – of trade, transport and energy – to move beyond the human-centered focus of blockchain technologies raising questions of control and agency for potential future value exchange between things (and people).

# GEOPACT

Exploring agreements between objects, people and space, using smart contracts and IoT

*Creating high fidelity system supports scrutiny of real issues for application*

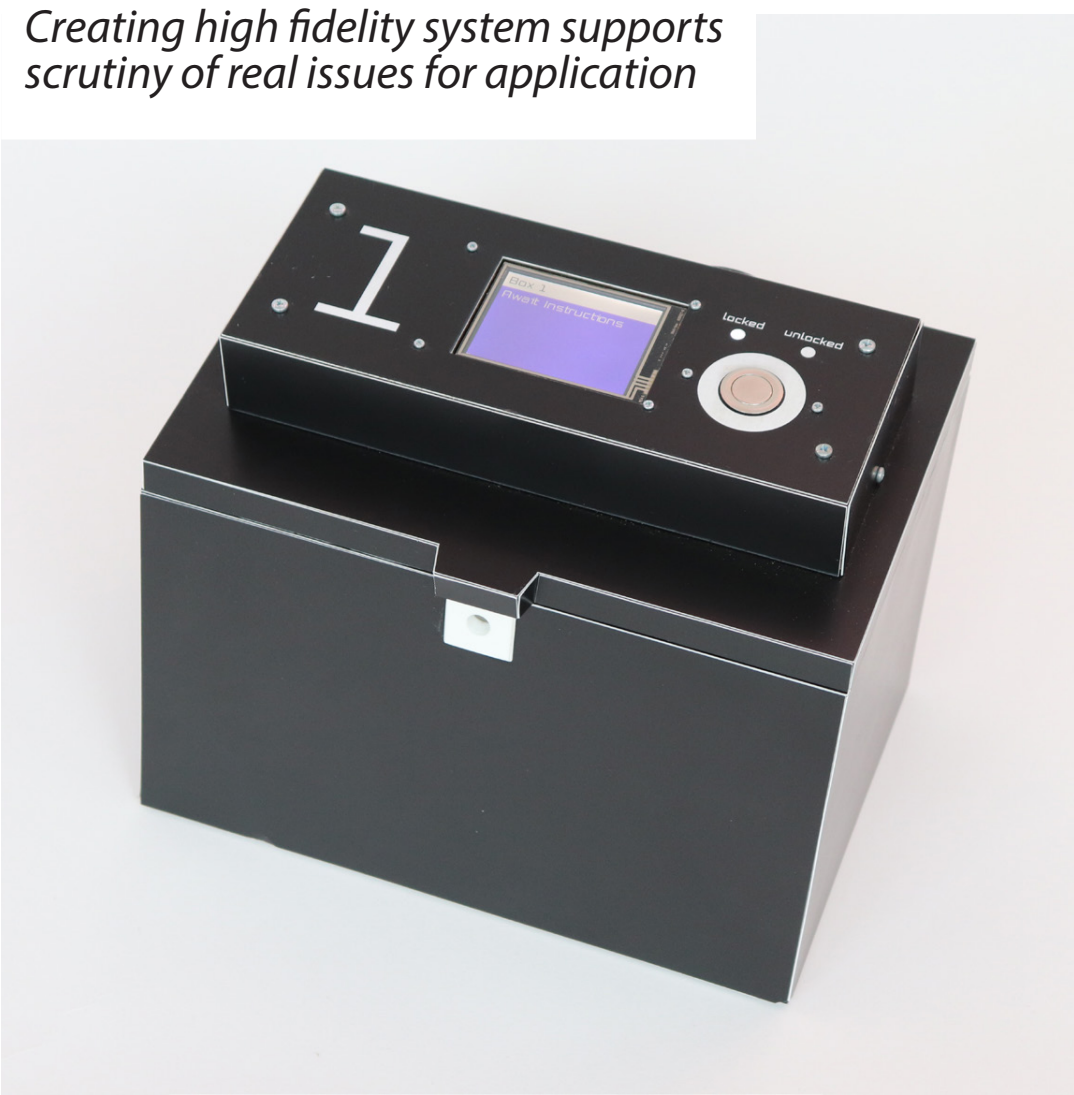

GeoPact lock-box. Image credit: Joe Revans

**Roleplay & Collaboration**

**Fidelity of Technology**

**Contextualisation**

*Concretising smart contracts in tightly curated scenarios helps demystify concepts*

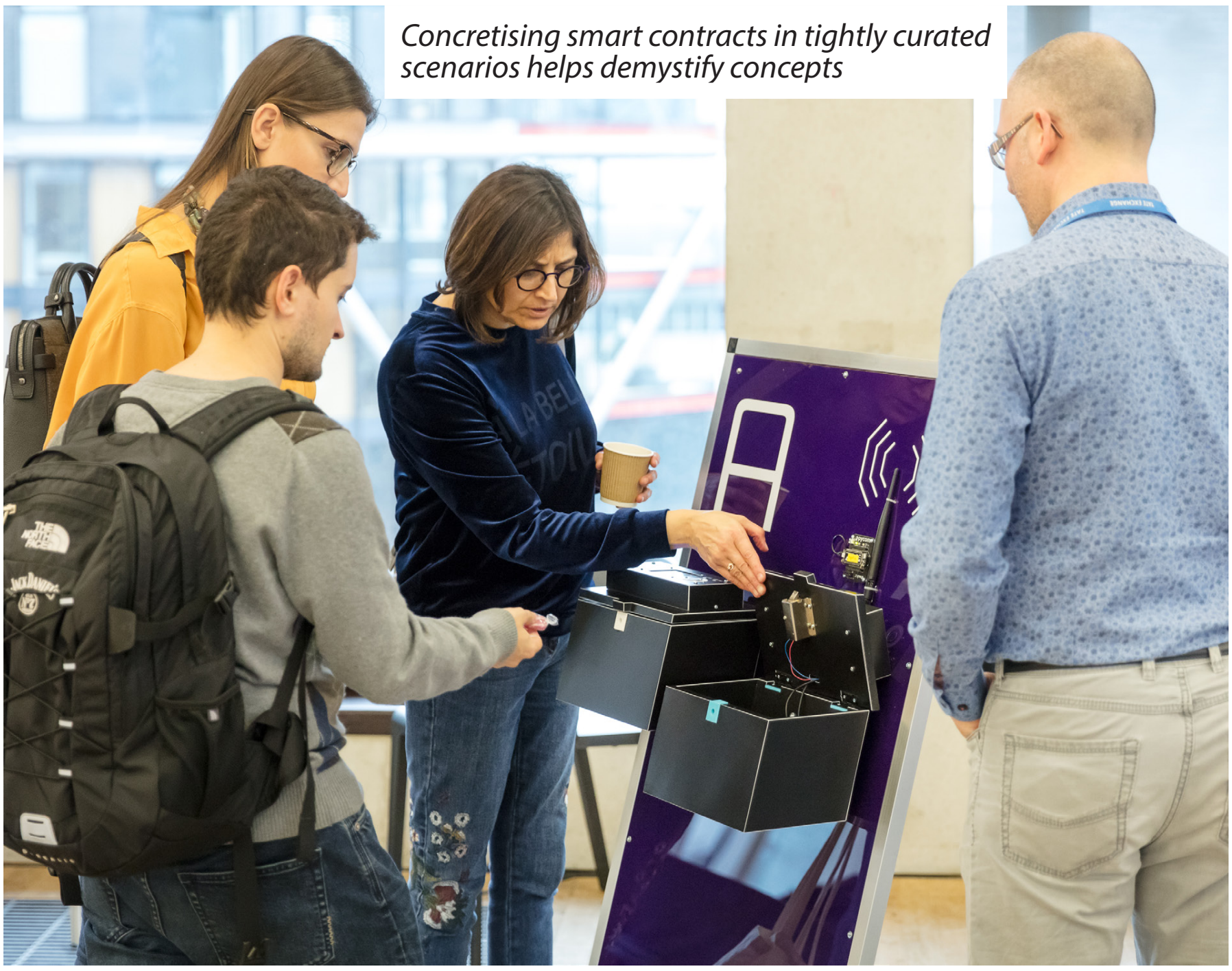

GeoPact in exhibition at the Tate Exchange. Image credit: Dan Weill

# *GeoPact - Details*

## Aims and Context

GeoPact is an infrastructure for creating location-based smart contracts. It uses a LoRa long range IoT network to provide verifiable location data, which is then stored securely and anonymously in a blockchain. Geolocated smart contracts use location and proximity to control the operation of smart objects such as lock boxes that open up when the right people are in the right place at the right time. Initially focussed on promoting low-carbon travel, GeoPact has enabled publics and transport experts to engage with smart contract enabled delivery scenarios.

The development of GeoPact arose from two threads. Firstly a combination of ethnographic and social science work to make sense of what is desired and needed from this kind of system - how do people relate to space, and what transactions would they like to enact?

Secondly, an attempt to put as much as possible of the logic onto the blockchain, so that interactions are fully defined using smart contracts, allowing a grounded design exploration of the technology..

## Experience

The system has been manifested as a series of demos around the UK in 2018 and 2019 that use smart 'lock boxes' as actors within smart contracts. Each box knows where it is, and whether other infrastructure is nearby, and can choose when it is open or closed. They can give participants instructions through a screen, and ask people to 'verify' that certain things have happened. Based on this, we created a set of scenarios in which we could create simple contracts for people to execute – for example, a supply chain, that required moving various parts from place to place in order to construct an electric car. However, we could attach different security considerations to each stage - does a human need to present to verify? Can the courier open the box they are carrying? Does the courier get to see what is inside the box? All of this activity was displayed on a large screen, indicating the current state of the contract and the next actions alongside a blockchain backed log of all of the events that had taken place. Based on this, the general public in open settings could participate in an experience lasting between one and five minutes, that conveyed the central aspects of smart contracts and blockchains, through tangible interactions.

# BITBARISTA

Using autonomous objects to understand relations between everyday habits and global supply chains, with a Bitcoin powered coffee machine.

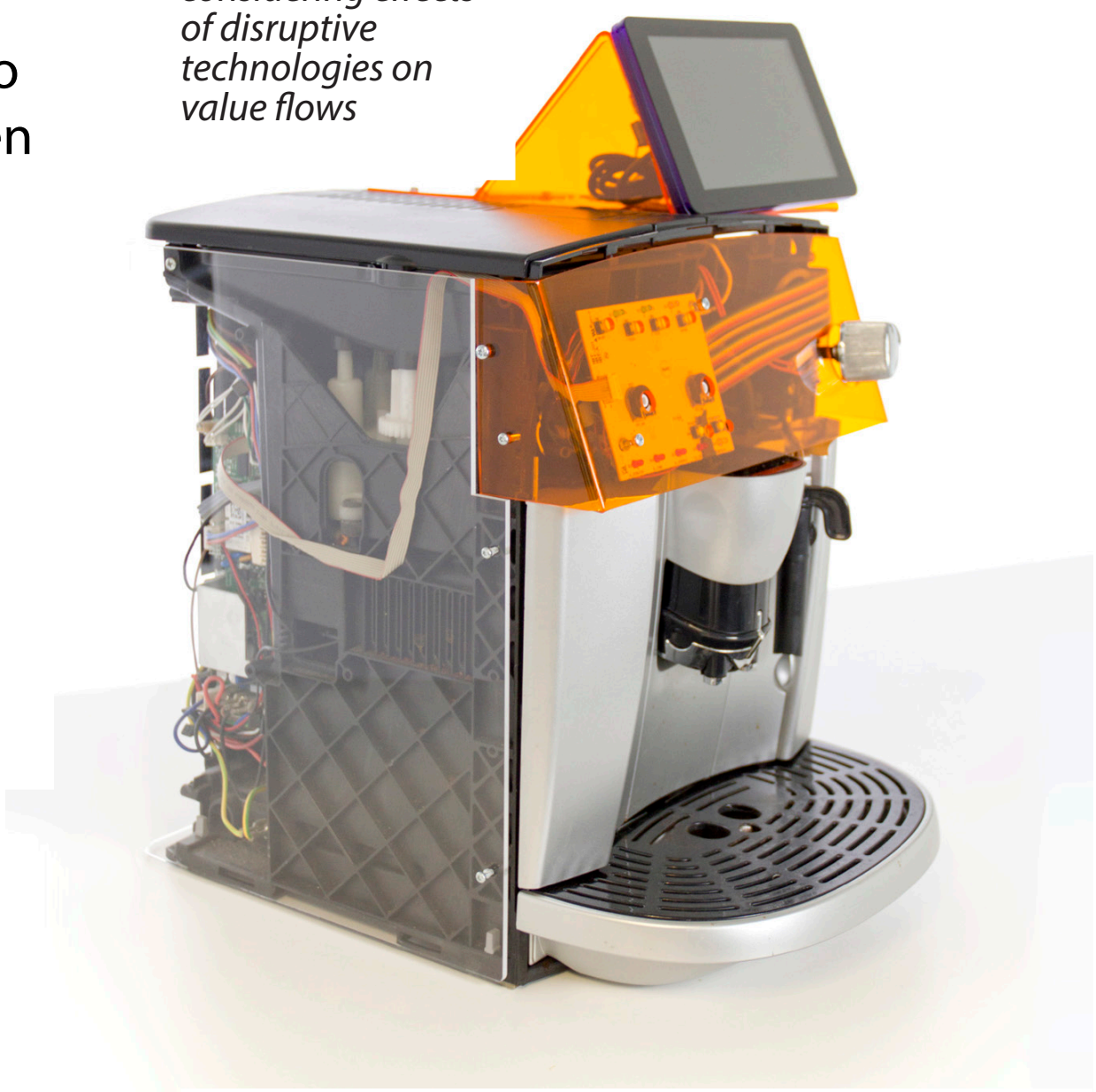

The Bitbarista, a hacked Delonghi home coffee machine.
Image credit: Mark Kobine

**Moments & Seams**

**People & Machines**

**Rethinking Society**

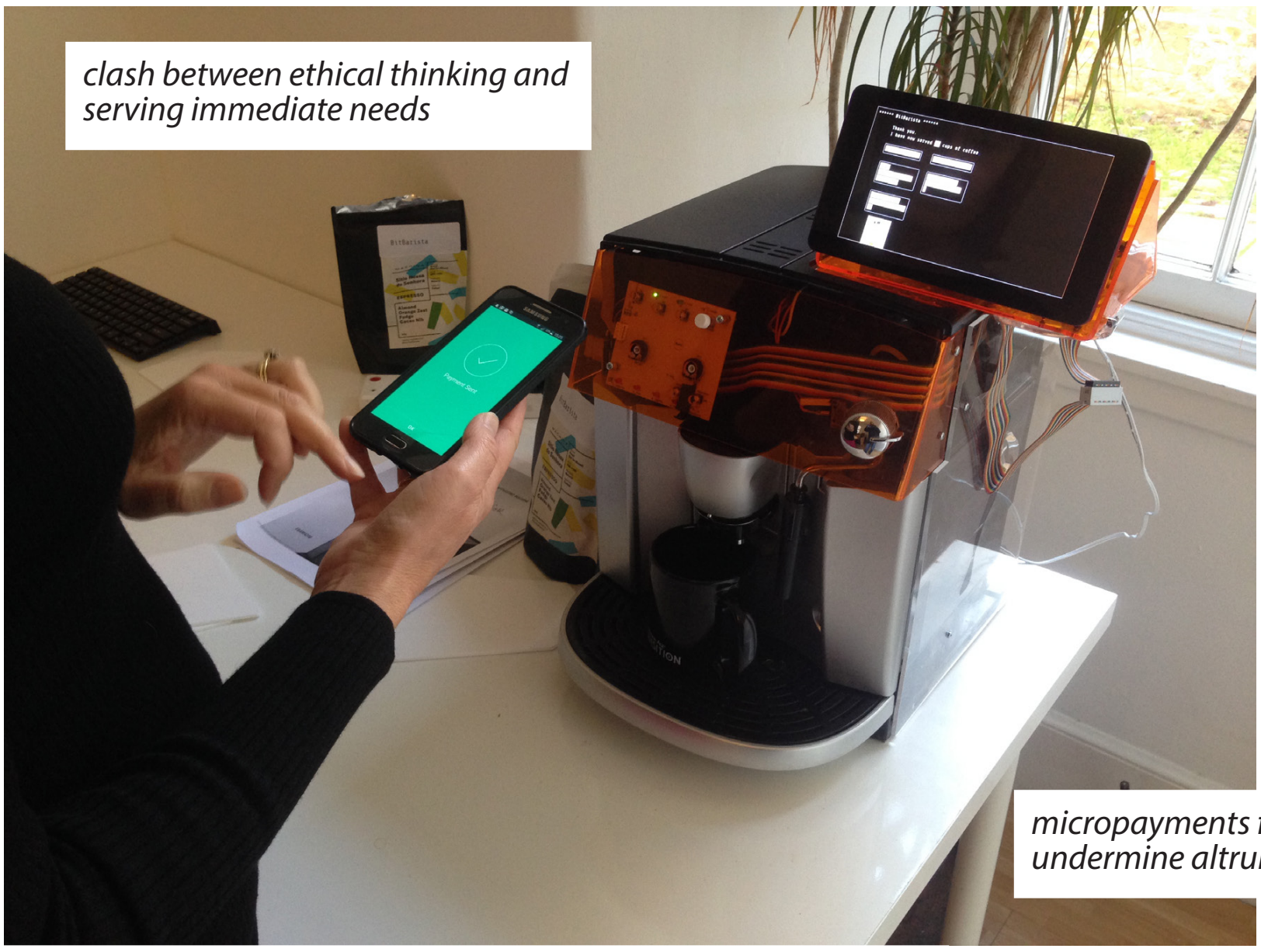

The Bitbarista in use in a co-working studio in Edinburgh.
Image: Ella Tallyn

# *Bitbarista - Details*

## Aims and Context

Bitbarista is a Bitcoin enabled coffee machine that connects consumers to producers through the autonomous machine and embodies the idea of radical disintermediation within the supply chain. Giving the machine a wallet allows for a greater sense of its autonomy, which provides a moment for coffee drinkers to notice the way that coffee comes to them. The sourcing of coffee beans through the voting process encourages customers to consider the origins of their coffee at the time of purchase, offering greater transparency of the value flow through supply chains and the impact of consumer practices on broader societal issues. Involving customers in the ongoing maintenance of the machine in return for micropayments is intended to stimulate thinking around the effects of Bitbarista's role in the flow of value at the consumer end of the supply chain, the potential impact of this on their individual experiences and activities, their immediate community, and that of wider society.

## Experience

The Bitbarista is used in studies of how people relate to financial technologies - it takes the place of an existing coffee machine, living with the users for a period of time.

The Bitbarista has its own Bitcoin wallet, that enables it to trade directly with customers, via their wallets on smart phones. It is presented as an autonomous object that buys its own coffee, but allows customers to vote for potential future coffee supplies. It displays data relating to the production of coffee to customers, such as labour conditions, geopolitics, quality and price, and purportedly analyses this to present choices of potential sources. The customer is able to vote for the future supply by selecting from these choices, but is offered the coffee voted for by previous customers. The price of the coffee served depends on the choice of future supply.

The Bitbarista also offers Bitcoin micro- payments, or free coffee, to customers in exchange for maintenance tasks, such as refilling the hopper with coffee beans, filling its water tank, and cleaning out used coffee grounds. Bitbarista has sensors that detects when these tasks are required, and is fitted with a small camera to scan QR codes from customers bitcoin wallets, in order to make these pay-outs.

# GIGBLISS

Exploring energy futures with speculative hairdryers connected to smart grids. GigBliss uses roleplay with actors to paint rich pictures of technology

**Moments and Seams**

**Roleplay & Collaboration**

**Rethinking Society**

*highlighting competing values and interests in distributed energy systems*

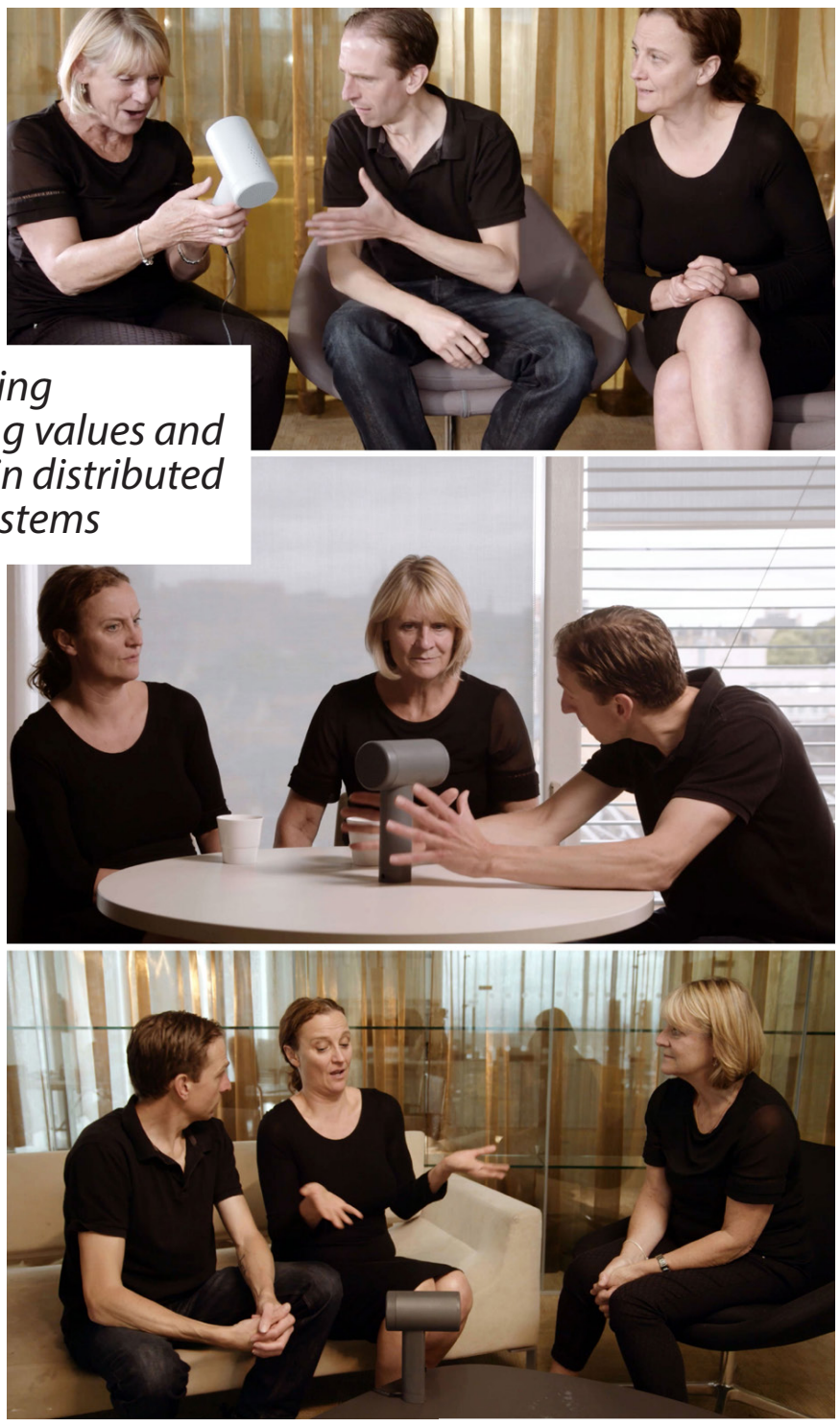

Actors improvising sketches based on the GigBliss prototypes

*Allowing participants to define what would work for them in distributed systems*

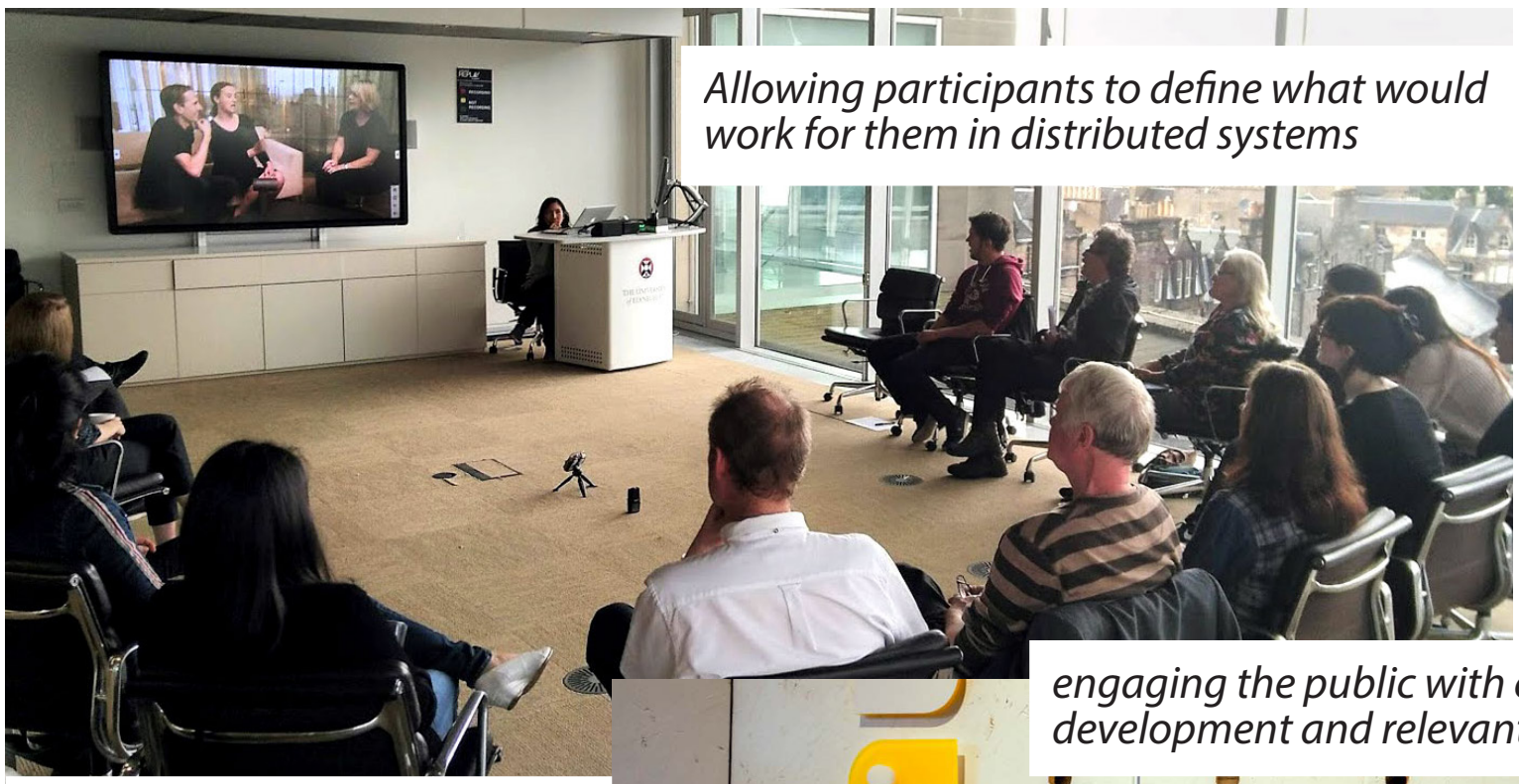

Workshop participants: using video and prototypes to support discussion of values behind energy systems

*engaging the public with complex technology development and relevant policies*

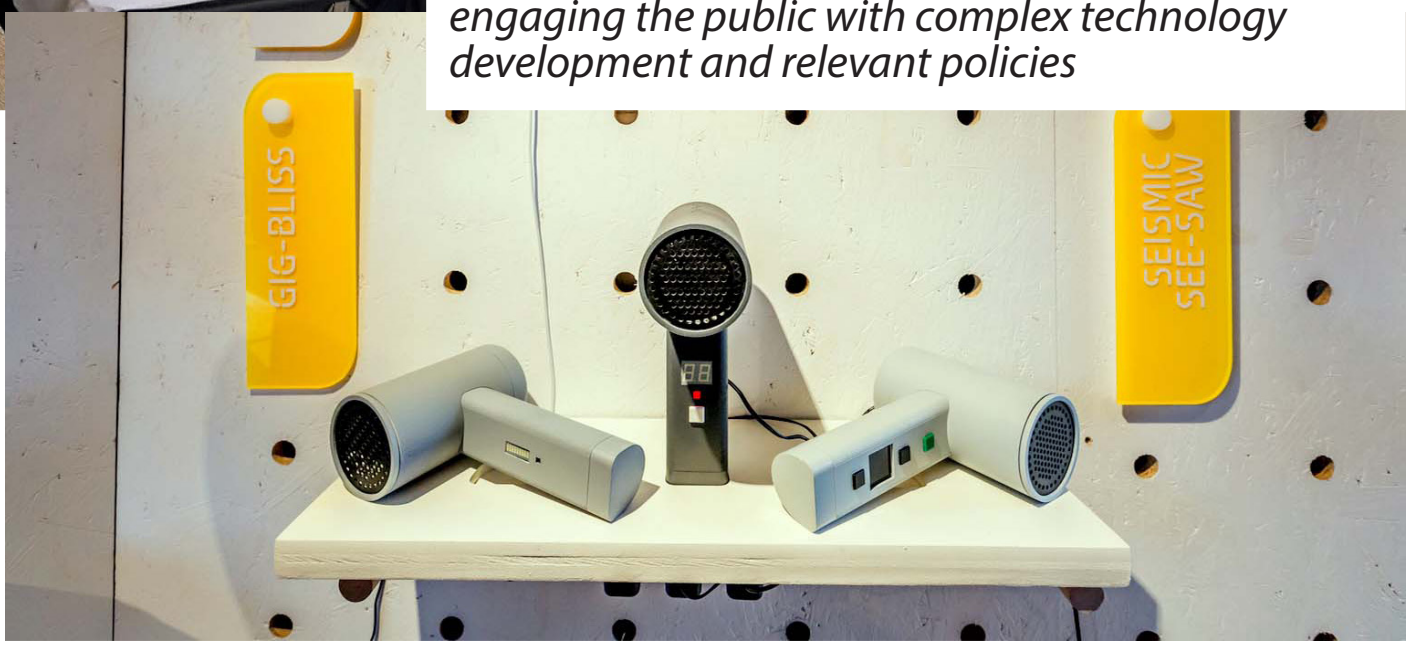

GigBliss prototypes in exhibition

# *GigBliss - Details*

## Aims and Context

The GigBliss hairdryers represent three scenarios that illustrate different levels of control in distributed energy systems. The hairdryers aimed to provide a tangible context to engage the general public and policy makers in discussions around algorithmic transactions that can mediate energy usage including issues of ownership, control, transparency, and predictive decision-making. The three models have decreasing levels of control to indicate that the design, not only of the device but the service and infrastructure behind it, could exacerbate existing power asymmetries between individual users, corporate bodies and governments. This raises questions regarding who defines algorithms that support autonomy, for whose benefit, what is the integrity of the data and what would be the social impact of different approaches. These systems also complicate relationships between stakeholders, disrupting traditional notions of value, control and ownership.

## Experience

The three concepts were developed into functioning prototypes, which worked based on simulated data stored in each device. In the workshops, professional actors were invited to perform critical sketches, also interacting with participants to mimic and discuss issues around control and autonomy in distributed energy systems . The the versions are:

- **GigBliss Plus** combinines  energy storage with the ability to track energy prices, letting owners control transactions to buy energy when prices are low and sell when they are high
- **GigBliss Balance** is based on a sustainable business model that allows consumers to host the hairdryer and return it to the GigBliss factory when the device is no longer needed. When inactive, it trades energy on the blockchain, but at peak times, the device might be busy trading energy and users might need to wait a few minutes to dry the hair.
- **GigBliss Auto** allows third parties to subsidise costs of both devices and electricity supply for a particular group or set of hairdryers. The device would be made available for free via local councils, community services and charities. With no buttons on its panel and no user control, the hairdryer can only be used at very specific times, potentially defining people’s actions rather than vice versa.

# KARMA KETTLES

Explore the future of energy transactions though a connected kettle. Will you pull energy or push it out for others to use? and will this improve your karma?

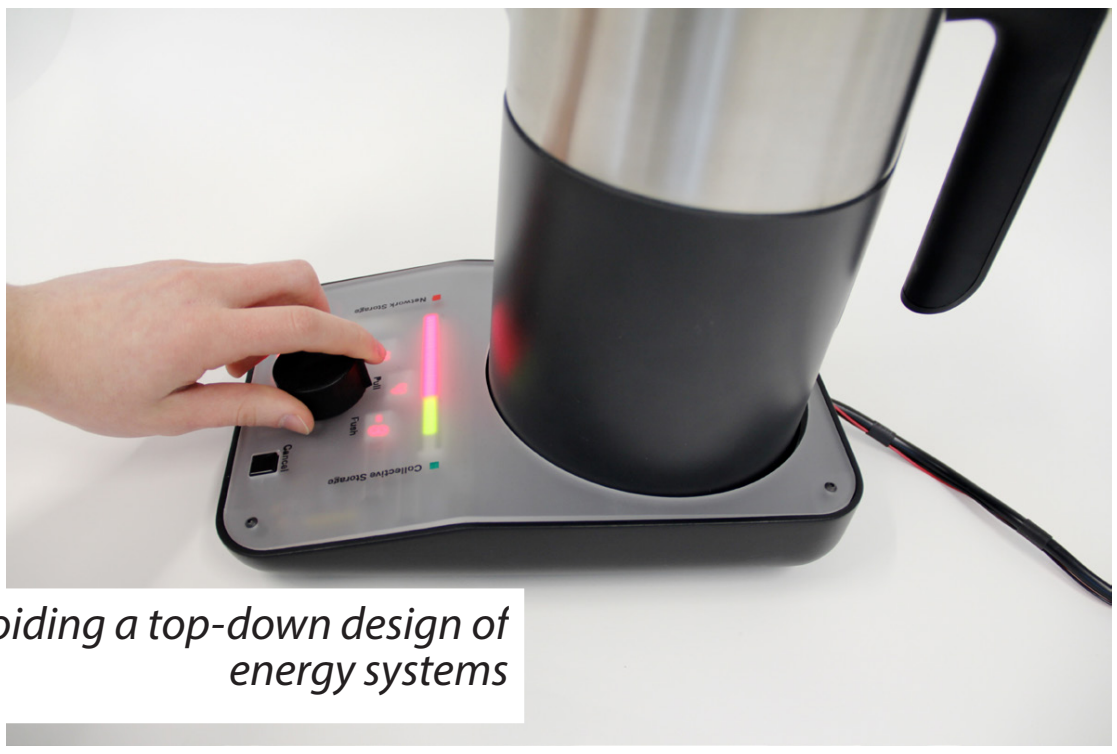

*Avoiding a top-down design of energy systems*

**Value Exchange**

**Tangibility**

**Rethinking Society**

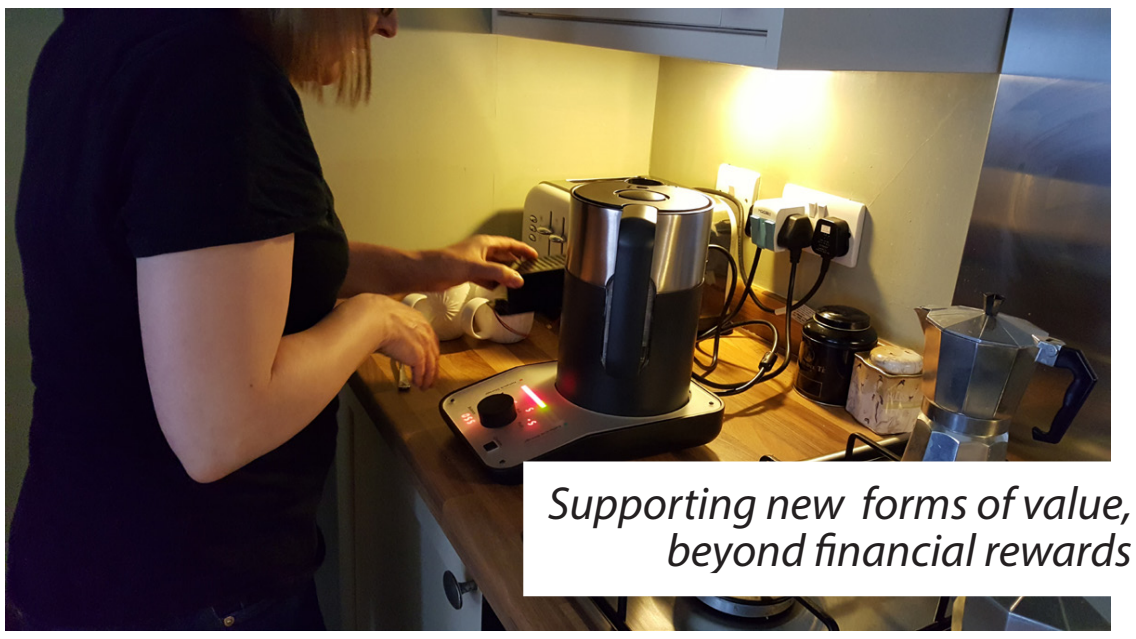

*Supporting new forms of value, beyond financial rewards*

User study: envisioning energy usage in a neighbourhood

*Helping participants consider energy requirements of a wider community.*

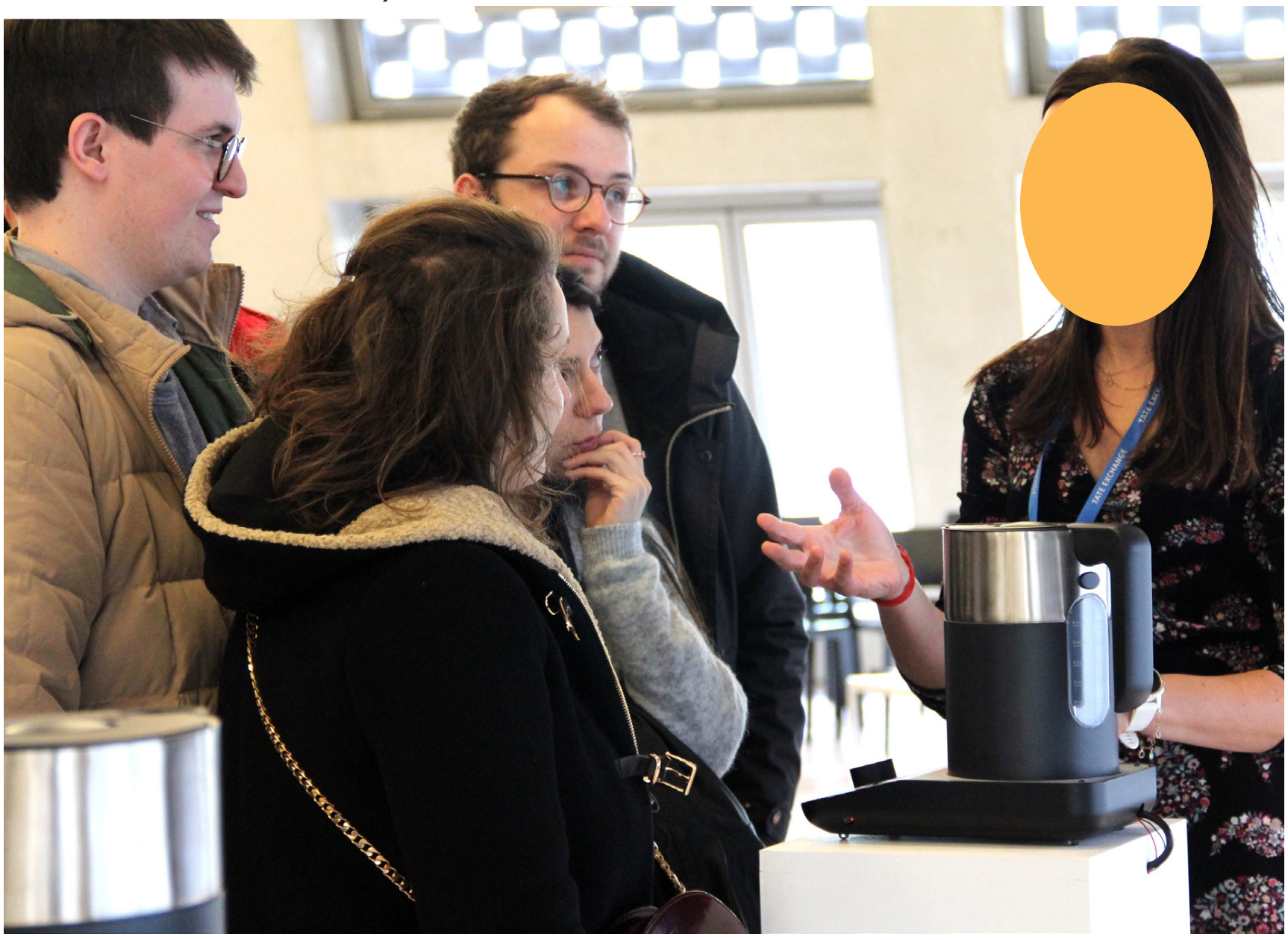

Karma Kettles exhibited at Tate Exchange. Image: Dan Weill

# *Karma Kettles - Details*

### Aims and Context

The Karma Kettles allow users to choose if they would like to take full control of energy transactions or delegate it to a predefined algorithm. If they decide to take control they can store the energy (pull), make it available to others (push), or use it to boil water. While distributed energy resources are often designed to operate without human intervention, the kettles allow people to experience a scenario where they have real-time information on energy availability (the state of the grid and their own energy resources) and can choose to have more control over energy transactions or delegating decisions and agency to algorithms that balance energy in the network.

They were designed to inspire discussion on levels of agency, and on how energy systems could be designed to promote more participatory approaches in distributed energy systems. They have been used to explore energy usage in different settings, including energy awareness games in exhibitions and museums, and in people's homes through a field study.

### Experience

The kettles are presented  as an interactive installation or through workshops. Users can take control, and the kettles allow them to a) store energy (pull), therefore contributing to increasing the amount of energy currently available in the regional storage, b) push it back into the grid, therefore making it available for others to use, or c) simply use the stored energy to boil water. Connected to each action is a certain number of karma points that indicate when an action benefits or hinders the health of the energy system – a scenario of high availability in the grid and low regional storage would give a high, positive score for those who decide to store energy, or a low, negative score for those who decide to push energy back into the grid. People who decide to use the engergy get a medium score which could be either slightly positive or slightly negative depending on how much energy there is in the grid.

The kettles included a BOT mode, which allowed people to shift agency to the system. The BOT mode intended to illustrate a scenario where the kettle pushes energy when the grid is critically low but still keeps some energy stored for usage in the device, therefore balancing the needs of the grid with the needs of the individual.

### 7.1 Common themes

The projects in this section focus more on the potential autonomous behaviours that objects can be programmed to enact and the meaning of such automated decision making by devices for everyday interactions, personal relationships and the way we see economic systems and businesses. These artefacts speculate about potential interactions with blockchain-enabled devices, services and infrastructures where the agency of artefacts and systems is often kept in the background of people's awareness.

As artefacts, they draw on the power of **tangibility** (Section 4.3) to support open investigation. They explore the balance of control between human and systems, where giving things more agency clashes with ideals of human freedom (Gigbliss), and when we humans become part of a distributed system (GeoPact). They look at what happens when value judgements are either automated or where seamfulness and resistance highlights them (Bitbarista).

Having touched on **technical fidelity** previously (more discussion in Section 4.5), several projects here build on and built in high fidelity systems which allowed for a closer investigation of how a system may interact with other infrastructures [138] or impact on peoples lives and behaviours [137]. Where GeoPact used real networks and blockchains, and Bitbarista had a functioning bitcoin wallet, GigBliss and Karma Kettles abstracted key qualities of the networks they described in support of creating experiences. GeoPact was structured as if it would be a 'real' technology and the investment in such a high fidelity system led to a range of follow up studies, allowing exploration of different aspects of the system and a range of reconfigurations.

### 7.2 Longer term engagements

Experiences of autonomous objects can take more time than is possible in a walk-up interaction – the understandings of autonomy come through as interactions unfold. This came through looking at relations to the Bitbarista – the initial study looked at buying a single coffee in lab conditions [118], while the follow up [136] looked at deployments in three offices, for a month each time. This highlighted the quotidian issues of maintenance once the technological excitement had subsided. It also showed how the machine disrupted existing rituals, particularly for those who used to gain social value from making coffee or looking after the equipment. The micro-payments offered to participants for small tasks were generally not enough to make up for the disruption, and it undermined a sense of altruism and community. The study of long term usage revealed a complex picture in which customers struggled to align the long-term thinking required when voting for future coffee supplies, with their immediate need for coffee. And although GeoPact wasn't employed in a longitudinal study, the potential of the 'real' infrastructure allowed for more real world testing and evaluation of smart contracts in the wild or outside the lab beyond the engagement of general public audiences.

### 7.3 Experiencing Value Flows

Building on previous themes on how values are encoded and explored through artefactys and systems, the autonomy of the objects in this section highlighted some of the ways that values flow and evolve as the systems are in use. With Bitbarista, whilst participants for the most part initially engaged with issues of ethics around their consumption, they tended to develop a strategy for voting rather than consider the issues afresh each time they bought a coffee, regardless of changing options. Interactions highlighted an implicit assumption that the Bitbarista's autonomy would make decisions in line with its ethos as an 'ethical' machine, providing benefits to themselves, the coffee growers and society more broadly. The GigBliss hairdryers provided a way to highlight competing values and interests in an experiential way, engaging the public with complex technology development and relevant policies through making everyday tradeoffs. Studies with professional bike couriers using the GeoPact system [137] showed how interaction with the technological system led to discussion of existing values, and how they were being negotiated and assessed in the face of increasing automation. A study with the Karma Kettles suggested that enabling people to make conscious decisions about when to pull, push, store and use energy can engage them in considering energy requirements of a wider community in a bottom-up rather than top-down manner.

### 7.4 Contextualisation

Contextualisation was key to several of the projects. Strongly grounded scenarios helped participants to engage the technological and the social at the same time, and being able to role-play through relatable scenarios helped with the experience. Working with actors and improvisation in GigBliss brought a high degree of reflexive contextualisation to bear, as they considered the way that the objects would fit into their lives. GeoPact we explored a multiple scenarios in various levels of detail: initially, the cryptographic architecture was presented in a simplified form, and then contextualised through a stylised supply chain where participants took on different role; later, high fidelity delivery infrastructure provided a real-world context for enactment with bike couriers. The more detailed contextualisations gave a clearer picture of how people would relate to the technology, navigating the space between previous work that attempted to communicate about the technology into using design as a way to map out unexpected technical and socio-political implications.

### 7.5 Humanising infrastructures

The GigBliss hairdryers were explored through methods of drama and deliberation. This provided a way to engage people in negotiating meaning, value and control in autonomous energy transactions. During workshops, participants discussed not only the implication of the different systems but also what would work for them and how they would be willing to compromise to balance concerns regarding lack of control and the convenience provided by autonomous systems. Part of this was supported through the ways that concepts were related to people. For example, with Karma Kettles, the creation of 'karma points' served as a reference to individual impact on the broader system. As well as humanising abstract concepts, it points out the need for careful consideration of reward mechanisms, considering gentle incentivisation without relying solely on financial motives, which may eclipse thinking around other forms of value. In GeoPact, by starting from a human concept of location, we could create ideas that are more generally accessible than a purely technological solution would be. Rather than being purely shaped by the kinds of location operations we can include in smart contracts, we could work towards the ways in which people experience location in their lives.

In each case, mapping the infrastructures into human experience allowed an exchange of ideas regarding what would work for them, and demonstrated the ways that deliberation can support critical design to achieve its societal aim.

### 7.6 Summary and Directions

Here the artefacts or things give us a different way to engage with people - tangible interactions can be immediate, and embedding things into daily life helps to understand the frictions, edges, specifics and so on of designing these systems. This is reflected in the associated study methods as well – the pieces support a longer term engagement, which helps to move beyond the 'wow' factor of speculative objects and into a more nuanced understanding of the technological propositions. When designing participatory work, having physical things can scaffold the kinds of engagement we are interested in - curious and engaged, connected without being buried under the details.

These projects all offer perspectives on one of the more interesting possibilities of DLTs - that anything can have a wallet, and hence the things we interact with can have a bit more agency and do a bit more without human intervention. They have helped publics to question how these agencies should be distributed and think through which decisions should be made by users and which by algorithms or other agencies. It is important to develop this broader picture of the ways people interact with autonomous things as giving the power to an object helps articulate the system's behaviours and choices, and make sense of the implications and wider effects.

## 8 DISCUSSION - COMMON THEMES

The core of this paper is a holistic presentation of a rich portfolio of related studio projects to "capture the family resemblances that exist" [13, 150] between them. The projects spanning more than half a decade, have various purposes, and touch on diverse aspects of blockchain technology. In presenting these projects based around their primary aims, we have already drawn out a number of key threads of our practice: approaches to supporting better understanding of blockchains, producing engaging public experiences, making new rules, and thinking through autonomous things.

In our discussion we take these analyses further, and indicate future directions for academic research and design. Our discussion begins pragmatically, reflecting core learning we have developed about how to design blockchain systems, before specifically discussing reflections on involving people in the design of blockchains. We then look more broadly to consider how designing with blockchains entails a more infrastructural turn for design in general. Finally, we offer some key trends and outstanding societal questions for blockchain applications.

### 8.1 Learnings: how to design blockchain systems

If there is in fact a single common theme in our projects here, it would be **the importance of situating and grounding blockchain technologies in real-world contexts of use**. While fascinated by the wide but abstract promises of decentralisation, autonomy and reprogramming economies, our primary concern was always to explore what this hyped technology could meaningfully *do* – whether to reconfigure collective coffee consumption, or program and automate new charitable donations. The lack of many significant mainstream blockchain applications highlights the challenges in working with blockchains in real-world contexts. Though often beginning with more abstract concepts such as 'escrow' or 'location-based smart contracts', to design with blockchains, we sought ways to articulate the broad promises of the technologies, making them concrete and digestible, preparing the ground for meaningful engagements.

In context, **it becomes clear that blockchains are not simply formalising abstract permissions, or even particular values or rights, but formalising relationships, between people and/or things**. This is most clear in a project like GeoPact, where the smart contracts governing the behaviour of a lockbox required defining, questioning and anticipating all of the relationships in a delivery service. The Seismic SeeSaws encode a very particular set of relations between charities, donors, beneficiaries and data providers; KASH Cups and the BitBarista produce and enforce new social relationships by constraining human actions around coffee production and drinking. Designing with and for these relationships requires a broader system-design perspective that places focus on the range of actors involved, and the relationships and trust between them. Attaching Strings offers one collaborative approach to doing this from a birds-eye view. Importantly, a focus on relationships beyond only values, demands attention to where power and control in such systems lie, whether with human or machine.

**A recurrent challenge in designing with blockchains concerns dealing with the rigid formality of computing systems**[36]. Blockchains are explicitly transactional protocols, designed to repeatedly and immutably follow particular rules. 'Smart' contracts are not in fact responsive, dynamic or adaptive as other 'smart' systems appear to be; instead they will sit inactive, until receiving specific inputs, and delivering very specific outputs. In this sense, blockchains starkly expose an inherent tension between the emergent, and unpredictable messiness of everyday life, and a pre-determined and idealised computer model. The various projects on 'Making the Rules' confronted participants with these tensions, and raised questions about how to practically manage the necessary inflexibility of blockchain technologies. It also raised questions about who 'commissions' such systems, and who is forced to use them. These systems could empower and enforce some individual users' very specific wishes or values, but viewed more broadly, they configure quite restrictive terms for complex interactions. The GigBliss hairdryers are particularly provocative, with one version completely dictating when people can dry their hair.

Such encodings aptly demonstrate how **applied, design-led approaches are able to highlight the limitations of blockchain applications** and bring into question when, how and where blockchain technologies might offer value. In addition to surfacing the inflexibility of smart contracts and transactional protocols, we encountered difficulties in managing temporality; people and contexts change, but contracts don't. In essence, **our various design projects have exposed in different ways how difficult it is to connect blockchains to the world in which they are situated**. Firstly in terms of what a blockchain understands as truths about the world: has an earthquake taken place? Where exactly is this lock box? Have the farmers made a profit yet? Several of the workshop approaches (IFTTW, Programmable Donations) highlight gaps between the intentions of the parties and what can actually be encoded [1, 104]. Secondly, blockchains can be disconnected in the ways users and stakeholders are able to meaningfully interact with them. **A crucial design question remains around how explicitly and to what level of detail it is necessary to expose the underlying functions of a blockchain to various groups of users**. Many of the promises of the technology hinge upon the potential for radical transparency and the possibility to verify transactions taking place; but the workings of this infrastructure are inevitably complex, and require trust in additional intermediaries. As research projects, we have tended to over-expose particular aspects of the technology, in order to direct participants attention and elicit their reflections.

In doing so however, we have identified the importance of a careful balance between seamless operation and integration [148] and seamfulness [20, 21] that exposes the workings and individuality of systems. Seams prompt questions and reflections from users. The moment in AfterMoney where you are asked to decide what currency to pay in prompts a discussion and speculation about currencies and their value. How does one weigh up a loss of data privacy versus spending a few minutes sweeping a floor? The BitBarista asks you to decide between the cheap but decent coffee, or the more expensive but more ethical beans, or when you have to rapidly rethink the value of imaginary resources in BlockExchange.

Some seams are a necessary component of engaging with the mess inherent in complex systems [6]. When GeoCoin does not immediately show transactions, or allows participants with newer phones to collect more currency, it highlights a messiness in access; when messages take a long time to reach a GeoPact box, visitors start to wonder what happens when the system goes wrong. Seams based on prompting decisions, highlighting mechanisms and exposing mess are useful ways to prompt reflection, and ultimately trust. However, they all require labour of some sort on behalf of those interacting with the system. Over longer term, everyday interactions; such as when the Bitbarista was situated in offices for extended periods of time [136], some users became frustrated with the neediness of the machine and its protocols, and became disengaged. Carefully balancing the seamfulness of blockchain technologies to properly expose their function, while remaining accessible and easy to use is hence a key design challenge.

### 8.2 Learnings: Involving people in the design of blockchains

An evident element of this portfolio of projects is to involve people in the design of blockchains; there are several reflections we can offer around how to do this. Firstly, it is clear that the inherent complexity of the technology can immediately be excluding. In many of these projects, we've therefore often **introduced what a blockchain does rather than what it is**. The Block-Exchange workshop begins as a trading game, gradually introducing features and characteristics of blockchains. Programmable Donations focuses participants attention on the qualities of conditionality, rather than the specifics of a 'financial escrow'. Happily Ever After (Bitcoin) starts from a traditional legal contract, before introducing 'smart contracts' as short-term partnerships between strangers. This project also reflects the value of **setting a familiar, quotidian and mundane context for blockchain technologies**. This offers familiar hooks and routines that are easily grasped by participants (having a coffee, taking a delivery, boiling a kettle, and drying your hair) but it also speaks to the infrastructural ambitions of blockchain technologies. By situating these technologies in such contexts, we're able to work with participants to understand the extent to which revolutionary ambitions of blockchains stand up to common needs and work. We've also found that **presenting examples of value exchange provide a human way to make sense of blockchains**. Transactions in After Money present opportunities to 'swap' one form of value – data, cryptocurrency, labour – for another, sweeties. BlockExchange begins as a trading game, where participants exchange various resources with the currency of lego blocks, before being prompted to understand how they might exchange the things they value without traditional forms of money. In our experience, we have found focusing on fundamental ways in which we exchange value, opens the doors for many participants to think very broadly about the potential of blockchains and distributed ledger technologies.

To develop an understanding of specific qualities or features of these technologies, **we have found roleplay and collaboration to be particularly effective approaches**. Playing through abstracted representations of technology lets participants build understandings of the concepts collaboratively and incrementally. Simple interactions provide a gateway to considering the possibilities of Blockchain systems whether exchanging a resource card for some Lego bricks, enacting a marriage or recording a new pizza making skill with a unique set of stickers. By focusing first on fulfilling a simple co-operative role, participants can feel their way into a networked experience.

Playing through a version of a system also provides an experiential approach, enabling participants to imagine interactions as part of their own lives and consider possible real moments, frictions and transitions. Collaboration enables participants to develop an understanding of the dynamics and processes while discussing and talking with others who are also developing this understanding. Projects like PizzaBlock also viscerally demonstrate the amount of labour and computation required for a distributed network to function. Several projects place an emphasis on the different roles and relationships within these emerging systems (Attaching Strings, GeoCoin, GeoPact), exploring how the different roles may play out and change existing social enterprises, leading to consideration of alternative forms of governance. Formal technical concepts quickly become humanised for participants through encountering their social implications.

Often to support roleplay, the projects in this portfolio frequently also **rely on designing carefully with regard to visibility and transparency**. Some projects were rigorously transparent – every person's credentials in PizzaBlock can be seen hanging on the line, although their personal identities are pseudonymised. Bitbarista reveals data from coffee production to consumers before they buy a cup of coffee. The Seismic SeeSaw visually demonstrates the money being held in escrow, along with the mechanism that will release it – although not the data that causes the unlocking. When people get married Happily Ever After (Bitcoin), there is a paper record for them to take away, and a public record visible for others.

Sometimes information was hidden or obfuscated either purposefully or accidentally from participants. Like many distributed ledgers, it is hard to look at the structure resulting from the 'mining' process in BlockExchange and understand what has gone into it. KASH Kups don't have active electronics in them, so they can't display credit levels on the cups – users often only discover that they are out of credit through the embarrassing moment of finding they can't get coffee. In each case - we sought to balance our aims to produce meaningful experiences, educate or engage participants around a particular feature of the technology, while ensuring an appropriate level of technical fidelity. For example GeoPact attempts to show the state of every object in the system, the current and future status of a smart contract, and a blockchain log of each event that happens. This level of fidelity produces a lot of information, but for the in-depth workshops with engaged stakeholders, it was important to be able to capture these different views on the system, and match up the informational perspective with the physical actions taking place in the world.

We often **made concepts or features of the technology visible to participants through making them tangible**. Tangible materials have been widely employed within participatory design practices to support participants in externalising their thinking [98] and distill complex concepts into simple, manageable forms. In these projects, materials work to produce abstractions that focus the participants – and researchers – on specific aspects of the technology and enable them to explore these in experiential activities. Often the materials used in the interactions create a common language, which has the benefit of leveling the playing field and enabling participants from different disciplines and levels of technological competency to work together. Several projects used objects that feel familiar as a way to introduce new concepts: the GigBliss hairdryers, the broom used in AfterMoney and the Bitbarista start from everyday objects and then introduce seams that disrupt the natural responses to handling those objects.

Tangible experiences are consistently powerful. Even members of the public who already had read a lot around cryptocurrency found things in our experience that helped them understand blockchain issues that they had previously ignored. As designers, producing highly-finished tangible objects also forced us to resolve ambiguities in our own understanding of the technology, or to bring focus to the concepts we wanted to play out with participants.

Whatever particular methods and approaches we took to engage publics with blockchain technologies, **we frequently found ourselves choosing to focus on specific qualities of the technology while abstracting others**. The Attaching Strings DAO workshops used very high level abstractions of the idea of a Distributed Autonomous Organisation without worrying about the details; the PizzaBlock workshops used technically grounded abstractions, but without delving into cryptography. The Seismic Seesaw focused strongly on the concept of an automated escrow, and was not required to run on a blockchain to discuss this with participants. With complex systems such as blockchains, there are many aspects that need to be understood in order to have a comprehensive technical discussion; at the same time, it is possible to have a well rounded engagement with implications, without unravelling the entire tapestry. This is often easier to do with an object, rather than a question – the object articulates a particular position and presents a coherent portion of the system which can be digested, and gives space for further debate.

### 8.3 Learnings: From designing blockchains to designing for algorithmic networked infrastructures

While much of the weight of blockchain research has historically been concerned with their functioning around cryptocurrencies, the work in this portfolio can also be extended to the more general question of **how do we do design in relation to networked digital infrastructures?**. These are complex situations, typically consisting of multiple technologies, concepts and ideologies that come together to support a range of overlapping activities. As such, they provide a particular challenge for design: much of the important functioning of such systems is encoded in algorithmic and network protocols, which are often complex, technologically opaque and counter-intuitive. There are often multiple concepts that need to be grasped before the system as a whole makes sense, and multiple viewpoints that have to be considered - designers, users, commissioners and so on. Views of a network are often partial and asymmetric and not nearly as stable as when considering the design of a physical product or interface. The interactions in GeoPact, for example, highlighted what happens when different networks rub up against each other – the formalised blockchain contracts with the more worldly IoT logistics infrastructure and the more socio-physical road and transport networks. Whether for low-level blockchain consensus or higher level smart contract design, looking at the algorithmic mediation of society throws up challenging questions about what algorithms are and how they relate to people. An algorithm can be seen as one of Barad's *agential cuts*[3], as being a device by which properties of interest can be constructed, yet at the same time as being so buried in details of implementation,materiality, and wider systems, that it can be hard to point to any part and say 'this is the algorithm that caused this' [32, 84].

Two threads of work come together here, as we look at algorithmic infrastructures for human interaction, around the term *institution*[126]. Within participatory and co-design, *insitutioning* has been highlighted as a move towards designing the institutions that create contexts for action [53, 73, 139]. This shift of attention from the micro-level of interactions to the meso-level of the sociopolitical context in which they happen is a part of several of the projects: Happily Ever After (Bitcoin) grew out of ideation around the social implications of designing a new form of currency. Here the introduction of blockchain functioned as a disruptor, that **created a space for critiquing social rules and norms, along with a handy tool for sketching new possibilities**.

In parallel, there is a thread of work within computer science around *electronic institutions* [31, 46, 47], as attempts to algorithmically formalise the rules and norms by which interactions take place, in order to create a basis for societies of artificial intelligent agents. Just as with blockchain systems, work continues on how to connect these kinds of formalisations of interaction within the messiness of human life [106], and how to open up the design of formal systems to a wider audience [105], in particular through the creation of 'social machines', in which network infrastructures support open and creative human interaction [112, 128, 129]. Two the key challenges these projects have addressed is **how to form and communicate connections between formal systems and the world**, as well as **how to support users to the point where they can meaningfully design rules for formal systems**. This difficulty of forming connections between formal systems and the world, and of getting users to the point where they can meaningfully design rules is the heart of many of the above projects, e.g. Programmable Donations and GeoPact. This can be seen in the rule creation of Programmable Donations, the ideation in GeoCoin and contract building in GeoPact and the explicit development of new kinds of institutions in Attaching Strings.

A challenge with network infrastructures is understanding how the micro-interactions relate to macro-scale properties. This has been part of the design of blockchain systems from the start – transforming individual needs around transactions through mining incentives into system-wide properties of trust and consensus. The **use of designerly strategies and provocative interventions helps to mediate between micro and macro scales** – the BitBarista uses microboundaries in the human experience of coffee making to connect to global supply chains, GigBliss uses discussion of living with speculative items as a connection to the wider world of smart energy grids, and GeoPact uses moments of exchange of physical goods to articulate algorithmic ideas of responsibility. This echoes Rosenberger's move to amalgamate a detailed, interactional viewpoint of individual experience with the connected, entwined view of the networks and constellations provided by Actor Network Theory [124]. There is also a move towards understanding the *multistability* of these emerging technologies, as their meanings and possibilities are constructed towards different purposes. Carrying out ideation workshops around given concepts or infrastructures is a form of Ihde's *variational analysis*[75], where the various possibilities of a technology are explored as it is brought to bear on different situations and contexts. Beyond this, **designing interactions and institutions used by different stakeholders explores the *multi-intentionality* of these network technologies**, as alternative desires, conceptualisations and behaviours come into tension, alignment and entanglement [121, 149].

### 8.4 Learnings: Agency and values in distributed ledger technologies

Lessons learnt from our projects indicate two key considerations for the design of distributed ledger technologies such as blockchains: the first is to consider agency beyond the human, and the second to situate transactions within a larger system of values.

Our projects indicate ways in which we can include people in discussions around the design of systems that have some level of autonomy. The idea that **non-human things can have wallets, and hence some form of enhanced transactional power**, gives them an obvious sense of autonomy which makes it easier to discuss implications of their participation in socio-technical systems and beyond. What roles would we like them to play in these systems? How neutral would these roles be? What are the implications of giving them more or less autonomy? Could we shift roles and regain agency when we wish? This has been grounds for thinking through ideas such as coffee machines that own and look after themselves (Bitbarista) or hairdryers and kettles that can trade energy (GigBliss and Karma Kettles). Such systems make visible that agency could be distributed among other actors or aspects of the environment, and the way it is distributed is a design choice. GeoCoin explored the idea that currency could be located in space. GeoPact develops this idea into systems of objects that can both direct humans and pay them for their services. All of these point towards questions of **what happens when we de-centre human agency** [26], and take a wider view of the implications of what we are designing . The objects here have been purposely designed to embody a specific complexity (e.g. around coffee production, energy, distribution of goods, etc) but they can also serve as starting points to question of what happens when objects are *fluid assemblages* [62, 121], becoming portals to a much larger complexity which constantly grows and changes with time. In these situations, the object itself may not be as important as the services, systems and actors that it mediates. A second important point of discussion regards the values behind these systems. Which values are they mediating? How flexible is this mediation? **Redesigning the way value and capital are represented opens up space for different values to be shared and performed**, exemplified by the radical imaginaries of Bitcoin and cryptocurrencies [17, 71, 135]. While designers have often been concerned with how values are embedded within certain technologies and systems [e.g. 12, 27, 56] our work here highlights the intriguing opportunities to redesign systems of value exchange themselves. Exploring how economy and values collide in social interactions can be uncomfortable terrain for designers [43, 130], but these projects illustrate how fertile a design space there is in questioning and redesigning representations of value and creative transactions [41].

Many of these systems change the relationship between people, labour and value, whether asking people to carry out tasks (GeoPact, Bitbarista, AfterMoney) or demonstrating a worker's value by recording their history (PizzaBlock). Several of the systems replace human action with contracts and scripts, and we highlight the critical questions of who programmed it? Who is controlling it? What are their values and can they change over time? What are the implications of contributing to the creation of these *algorithmic cultures* [132] for wider society? By making the possibilities of decentralisation and disintermediation vividly apparent, the projects here create a space to surface critical issues and potentially rethink assumptions behind these relationships. By exploring the imaginaries behind these systems and their effects on society, these projects shift the building blocks from which we make our possible worlds. By giving people the tools to engage with emerging infrastructures, we hope that this combination of design and HCI supports the scaling up of action in a changing world [54, 91] and a move from simply considering users to the consideration of people and communities acting towards a shared system of values and citizenship [52].

## 9 CONCLUSIONS

This annotated portfolio has described a collection of projects that use design led approaches to question and engage around blockchains and related systems. Dealing with these complex systems – that combine cryptography and distributed computing with questions of trust, economics and governance – requires multiple viewpoints, and we hope the value of bringing together these varied perspectives and approaches is clear.

Algorithms play an increasing role in daily life - social, economic and political; there are pulls towards both increasing autonomy and decentralisation, and towards centralisation and concentration of power. Blockchains, distributed ledgers and smart contracts are part of this transformation, not only in terms of decentralised finance and economies, but also towards social and organisational change. The projects showcased here are ways to think through the issues involved, for researchers, engaged stakeholders and for a broad, interested public. They look at how blockchains are linked to the physical world by building functionality into artefacts; they look at the interactional challenges of using blockchains; they look at the social embeddings, and how distributed ledgers or smart contracts affect existing human relationships and practices; they look at how we can navigate the hype around emerging technology, and communicate a sense of possibilities and limitations.

This collection has worked through a range of projects that engage the public, develop understanding and explore future possibilities. It illustrates how design and HCI methods can be applied to complex network infrastructures in service of speculation within and beyond the technology in detailed workshops, rapid public experiences and through developing speculative objects and systems. By reading across the various projects, we have articulated approaches to design blockchain systems, as relationships and values grounded in use, rather than abstract formal structures. We have highlighted how connecting the new technology to the everyday and quotidian – through roleplay and physical artefacts – offers a way to engage publics in otherwise impenetrable spaces. The variety of projects, and in particular bringing together ideation with technical provocations has articulated a connection to innovation, infrastructuring and institutioning based on the possibilities of distributed ledger technology, and opened up a view of the algorithmic cultures they may create.

Taken together, we hope that this collection projects offers a provocative yet grounded exploration of the possible spaces for blockchains in the coming years, for HCI and design researchers in particular. Using experiential methods, defamiliarisation and role play alongside technical grounding and research through design, we hope to open up new ways to think about the future, creating new imaginaries around distributed ledger technologies.

## ACKNOWLEDGMENTS

This is a short text to acknowledge the contributions of specific colleagues, institutions, or agencies that aided the efforts of the authors.

## 10 BLOCKCHAIN OVERVIEW

This appendix provides an extremely brief overview of blockchain technology to help a reader make sense of the technical aspects of this paper.

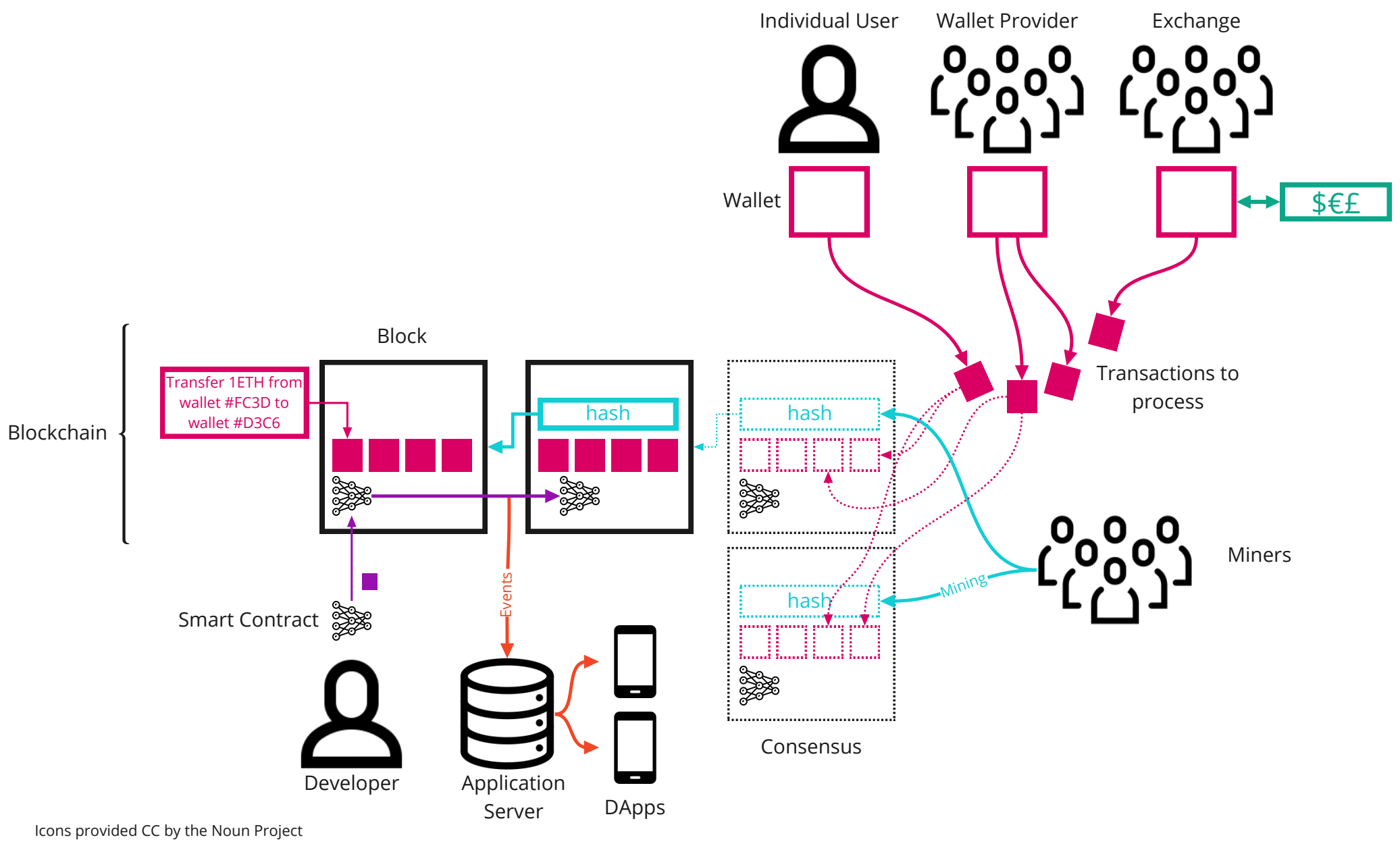

Fig. 2. Simple overview of the Ethereum blockchain. Transactions represent transfers of cryptocurrency from one wallet to another, identified by unique IDs. Transactions are combined to form Blocks, and each Block has a cryptographic hash of the previous block in the chain - this forms the Blockchain, and provides proof against tampering with historic data. Transactions may be created by individuals, who control their own wallets, or for most users by organisations that manage multiple wallets on their behalf and allow exchange with traditional currencies. Miners attempt to create the next block out of the pending transactions, typically for some reward in cryptocurrency. This is governed by a distributed consensus algorithm that decides who gets to make the next block - in current Ethereum, this is based on 'proof of work', where miners compete to solve the arbitrary difficulty problem of finding a 'magic number' that means the output of the hash function is in a certain range. In the bottom of the diagram, developers create Smart Contracts, and deploy them to the blockchain, typically with payment in a secondary currency ("Gas" in the case of Ethereum). The Smart Contracts respond to events on the chain or ones fed in by connected applications, and changes to their state are recorded on each block. Distributed applications (DApps) allow users to interact with these smart contracts.

Table 1. Glossary of terms used in the paper related to cryptocurrency, blockchain and distributed ledgers.

- **Blockchain** A data structure that provides a tamper-proof record of data in a decentralised manner, by cryptographically linking new data to the existing record. As new blocks of data are added to the front of the chain, it becomes increasingly difficult to alter any of the previous blocks.
- **DLT** Distributed Ledger Technologies. A ledger is a way to record transactions so that they cannot later be altered - for example, an accounting book, where each transaction is recorded along with the new account balance, so that a historic transaction could not be altered without changing every subsequent transaction. Distributed ledgers do this in a decentralised manner, so that many people can verify correct operation and add data to the ledger. Blockchains are a common implementation of this concept.
- **Smart Contract** A way to write code whose execution can be independently verified, allowing it to have financial actions. Typically, they are executed on a blockchain or other DLT.
- **Bitcoin** A cryptocurrency, based on the Bitcoin blockchain. Probably the best known cryptocurrency, although it has technical limitations compared to subsequent projects: limited possibility for smart contracts and a long time between each block.
- **Ethereum** A cryptocurrency, based on a blockchain. More modern than Bitcoin, it allows for smart contracts, written in Solidity or other high level languages. It is popular with developers, as it is easy to run a private, local blockchain with the same technology, and development is relatively rapid.
- **Fiat Currency** Most standard currencies at present. A currency used as money that is not directly convertible into anything else (i.e. no direct equivalence to gold) that gains its value from exchange and taxation.
- **Altcoin** An alternative currency, often used to refer to any cryptocurrency except bit coin, and particularly for currencies with specific purposes or affordances.
- **ICO** Similar to floating a company on the stock exchange, in an Initial Coin Offering, some amount of a cryptocurrency is made available, in exchange for either fiat currency or other, more established currencies.
- **Escrow** A financial structure where money is held in trust until certain conditions are met - for example, a payment may be held in escrow until goods are delivered. This is designed to ensure that the buyer cannot avoid paying, and the seller cannot get paid without delivering.
- **On-chain and Off-chain** When transactions on a blockchain are slow or costly, 'off-chain' solutions can be used, where exchanges are made rapidly and cheaply in another network, and occasionally written back to the main blockchain. For example, Bitcoin transactions take at least 10 minutes to be visible, longer to be sure of, and cost a significant fraction of a dollar to carry out. The Lightning Network is intended to run alongside the Bitcoin blockchain, rapidly and cheaply carrying out transactions, and only committing them to the blockchain when necessary.
- **Attestation** An assurance that a claim is true, for example attesting that a person has a particular credential or certification
- **Mining** The process of creating new blocks on a blockchain by a) collecting pending transactions b) satisfying a cryptographic function that requires randomly guessing from a huge number of possibilities. This is a known as 'proof of work' - it means that each person's chance of creating a new block is dependent on the computational power and energy that they use. Creating a new block typically gives the creator an amount of cryptocurrency, so there is a correlation between energy input and value.
- **Self Sovereign Identity** A form of identity that does not require a centralised authority, often backed by a blockchain.